\documentclass[pdflatex,sn-mathphys-num]{sn-jnl}

\usepackage{amssymb,amsfonts}%
\usepackage{amsthm}%
\usepackage{mathrsfs}%
\usepackage[title]{appendix}%
\usepackage{textcomp}%
\usepackage{manyfoot}%
\usepackage{booktabs}%
\usepackage{algorithm}%
\usepackage{algorithmicx}%
\usepackage{algpseudocode}%

\theoremstyle{thmstyleone}%

\theoremstyle{thmstyletwo}%

\theoremstyle{thmstylethree}%

\usepackage{siunitx}
\usepackage{amsmath} 
\usepackage{bm}
\usepackage{geometry}
\usepackage[x11names]{xcolor}

\usepackage{graphicx}
\graphicspath{ {DGraphics/} }
\usepackage[outdir=DGraphics/]{epstopdf}
\usepackage{caption}
\DeclareCaptionLabelSeparator{bar}{ | }
\DeclareCaptionLabelSeparator{dot}{\bf{.} }
\usepackage[font={large}]{subfig} 
\usepackage{wrapfig}
\usepackage[export]{adjustbox}
\usepackage{afterpage}

\usepackage{relsize}
\usepackage{multirow}
\usepackage{dcolumn}
\usepackage{hhline}

\usepackage{tikz}
\usetikzlibrary{shapes,arrows}
\usetikzlibrary{decorations.pathreplacing}

\tikzstyle{io} = [trapezium, trapezium left angle=70, trapezium right angle=110, minimum height=0.6cm, text badly centered, draw, fill=red!20]
\tikzstyle{decision} = [diamond, draw, fill=blue!20, text width=4.5em, aspect=2, text badly centered, node distance=3cm, inner sep=0pt]
\tikzstyle{block} = [rectangle, draw, fill=green!20, text width=5em, text centered, rounded corners, minimum height=4em]
\tikzstyle{section} = [rectangle, draw, fill=red!40, text width=5em, text width=8em, text centered, rounded corners, minimum height=4em, minimum width=6em]
\tikzstyle{line} = [draw, -latex', line width=1.5pt] 
\tikzstyle{cloud} = [draw, ellipse,fill=red!20, text centered]

\newcommand{\Figref}[2][Fig.~]{#1\ref{#2}}
\newcommand{\figref}[2][Fig.~]{#1\ref{#2}}
\newcommand{\tabref}[2][Table~]{#1\ref{#2}}
\newcommand{\secref}[2][Section~]{#1\ref{#2}}
\renewcommand{\eqref}[2][Equation~]{#1\ref{#2}}

\newcommand{\suppcol}[1]{{\color{black}{#1}}}

\renewcommand{\|}{$|$}
\newcommand{\code}[2][]{\texttt{#2}}

\newcommand*{\citen}[1]{%
  \begingroup
    \romannumeral-`\x 
    \setcitestyle{numbers}%
    \cite{#1}%
  \endgroup   
}

\usepackage{soul}

\begin{document}

\title[Article Title]{RAFFLE: Active learning accelerated interface structure prediction}

\author*[1]{\fnm{Ned Thaddeus} \sur{Taylor}}\email{n.t.taylor@exeter.ac.uk}

\author[2]{\fnm{Joe} \sur{Pitfield}}

\author[1]{\fnm{Francis Huw} \sur{Davies}}

\author*[1]{\fnm{Steven Paul} \sur{Hepplestone}}\email{s.p.hepplestone@exeter.ac.uk}

\affil*[1]{\orgdiv{Department of Physics and Astronomy}, \orgname{University of Exeter}, \orgaddress{\street{Stocker Road}, \city{Exeter}, \postcode{EX4 4QL} \country{United Kingdom}}}

\affil[2]{\orgdiv{Center for Interstellar Catalysis, Department of Physics and Astronomy}, \orgname{Aarhus University}, \orgaddress{\postcode{DK-8000}, \city{Aarhus C}, \country{Denmark}}}

\abstract{%
Interfaces between materials play a crucial role in the performance of most devices.
However, predicting the structure of a material interface is computationally demanding due to the vast configuration space, which requires evaluating an unfeasibly large number of highly complex structures.
We introduce RAFFLE, a software package designed to efficiently explore low-energy interface configurations between any two crystals.
RAFFLE leverages physical insights and genetic algorithms to intelligently sample the configuration space, using dynamically evolving 2-, 3-, and 4-body distribution functions as generalised structural descriptors.
These descriptors are iteratively updated through active learning, which inform atom placement strategies.
RAFFLE’s effectiveness is demonstrated across a diverse set of systems, including bulk materials, intercalation structures, and interfaces.
When tested on bulk aluminium and MoS$_{2}$, it successfully identifies known ground-state and high-pressure phases.
Applied to intercalation systems, it predicts stable intercalant phases.
For Si\|Ge interfaces, RAFFLE identifies intermixing as a strain compensation mechanism, generating reconstructions that are more stable than abrupt interfaces.
By accelerating interface structure prediction, RAFFLE offers a powerful tool for materials discovery, enabling efficient exploration of complex configuration spaces.
}


\maketitle

\section{Introduction}
\label{sec:intro}

Modern devices, such as semiconductor transistors, lithium-ion batteries, and solar cells, rely heavily on the unique physics of their constituent interfaces~\cite{Hwang2012EmergentPhenomenaOxide,Ince2023OverviewEmergingHybridComposite,Wang2021DeterminationEmbeddedElectronic} to either enhance or suppress carrier flow, through means such as scattering~\cite{AlAmiery2024InterfacialEngineeringAdvanced,Schmidt2019RecentAdvancesApplications,Kanevce2017RolesCarrierConcentration} or create local field effects~\cite{Davies2021BandAlignmentTransition,Nangoi2024FirstPrinciplesStudies}.
Despite their importance, device research and development tends to prioritise the bulk properties of materials over the interfaces they form.
Thus, interfaces remain a relative unknown in comparison.
This presents a significant challenge, as many properties that dictate the capabilities and limitations of a device are derived from the physics at their boundaries.

Bulk materials have been the main focus of device optimisation until now due to their relative ease of modelling, which often exhibit high symmetry and can be represented by small unit cells within a periodic space.
In contrast, interfaces are large, disordered systems with comparatively low symmetry, making them computationally expensive to model.
Furthermore, interfaces are not merely simple connections between materials, but are complex regions that often undergo significant atomic reconstruction due to internal strains~\cite{Martin2024AtomicReconstructionAu}.
Accurately capturing these effects requires structure search methods to identify the interface phase when joining two materials.
Before materials modelling (and especially that of interfaces) was possible, experimental observation was the only method to understanding them~\cite{Smith1971MethodReconstructionInterface}; but this is still fraught with difficulty due to the challenges in reliably synthesising and characterising such small and messy systems as interfaces.

The search for new molecular, bulk, and amorphous material phases has been a central research focus for decades~\cite{Oganov2019StructurePredictionDrives}.
In the 1970s and 80s, the development of empirical potentials enabled accurate modelling of experimentally observed material structures~\cite{Muser2022InteratomicPotentialsAchievements}.
When combined with techniques such as random atomic placement and genetic algorithms, these potentials facilitated the exploration of chemical energetic landscapes~\cite{Woodley2004PredictionCrystalStructures}.  
Advancements in computational power and the accessibility of density functional theory (DFT) have since led to significant improvements in structure prediction~\cite{Oganov2019StructurePredictionDrives,Schusteritsch2015FirstPrinciplesStructure,Putungan2016MetallicVS2Monolayer}.
Structure prediction has been applied to interfaces in recent years, but only to A\|A interfaces~\cite{Schusteritsch2014PredictingInterfaceStructures} (i.e. grain boundaries between two crystals of the same material), but more complex A\|B interfaces are still fraught with difficulties.
The recent rise of machine-learned potentials, including Gaussian process regression and neural networks~\cite{Butler2018MachineLearningMolecular,Behler2007GeneralizedNeuralNetwork}, has further reduced the cost of random structure search by improving predictive accuracy and enabling a more thorough exploration of high-dimensional potential energy surfaces.
These approaches are now being extended to interface structure prediction~\cite{Oganov2019StructurePredictionDrives,Kiyohara2016PredictionInterfaceStructures,Bisbo2020EfficientGlobalStructure,Pitfield2024PredictingPhaseStability}.

While tools such as AIRSS~\cite{Pickard2011AbInitioRandom}, CALYPSO~\cite{Wang2012CALYPSOMethodCrystal}, and USPEX~\cite{Oganov2006CrystalStructurePrediction} facilitate random structure searches for bulk materials,
and machine learning-enhanced structure searches, such as GOFEE ~\cite{Bisbo2022GlobalOptimizationAtomic} and BEACON~\cite{Kaappa2021GlobalOptimizationAtomic}, enable searches spanning the space of surface reconstructions and clusters, accurately and efficiently predicting interface structures remains a challenging task.

Where the complexity of perfect crystal searches is reduced by symmetries, and surface reconstructions by size and starting setup, interfaces are afforded none of these conveniences.
Factors including surface termination, intermixing, growth conditions, local strains, rogue particles, and defects all influence the interface between two materials~\cite{Taylor2020ARTEMISAbInitio}.
Additional complications arise from lattice mismatches and the large size of unit cells.

In increasing cases, machine learning techniques are mimicking the accuracy of first-principles methods whilst maintaining similar speed of empirical approaches~\cite{Butler2018MachineLearningMolecular}.
In structure prediction, machine learning typically uses the chemical structure as an input to predict properties such as energy or forces.
Applying machine learning within materials science requires a set of descriptors (also known as fingerprints and representations) that are invariant (or equivariant) under symmetry transformations.
In practice, uniform descriptors are spatially invariant, ensuring that equivalent systems are represented consistently.
Spherically invariant bond length distribution functions, which highlight interatomic distances, often serve as effective descriptors.
Higher-order n-body descriptors, such as those capturing bond angles (also known as Keating functions)~\cite{Keating1966RelationshipBetweenMacroscopic} and dihedral angles, are also commonly employed~\cite{Gale2005GULPCapabilitiesProspects}.
Recently, equivariant descriptors have gained popularity for their success in predicting material properties, though they too face limitations similar to invariant descriptors~\cite{Schutt2021EquivariantMessagePassing}.

In this article, we introduce the RAFFLE software package, a large-scale method for predicting the structures of interface materials.
This represents a formal implementation of a previously proposed methodology discussed by the authors~\cite{Pitfield2024PredictingPhaseStability}.
By providing an initial database of structures and corresponding energies, 2-, 3-, and 4-body distribution functions ($n$-body contributions to the generalised descriptor) are generated.
Key structural features are identified in the descriptors that correspond to improved energetic favourability, which are used to inform the placement of atoms within a host structure (i.e. the two bulk materials separated by a certain amount).
This allows for an efficient sampling of configuration space, guided by knowledge of local environments from existing energetically favourable structures.
A set of examples are outlined to highlight capabilities and robustness of the package, including one on Si\|Ge interfaces, highlighting how RAFFLE can be used to identify interfaces more energetically favourable than traditionally abrupt interfaces.
The RAFFLE software makes interface structure prediction more accessible and faster through automated generation of structures, energy-guided predictions, and reducing human bias.
Benchmarks are presented to show scaling with parameter choice, whilst example cases are presented to highlight reliability and features in known sampling spaces.

\section{Results}
\label{sec:results}

\subsection{Overview}
\label{sec:overview}

\begin{figure}[h]
    \renewcommand{\baselinestretch}{1.0}
    \centering
    \resizebox{0.5\linewidth}{!}{%
        \begin{tikzpicture}[node distance = 2.5cm, auto]
            \node [io] (host) {Host structure};
            
            \node [block, text width=2.5cm, below of=host, yshift=0.8cm] (populate) {Place atoms in host structure};
            \node [io, text width=2.2cm, xshift =0.8cm, right of=populate] (stoichiometry) {Placement\\ stoichiometry};
            \node [block, text width=2cm, below of=populate, yshift=0.5cm] (relax) {Structural relaxation};
            \node [block, text width=2.5cm, below of=relax, yshift=0.5cm] (rank) {Rank systems};

            \node [block, text width=2.25cm, xshift =-0.8cm, left of=populate] (descriptor) {Learn\\ generalised\\ descriptor};
            \node [block, text width=2.5cm, xshift =-0.8cm, left of=rank] (df) {Generate\\ distribution\\ functions\\ for each system};
            \node [io, text width=1.5cm, xshift=-0.8cm, left of=df] (profiles) {Materials database};
            
            \node [decision, text width=2.4cm, aspect=2, below of=rank, yshift=0.5cm] (en_check) {Multiple instances of lowest energy system?};
            
            \node [io, below of=en_check, yshift=0cm, text width=6cm] (success) {Lowest energy interface material identified};
            
            \path [line] (host) -- (populate);
            \path [line] (stoichiometry) -- (populate);
            \path [line] (populate) -- (relax);
            \path [line] (relax) -- (rank);
            
            \path [line] (df) -- (descriptor);
            \path [line] (profiles) -- (df);
            \path [line] (descriptor) -- (populate);
            
            \path [line] (rank) -- (en_check);
            \path [line] (en_check) -| node[pos=0.5, xshift=0.5cm, yshift=-0.25cm] {no} (df);
            \path [line] (en_check) -- node[pos=0.5, xshift=0.1cm] {yes} (success);
            
        \end{tikzpicture}
    }
    \vspace{1em}
    \caption{
        An overview of the workflow process of the RAFFLE package, outlining its inputs, outputs, and active learning approach.
    }
    \label{fig:flowchart}
\end{figure}

The workflow of RAFFLE is outlined in \figref{fig:flowchart}.
The library requires three inputs:
1) a host structure in which atoms will be placed,
2) a stoichiometry (a number for each element) of atoms to be placed by the workflow, and
3) a database of systems and their energies from which the software can determine preferential site placements.
The workflow then consists of five steps:
1) element-dependent structural descriptors will be populated from the information available in the provided database,
2) placement rules will be built based on the aforementioned descriptors,
3) atoms are placed within the host structure to satisfy the placement rules, where the atoms are chosen to match the provided stoichiometry,
4) the generated structures are output to allow for energetic calculation,
5) the evaluated structures are fed back into the generator to update the descriptors via active learning, and the process repeats from step 3 until convergence is achieved (i.e. the lowest energy structure is recovered multiple times).

Extended documentation and tutorials for the library can be found on the associated GitHub repository and Read\textit{the}Docs documentation~\cite{RAFFLEDocs2025}.

The Python wrapper handles atomic structure inputs and outputs using the atomic simulation environment (ASE) Atoms object~\cite{ase-paper}.
The Fortran-to-Python wrapper is generated with the help of \texttt{f2py} and \texttt{f90wrap}~\cite{Kermode2020-f90wrap}.
The auto-generated wrapper files have then been edited to improve and streamline functionality.

\subsection{Host structure}
\label{sec:host}

\begin{figure}
    \centering
    \includegraphics[width=0.9\linewidth]{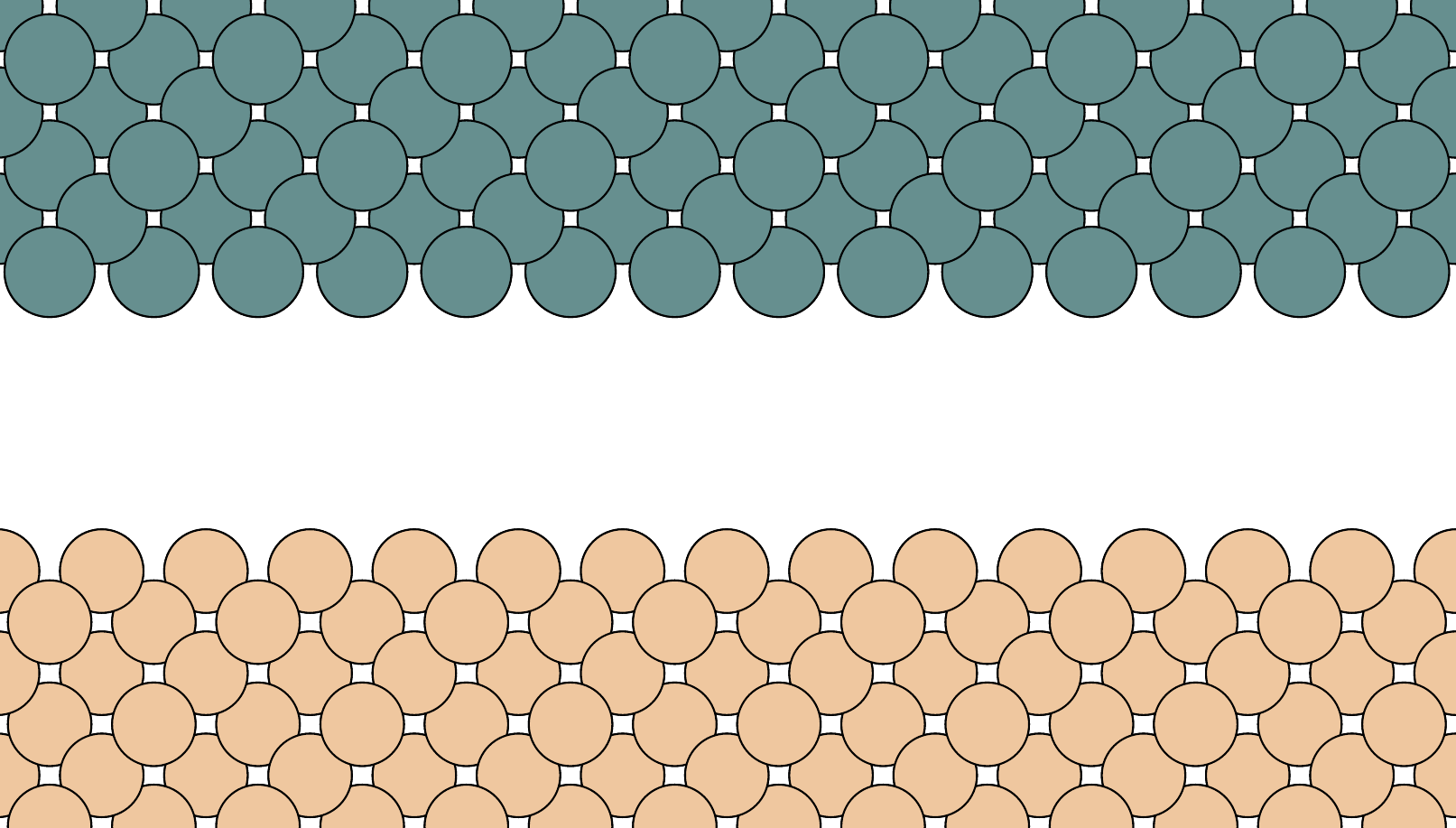}
    \caption{
        An example host structure, where circles represent atoms.
        A vacuum gap exists between two crystal surfaces, providing space for additional atoms.
        This setup allows for the identification of new material phases at the interface.
    }
    \label{fig:host}
\end{figure}

There exist many software packages and libraries to generate interface structures, some examples include ARTEMIS~\cite{Taylor2020ARTEMISAbInitio}, INTERFACER~\cite{Karasulu2023INTERFACERcode}, InterMat~\cite{Choudhary2024InterMatAcceleratingBand}, and
ASE~\cite{ase-paper}.
As such, host structure generation is not performed directly by RAFFLE and, instead, needs to be provided as a supported structure file (such as \code{POSCAR} or \code{extxyz}) or as an ASE Atoms object.
Additionally, the user can specify a \textit{bounding box} in which atoms can be placed within the cell.
If left undefined, the bounding box is set to the entire cell.
An example of a host structure is presented in \figref{fig:host}.

\subsection{Distribution functions}   
\label{sec:distribution_functions}

First, the software uses a set of \textit{distribution functions} to identify structural characteristics unique to each system; these are defined such that they remain invariant to transformation operations that are applied uniformly to the entire cell, such as translations and rotations.
These will be used to identify the probability (or energetic preference) of a certain bonding environment for a set of atomic species; as such, these functions will represent the preferential locations/arrangements of atoms within a system, and will be used to inform the placement of new atoms.
These functions are employed both for characterisation, and to aid in the inverse design of energetically favourable interface structures.
The set of distribution functions are used to define the RAFFLE generalised descriptor (termed \textit{descriptor}).

In this work, 2-, 3-, and 4-body distribution functions are employed as atomic environment descriptors, chosen to mimic classic empirical potentials.
These probabilistic functions are physically appealing: the 2-body function is commonly used to characterise amorphous structures~\cite{Wright2000DefectFreeVitreous,Kimmel2009DoublyPositivelyCharged}, the 3-body follows the well-established Keating potential~\cite{Keating1966RelationshipBetweenMacroscopic}, and the 4-body function corresponds to an improper dihedral angle~\cite{Steinmann2021AtomisticTwoThree}, which has been incorporated into frameworks like GULP~\cite{Gale2005GULPCapabilitiesProspects,Pearlman1995AMBERPackageComputer}.

Inspired by the success of empirical potentials in describing structural energetics, these simplistic descriptors allow for more intensive sampling than conventional approaches.
The choice of empirical model also informs the selection of cutoffs used throughout this work~\cite{Gale2005GULPCapabilitiesProspects}.
Furthermore, the modular structure of the source code enables users to customise the software with alternative empirical-inspired descriptors, with future developments planned to support fully custom descriptor implementations.

To generate the distribution functions for a defined structure, we sample each pair, triplet, and quartet of atoms with positions  ($\vec{r}_i,\vec{r}_j$), ($\vec{r}_i,\vec{r}_j,\vec{r}_k$), ($\vec{r}_i,\vec{r}_j,\vec{r}_k,\vec{r}_l$), respectively.
Gaussians are employed to smooth out structural features, allowing identification of similarity between structures based on functional overlaps.
The Gaussians are defined uniquely for 2-, 3-, and 4-body distribution functions by their standard deviations $\sigma_2$, $\sigma_3$, and $\sigma_4$, respectively.
 
The 2-body distribution function $\mathrm{A}_{\alpha\beta}(x)$ is defined uniquely for each pair of elements, or species, ($\alpha\beta$) and is a function over distance $x$, which represents the atom pair separation distance,

\begin{equation}
    \begin{aligned}
        \mathrm{A}'_{\alpha \beta}(x) &= \frac{1}{\sqrt{2\pi\sigma_{2}^2}} \sum_{i \neq j} \frac{1}{|\vec{r}_{ij}|} \mathrm{exp}\bigg(-\frac{(x-|\vec{r}_{ij}|)^2}{2\sigma_{2}^2}\bigg)  \mathrm{c}_{ij} \delta_{s_i,\alpha}\delta_{s_j,\beta},\\
        \mathrm{A}_{\alpha \beta}(x)  &=  \frac{\mathrm{A}'_{\alpha\beta}(x)}{\int_{R_\mathrm{cut}} dx \mathrm{A}'_{\alpha\beta}(x)}.
    \end{aligned}
\end{equation}

\noindent
The element of atom $i$ is denoted as $s_i$.
The $i$ and $j$ variables sum over atoms, whilst $\delta_{s_i,\alpha}$ forces this to only the specified elements.
A normalisation is applied to account for regions of varying neighbour density, resulting in an average descriptor experienced by each atom in the system.
Here, $\vec{r}_{ij}$ is the vector pointing from atom $i$ to atom $j$, 

\begin{equation}
    \vec{r}_{ij} = \vec{r}_{i} - \vec{r}_{j},
\end{equation}

\noindent
and $\mathrm{c}_{ij}$ is a cutoff function,

\begin{equation}
    \mathrm{c}_{ij} =
    \begin{cases}
        \dfrac{1}{|\vec{r}_{ij}|^2}, & \text{if } |\vec{r}_{ij}| \leq R_{\mathrm{cut}}, \\[8pt]
        0, & \text{otherwise}.
    \end{cases}
\end{equation}

\noindent
The decay is used to account for the square power law that neighbour number grows by.
The cutoff is a computational shortcut to reduce the number of calculations required. 
For the 2-body function, the increase in cutoff scales roughly as $R_{\mathrm{cut}}^{3}$, where the number of neighbours being summed can be averaged as taking the form of

\begin{equation}
    N_{\mathrm{neigh}}(R_{\mathrm{cut}}) \approx \rho\frac{4}{3}\pi R_{\mathrm{cut}}^{3},
\end{equation}

\noindent
where $\rho$ is the averaged density of the system.
As such, $R_{\mathrm{cut}}$ has the scaling $\mathcal{O}(R_{\mathrm{cut}})$.
The cutoff function $\mathrm{c}_{ij}$ is implemented to ensure parity with the behaviour of Coulomb-like 2-body interactions.

A similar form is adopted for the 3- and 4-body characterisations.
The 3-body distribution function, unique for each element, $s_a$, is a function over angle $\theta$ between $0$ and $\pi$ and is defined as

\begin{equation}
    \begin{aligned}
        \mathrm{B}'_{\alpha}(\theta) &= \frac{1}{\sqrt{2\pi\sigma_{3}^2}} \sum_{i \neq j \neq k} \mathrm{exp}\bigg(-\frac{(\theta-\theta_{ijk})^2}{2\sigma_{3}^2}\bigg)  \mathrm{c}_{ijk} \delta_{s_i,\alpha},\\
        \mathrm{B}_{\alpha}(\theta) &= \frac{\mathrm{B}'_{\alpha}(\theta)}{\int d\theta \mathrm{B}'_{\alpha}(\theta)},
    \end{aligned}
\end{equation}

\noindent
where

\begin{equation}
    \theta_{ijk} = \mathrm{acos}\bigg(\frac{\vec{r}_{ij}\cdot\vec{r}_{ik}}{|\vec{r}_{ij}||\vec{r}_{ik}|}\bigg)
\end{equation}

\noindent
is the acute angle subtended by the path $\vec{r}_j \rightarrow \vec{r}_i \rightarrow \vec{r}_k$.
The angle is taken as the acute value, meaning, any angles beyond $\pi$ are mapped back into the range of $0$ -- $\pi$ through the transformation $\theta_{\mathrm{mapped}} = 2\pi - \theta$.
The 3-body cutoff function is defined as

\begin{equation}
    \mathrm{c}_{ijk} =
    \begin{cases}
        \dfrac{1}{(|\vec{r}_{ij}| |\vec{r}_{ik}|)^2}, & 
        \text{if } R_{3,s_i,s_{j/k}}^{\mathrm{lw}} \leq |\vec{r}_{ij}| \text{ and } |\vec{r}_{ik}| \leq R_{3,s_i,s_{j/k}}^{\mathrm{up}}, \\[8pt]
        0, & \text{otherwise}.
    \end{cases}
\end{equation}

\noindent
The parameters $R_{3,s_i,s_{l}}^{\mathrm{lw}}$ and $R_{3,s_i,s_{l}}^{\mathrm{up}}$ define the minimum and maximum bond lengths considered for the three-body function.
These cutoffs restrict interactions to shells of neighbouring atoms around the test atom.
They are element-dependent and, by default, set to $1.5$ and $2.5$ times the average covalent radius of elements $s_i$ and $s_l$ (covalent radii obtained from reference~\citen{wolfram_2024_elementdata}).
Both the fractions and radii can be set by the user.

Finally, the 4-body distribution function, unique for each element, $s_a$, is a function over angle $\phi$ between $0$ and $\pi$ and is defined as

\begin{equation}
    \begin{aligned}
        \mathrm{D}'_{\alpha}(\phi) &= \frac{1}{\sqrt{2\pi\sigma_{4}^2}} \sum_{i \neq j \neq k \neq l} \mathrm{exp}\bigg(-\frac{(\phi-\phi_{ijkl})^2}{2\sigma_{4}^2}\bigg)  \mathrm{c}_{ijkl} \delta_{s_i,\alpha},\\
        \mathrm{D}_{\alpha}(\phi) &= \frac{\mathrm{D}'_{\alpha}(\phi)}{\int d\theta \mathrm{D}'_{\alpha}(\phi)},
    \end{aligned}
\end{equation}

\noindent
where 

\begin{equation}
    \theta_{ijkl} = \mathrm{acos}\bigg(\frac{
            ( \vec{r}_{ij} \times \vec{r}_{ik} ) \cdot 
            ( \vec{r}_{ik} \times \vec{r}_{il} ) 
    }{
            | \vec{r}_{ij} \times \vec{r}_{ik} |
            | \vec{r}_{ik} \times \vec{r}_{il} |
    }\bigg)
\end{equation}

\noindent
is the acute improper dihedral angle formed by four atoms, with the central atom being atom $i$.
The cutoff function is

\begin{equation}
    \mathrm{c}_{ijkl} =
    \begin{cases}
        \dfrac{1}{(|\vec{r}_{ij}||\vec{r}_{ik}||\vec{r}_{il}|)^2}, & 
        \text{if } R_{3,s_i,s_{j/k}}^{\mathrm{lw}} \leq |\vec{r}_{ij}| \text{ and } |\vec{r}_{ik}| \leq R_{3,s_i,s_{j/k}}^{\mathrm{up}} \\
        & \quad\text{ and } R_{4,s_i,s_{l}}^{\mathrm{lw}} \leq |\vec{r}_{il}| \leq R_{4,s_i,s_{l}}^{\mathrm{up}}, \\[8pt]
        0, & \text{otherwise}.
    \end{cases}
    \label{eq:cutoff:2body}
\end{equation}

\noindent
where parameters $R_{4,s_i,s_{l}}^{\mathrm{lw}}$ and $R_{4,s_i,s_{l}}^{\mathrm{up}}$ define the lower and upper limits for pair separation in the four-body distribution.
As with the three-body cutoffs, these limits are element-dependent and, by default, set to $3.0$ and $6.0$ times the average covalent radius of elements $s_i$ and $s_l$.
Both the fractions and radii can be modified by the user.
This cutoff is included to identify plane alignment features.
This distribution function aids significantly with 2D systems and with interface surfaces, allowing more physical structures to be rapidly realised.
This is highlighted in~\figref{fig:descriptors} where the interplanar distance (or perpendicular requirement of the layer) is shown at $\pi/2$.

Triplet and 4-body functions for both computational efficiency and due to the nature of the Keating potential~\cite{Keating1966RelationshipBetweenMacroscopic} only consider the species of the atom which the bond-bending occurs around.
The species of the two neighbours chosen to calculate this term are less relevant for a bond bending potential, thus this information is discarded.
It is beneficial to include as many potentially relevant datapoints as possible to improve generalisability instead of overfitting to specific atomic arrangements.
Additionally, angle between atoms (and even more so for dihedral angles, consider van der Waals structures as an example) are less dependent on the species than the bond length.

\subsection{Generalised descriptor}
\label{sec:generalised}

With the distribution functions for a single structure now defined, we next need to combine those of multiple systems to obtain a set of generalised descriptors (or generalised distribution functions).
The purpose of such a combined descriptor is to identify features that a set of chemical elements are likely to form.
However, comparing distribution functions from multiple stoichiometries becomes difficult due to the unclear definition of chemical favourability.

Angular bonding plays a more dominant role in the system than precise bond lengths.
This can be understood in terms of symmetry: bond angles determine the space group, while bond lengths primarily influence expansion or compression.
As a result, enforcing only favourable bond angles strongly biases the system towards known atomic arrangements.

\subsubsection{Distribution function weighting methods}
\label{sec:weight}

With the distribution functions for a single structure (labelled $a$) now defined, we need to determine how to appropriately compare those of different systems, particularly those with drastically different energies of formation.
Such considerations become even more important when combining distribution functions to describe features across a range of chemical environments.
For example, how should one compare the bonding environment of carbon in diamond with that of carbon in lithium-intercalated graphite?
To mitigate the effects of combining a variety of systems, we apply a weighting to the distribution functions of each system, which is dependent on their formation energies.
The two methods currently available within RAFFLE are the \textit{convex hull} and the \textit{empirical} methods.

\paragraph{\textbf{Convex hull weighting method}}
\label{par:convex_hull}

The first method provided uses the convex hull of the configuration space.
Here, the weighting for structure $a$ is

\begin{equation}
\begin{aligned}
    w_{a} &= \mathrm{exp}\bigg(\frac{\Delta E_{a}}{k_{B} T}\bigg),\\
    w_{a} & = w_{a,\alpha} = w_{a,\alpha\beta} \quad \mathrm{for} \quad \alpha,\beta \in \mathbb{Z}^{+}
    \label{eq:weight:hull}
\end{aligned}
\end{equation}

\noindent
where $\Delta E_{a}$ is its energy above the convex hull.
The weighting is species-independent and, as such $\alpha$ and $\beta$ are free variables.
Thus, structures on the hull will equally contribute to the generalised descriptor as their weighting $w_{a}$ will equal~$1$.

\paragraph{\textbf{Empirical weighting method}}
When the convex hull cannot be accurately determined due to insufficient data or complexity, an alternative method is provided to weight individual contributions to the generalised descriptor.
While similar to the convex hull weighting approach, this method is adapted for data-sparse scenarios, such as interface exploration, ensuring that structure search remains viable.
This method entails first determining formation energy of structures from reference energies for their constituent elements and then attributing that formation energy to the elements and element-pairs in the structure.
The most favourable proportional formation energy is stored for each element and element pair, denoted as $\Xi_{\mathrm{f},\alpha}$ and $\Xi_{\mathrm{f},\alpha\beta}$, respectively.
The element-pair (or 2-body) weighting for structure $a$ is defined as

\begin{equation}
    w_{a,\alpha\beta}=\mathrm{exp}\bigg(\frac{\Xi_{\mathrm{f},\alpha\beta}-E_{\mathrm{f},a,\alpha\beta}}{k_{B} T}\bigg),
    \label{eq:weight:empirical}
\end{equation}

\noindent
where, similarly, $E_{\mathrm{f},a,\alpha\beta}$ is the fraction of formation energy of system $a$ attributed to element pairs $\alpha\beta$.
This fraction of formation energy is taken as

\begin{equation}
    E_{\mathrm{f},a,\alpha\beta} = E_{\mathrm{f},a} \frac{N_{\mathrm{2-body},a,\alpha\beta}}{N_{\mathrm{2-body},a}},
\end{equation}

\noindent
where $N_{\mathrm{2-body},a}$ is the total number of 2-body pairs summed over when generating the distribution functions for system $a$, and $N_{\mathrm{2-body},a,\alpha\beta}$ is the number of $\alpha\beta$ 2-body pairs summed over for system $a$.

The element (or 3-body and 4-body) weighting for structure $a$ is defined as

\begin{equation}
    w_{a,\alpha}=\mathrm{exp}\bigg(\frac{\Xi_{\mathrm{f},\alpha}-E_{\mathrm{f},a,\alpha}}{k_{B} T}\bigg),
    \label{eq:weight}
\end{equation}

\noindent
where $E_{\mathrm{f},\alpha}$ is the fraction of formation energy attributed to species $\alpha$.
Derivation of the formation energy and how it is attributed to certain elements and element-pairs is outlined in \secref{sec:formation_energy}

\subsubsection{Learning the descriptor}
\label{sec:combining}

To correctly scale structure features based on the 
energetic favourability of the systems they appear in, we apply one of the aforementioned weighting methods.
The generalised descriptors for 2-, 3-, and 4-bodies are generated sequentially, i.e. analysing one system at a time and adding its contributions to the existing descriptor.
The starting 2-, 3-, and 4-body descriptors are initialised to zero:

\begin{equation}
\begin{aligned}
    \mathrm{G}^{0}_{\alpha\beta} (x) &= 0,\\
    \mathrm{H}^{0}_{\alpha} (\theta) &= 0,\quad\mathrm{and}\\
    \mathrm{J}^{0}_{\alpha} (\phi) &= 0.
\end{aligned}
\end{equation}

The descriptors are then updated by adding in the features unique to structure $a$, which is represented mathematically using the set difference between two functions $A(x)$ and $B(x)$ (with a minimum allowed value of 0):

\begin{equation}
    S\bigg(A(x), B(x)\bigg) = \max \bigg( A(x) - B(x) , 0 \bigg).
\end{equation}

Finally, for RAFFLE active learning iteration step $n$, the 2-, 3-, and 4-body generalised descriptors are

\begin{equation}
\begin{aligned}
    \mathrm{G}^{n}_{\alpha\beta} (x) &= S \Bigg( 
    \frac{w_{a,\alpha\beta}\mathrm{A}_{\alpha\beta}^{a}(x)}{\max_{x}\big(\mathrm{A}_{\alpha\beta}^{a}(x)\big)} , \mathrm{G}^{n-1}_{\alpha\beta}(x) \Bigg),\\[8pt]
    \mathrm{H}^{n}_{\alpha} (\theta) &= S \Bigg( \frac{w_{a, \alpha}\mathrm{B}_{\alpha}^{a}(\theta)}{\max_{\theta}\big(\mathrm{B}_{\alpha}^{a}(\theta)\big)} , \mathrm{H}^{n-1}_{\alpha}(\theta) \Bigg),\quad\mathrm{and}\\[8pt]
    \mathrm{J}^{n}_{\alpha} (\phi) &= S \Bigg(
    \frac{w_{a, \alpha}\mathrm{D}_{\alpha}^{a}(\phi)}{\max_{\phi}\big(\mathrm{D}_{\alpha}^{a}(\phi)\big)} , \mathrm{J}^{n-1}_{\alpha}(\phi) \Bigg).
\end{aligned}
\end{equation}

\begin{figure*}
    \centering
   \subfloat[2-body]{\includegraphics[width=0.31\linewidth]{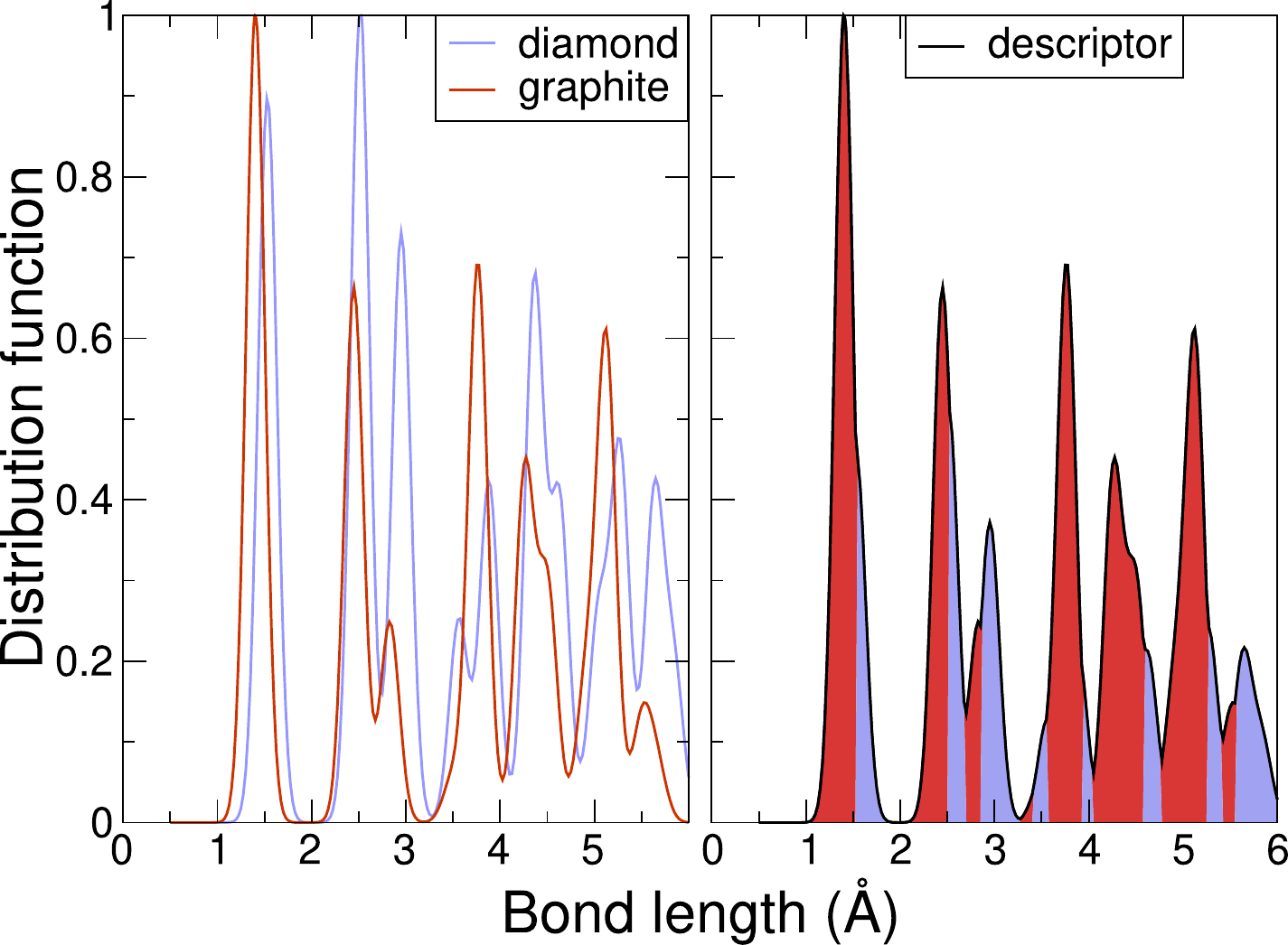}\label{fig:descriptors:2-body}}%
   \hspace{1em}%
   \subfloat[3-body]{\includegraphics[width=0.31\linewidth]{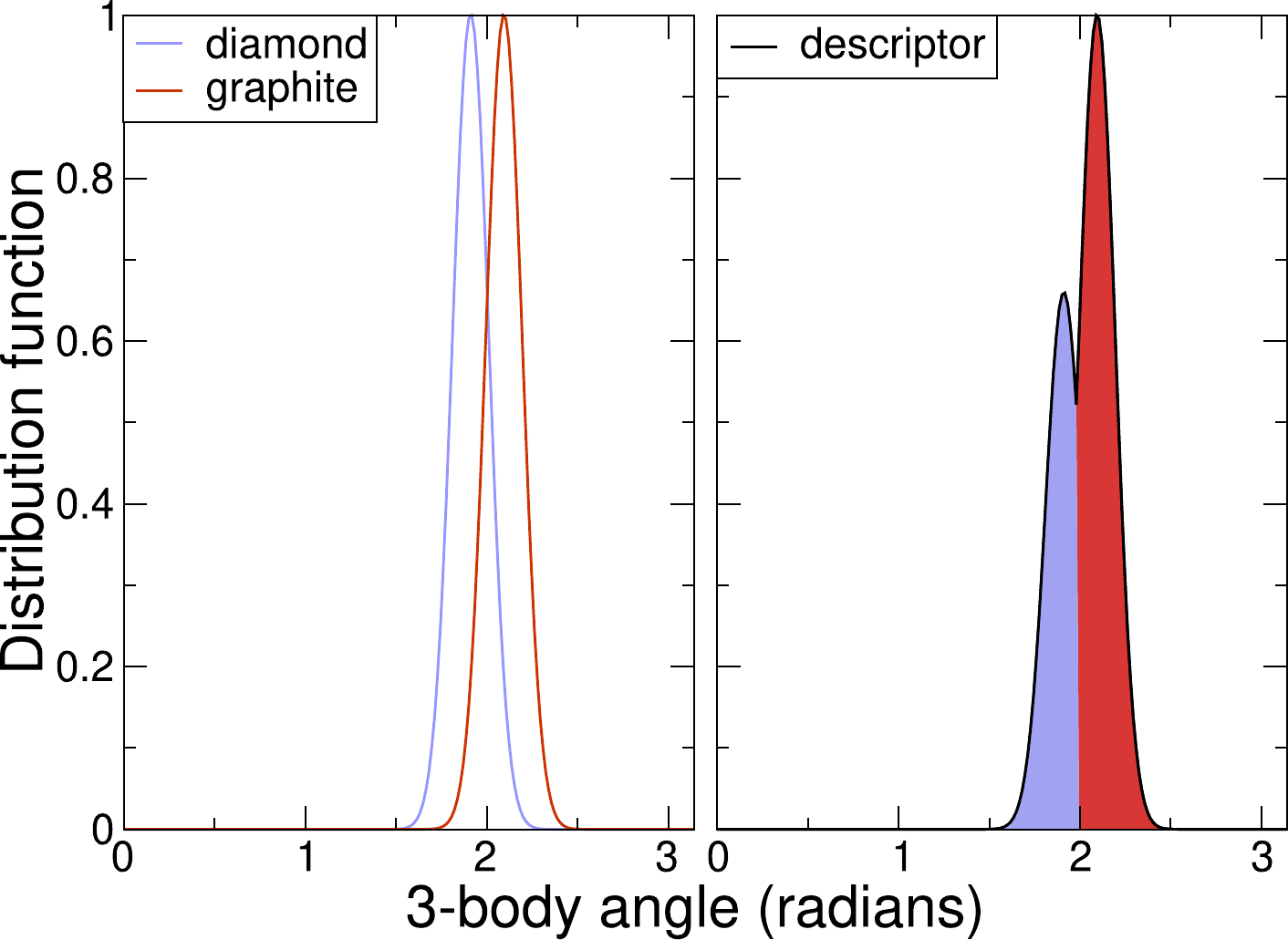}\label{fig:descriptors:3-body}}%
   \hspace{1em}%
   \subfloat[4-body]{\includegraphics[width=0.31\linewidth]{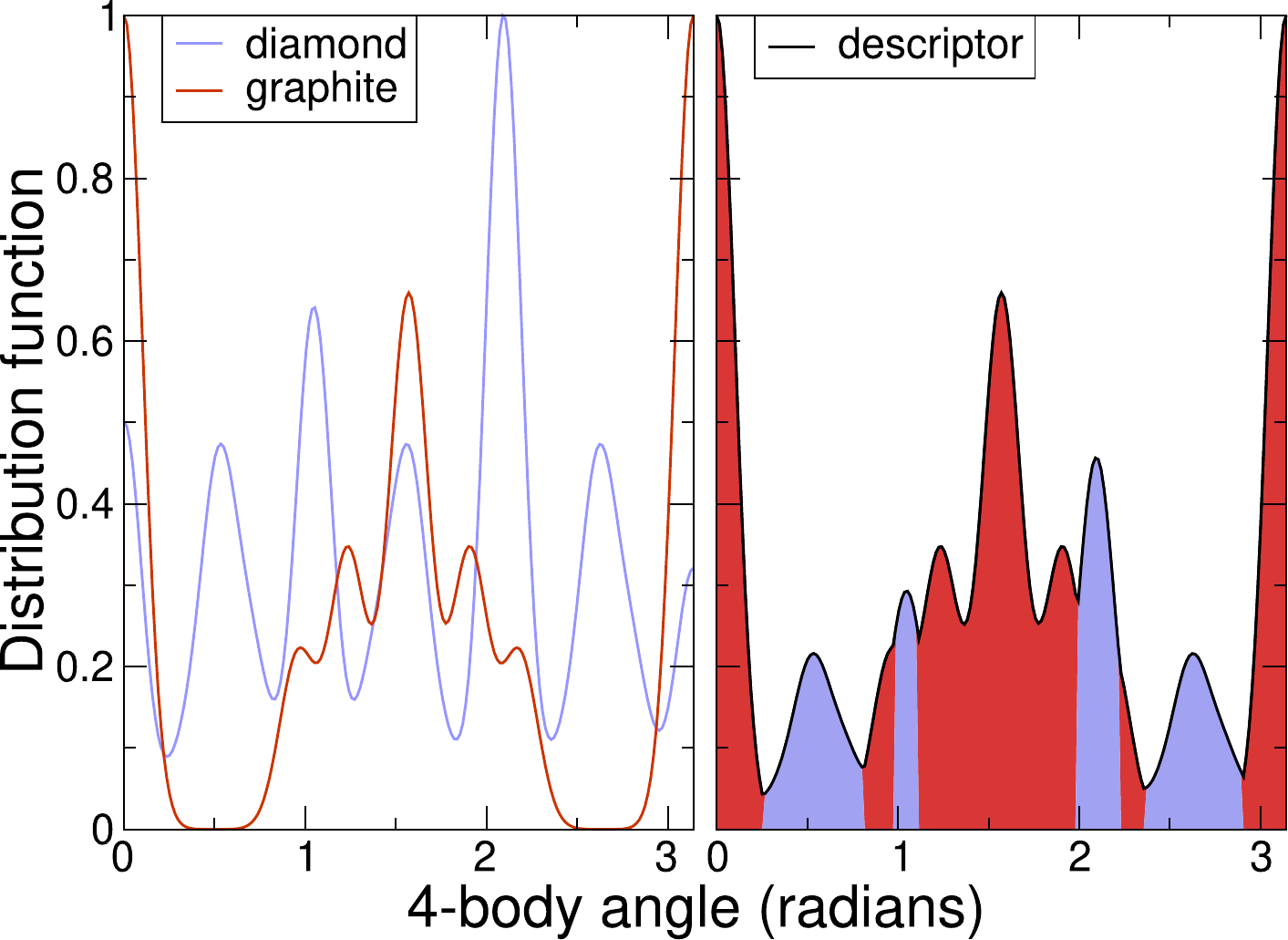}\label{fig:descriptors:4-body}}%
    \caption{
        RAFFLE descriptors.
        Example \protect\subref{fig:descriptors:2-body} 2-, \protect\subref{fig:descriptors:3-body} 3-, and \protect\subref{fig:descriptors:4-body} 4-body descriptors generated for carbon: (left) distribution functions for bulk diamond and bulk graphite (AB stacking), and (right) the generalised descriptor (or generalised distribution function) obtained by combining diamond and graphite descriptors using energetic weighting.
        Shaded regions of the generalised descriptor indicate the system contributing to each feature.
        Graphite is energetically more favourable than diamond by $0.107$~\si{\electronvolt}/atom (calculated using CHGNet~\cite{Deng2023CHGNETPretrainedUniversalNeural}).
    }
    \label{fig:descriptors}
\end{figure*}

An example set of distribution functions and generalised descriptors are presented for carbon atoms is presented in \figref{fig:descriptors}.
The individual distribution functions for carbon in diamond and graphite (AB stacking) are presented, along with the generalised descriptors generated from a database containing the two systems only.
The figure highlights how features are combined through energetic weighting considerations.
The corresponding notebook can be found in \texttt{./tools/descriptors.ipynb}, which details the diamond and graphite lattice constants used.
This process represents the active learning component of RAFFLE.

For bulk materials, the learned RAFFLE descriptors should eventually converge after a certain number of iterations, depending on the complexity of the system and the effectiveness of global minimum sampling.
This occurs because the generalised descriptors remain unchanged unless new, energetically favourable features are introduced that significantly alter them.
In contrast, for interfaces, the vast number of possible configurations makes full convergence to stable descriptors highly unlikely.
Instead, it is up to the user to analyse the formation energies generated and assess the variability in the results and compare these with the abrupt interface.

\subsection{Viability evaluation}
\label{sec:viability}

RAFFLE evaluates site viability by comparing the distribution functions of a test atom with the RAFFLE generalised descriptors.
Higher overlap indicates a more suitable placement site, guiding atomic placement.
To assess the viability of a site for a test atom of element $\alpha$ at position $\vec{r}$ in the host cell, contributions from neighbouring atoms are evaluated and combined into an overall probability (or viability).
This probability determines the likelihood of the test atom occupying that site and takes the form

\begin{equation}
    P_{\alpha} (\vec{r}) = V_{\mathrm{2}, \alpha}(\vec{r}) V_{\mathrm{3}, \alpha}(\vec{r}) V_{\mathrm{4}, \alpha}(\vec{r}),
\end{equation}

\noindent
where $P_{\alpha}$ ranges from $0$ to $1$ and $V_{n,\alpha}(\vec{r})$ is the $n$-body viability contribution to the overall probability.
The 2-body form is

\begin{equation}
    V_{\mathrm{2}, \alpha}(\vec{r}) = \frac{1}{N_2(\vec{r})} \sum_i G^{n}_{\alpha,s_i}(| \vec{r} - \vec{r}_{i} |),
\end{equation}

\noindent
where $i$ iterates over all atoms within $R_\mathrm{cut}$ of the query point $x$ (i.e. $1$ to $N_2(\vec{r})$), and $N_2(\vec{r})$ is the number of 2-body interactions summed over at point $\vec{r}$.
The denominator of the summation is the 
The 3-body viability contribution is

\begin{equation}
    V_{\mathrm{3}, \alpha}(\vec{r}) = \frac{1}{N_3(\vec{r})} \prod_{i \neq j} \mathrm{C}_{3,ij}(\vec{r}) H^{n}_{\alpha}( \theta_{ij}(\vec{r})),
\end{equation}

\noindent
where $i$ and $j$ cycle over all atoms between a distance of $R_3^{\mathrm{lw}}$ and $R_3^{\mathrm{up}}$ of position $\vec{r}$, $\theta_{ij}(\vec{r})$ is the angle subtended by path $\vec{r}_i$--$\vec{r}$--$\vec{r}_j$, and $N_3(\vec{r})$ is the number of 3-body interactions multiplied over at point $\vec{r}$.
The four-body viability contribution is

\begin{equation}
    V_{\mathrm{4}, \alpha}(\vec{r}) = \frac{1}{N_4(\vec{r})} \prod_{i\neq j\neq k} \mathrm{C}_{3,ij}(\vec{r}) \mathrm{C}_{4,ik}(\vec{r}) J^{n}_{\alpha}( \phi_{ijk}(\vec{r})),
\end{equation}

\noindent
where $i$ and $j$ cycle over all atoms between a distance of $R_3^{\mathrm{lw}}$ and $R_3^{\mathrm{up}}$ of position $\vec{r}$, whilst $k$ cycles over all atoms within a distance of $R_4^{\mathrm{lw}}$ and $R_4^{\mathrm{up}}$ of position $\vec{r}$.
$\phi_{ijk}(\vec{r})$ is the improper dihedral angle made between atoms $i$, $j$, $k$, and position $\vec{r}$ (where $\vec{r}$ is at the centre).
$N_4(\vec{r})$ is the number of 4-body interactions multiplied over at point $\vec{r}$.
The atomic neighbours considered to contribute to the viability of a test site is mathematically handled via the following cutoff function:

\begin{equation}
    \mathrm{C}_{m,ij}(\vec{r}) = 
    \begin{cases}
        1, & \text{if } R_{m,s_{i},s_{j}}^{\mathrm{lw}} \leq |\vec{r}_{i} - \vec{r}| 
        \text{ and } |\vec{r}_{j} - \vec{r}| \leq R_{m,s_{i},s_{j}}^{\mathrm{up}}, \\[8pt]
        \overline{H^{n}_{\alpha}}, & \text{if } m = 3 \text{ and } |\vec{r}_{j} - \vec{r}| > R_{m,s_{i},s_{j}}^{\mathrm{up}}, \\[8pt]
        \overline{J^{n}_{\alpha}}, & \text{if } m = 4 \text{ and } |\vec{r}_{j} - \vec{r}| > R_{m,s_{i},s_{j}}^{\mathrm{up}}, \\[8pt]
        0, & \text{otherwise}.
    \end{cases}
\end{equation}

\noindent
For any point where an atom is closer than $R_{3,s_{i},s_{j}}^{\mathrm{lw}}$, the site is considered unviable and the probability of placement is set to $P_{\alpha} = 0$.
This enforces a minimum bond distance, preventing atoms from being placed too close to each other, and is element pairwise dependent, similar to that found in AIRSS and other methods~\cite{Lu2021AbInitioRandomStructure}.
For the 3- and 4-body viability functions, if there exist no atoms to evaluate within the limits (excluding those removed by the aforementioned close limit), then a default value is used.
The default values for the 3- and 4-body viability functions are the average of the respective generalised descriptor, $\overline{H^{n}_{\alpha}}$ and $\overline{J^{n}_{\alpha}}$.

Multiplication of the $n$-body viability functions (and specifically within the angular viability functions) is done to quickly bias away from unfavourable bonding environments.
Angular bonding is found to be a more influential part of the system than exact bond lengths (this can be considered that angles define the space group, whilst bond lengths define the expansion/compression of the system), so ensuring only favourable bond angles occur heavily preferences towards known atomic arrangements.

\begin{figure*}[t]
    \centering
    \subfloat[]{\includegraphics[height=0.25\linewidth]{DGraphics/diamond_host_legend}\label{fig:viability:diamond:host}}%
    \hspace{1em}%
    \subfloat[]{\includegraphics[height=0.24\linewidth]{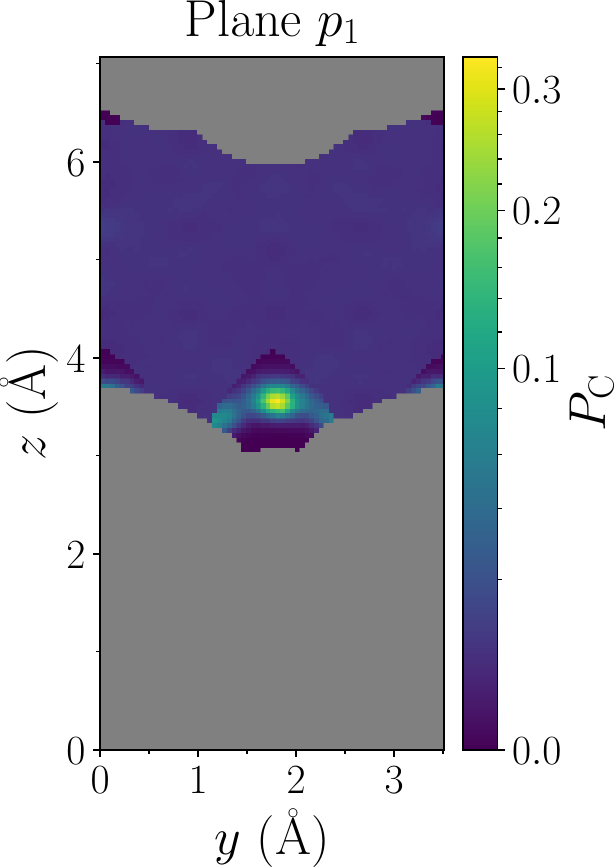}\label{fig:viability:diamond:C:a}}%
    \hspace{1em}%
    \subfloat[]{\includegraphics[height=0.24\linewidth]{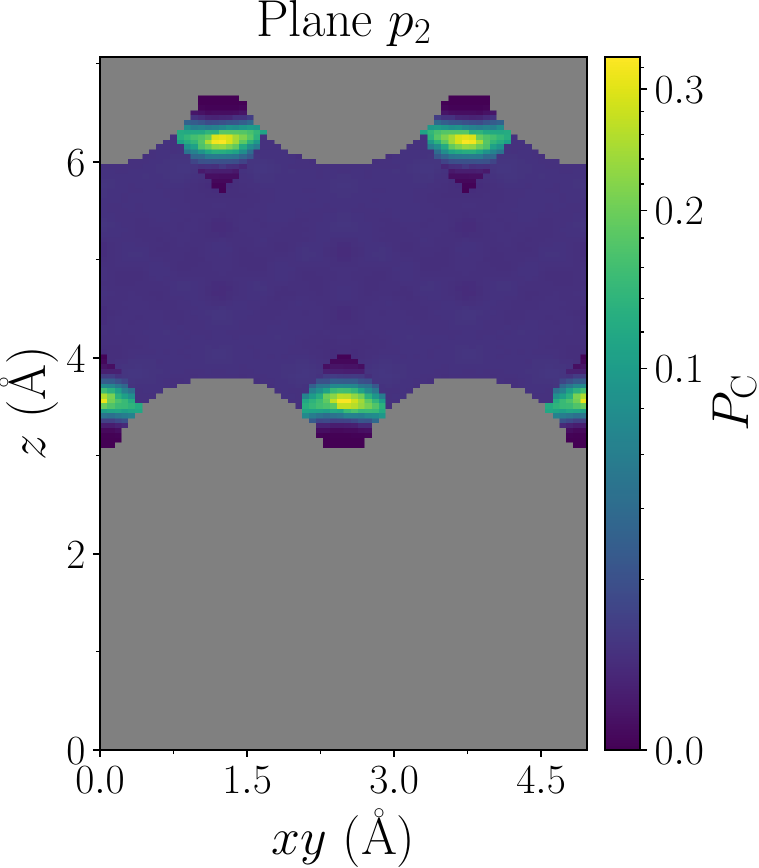}\label{fig:viability:diamond:C:ab}}%
    \hspace{1em}%
    \subfloat[]{\includegraphics[height=0.24\linewidth]{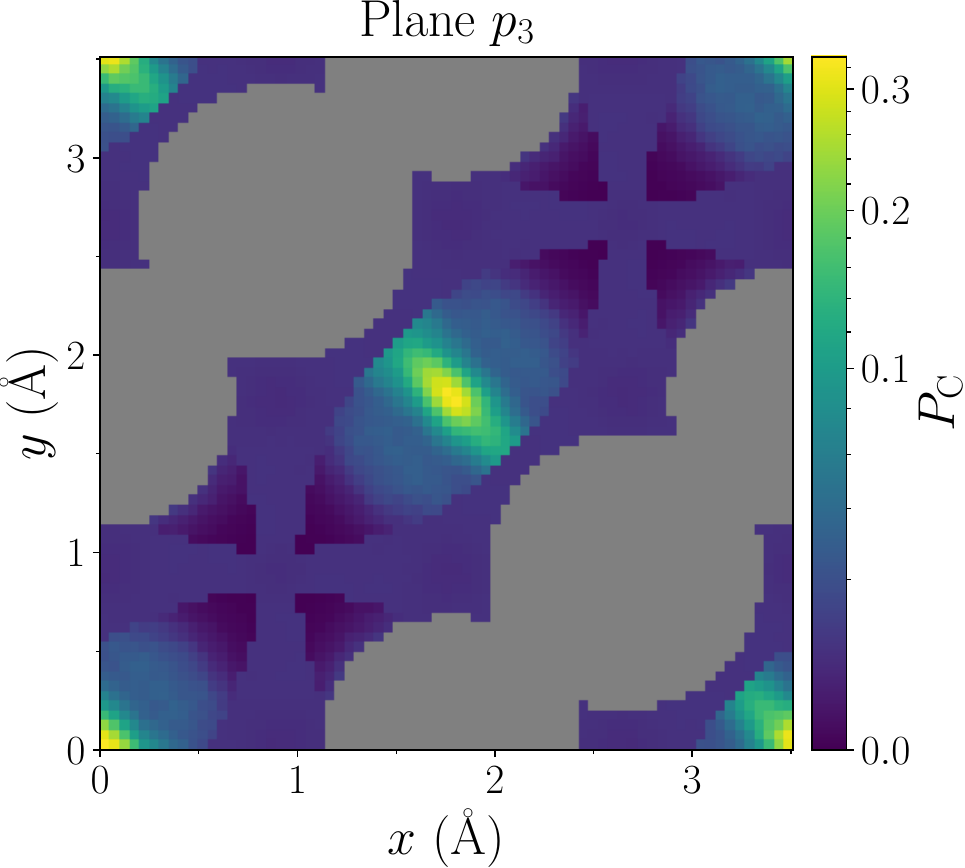}\label{fig:viability:diamond:C:c}}%
    \caption{
        Diamond site viability test.
        \protect\subref{fig:viability:diamond:host} Ball-and-stick representation of the host structure, with planes $p_1$ ((100), intersecting $0.5a$), $p_2$ ((110), intersecting ($0.5a$, $0.5b$)), and $p_3$ ((001), intersecting $0.5c$) represented with translucent planes.
        Brown spheres represent carbon.
        Viability heatmaps for carbon along \protect\subref{fig:viability:diamond:C:a} $p_{1}$, \protect\subref{fig:viability:diamond:C:ab} $p_{2}$, and \protect\subref{fig:viability:diamond:C:c} $p_{3}$, respectively.
        Grey regions indicate sites too close to other atoms to be viable.
    }
    \label{fig:viability:diamond}
\end{figure*}

Here, we outline two tests cases to show the ability for the generalised descriptors to be used to reconstruct heavily defected cells, highlighting the structural understanding encoded into both the generalised descriptors and the viability evaluation.
For the two test cases, the database provided to build the generalised descriptors are just the systems' pristine bulk cells.
The test case of an 8-atom diamond unit cell extended to a $1\times1\times2$ supercell of diamond with the top 8 atoms removed is used to visualise this probability function, seen in \figref{fig:viability:diamond}.
It can be seen that the regions of highest probability correspond to those of the two missing carbon atoms at either surface.
Another test case is the 5-atom BaTiO$_3$ tetragonal primitive cell extended to a $1\times1\times2$ supercell with the top 5 atoms removed, seen in \figref{fig:viability:BTO}.
It should be noted that the probability value associated with the Ba, the Ti, and the two upper O placements are more prominent than the lower O.
However, the lower O site is still recovered as its site viability increases as more atoms are reintroduced to the system.

\afterpage{%
\begin{figure*}[t]
    \centering
    \subfloat[]{\includegraphics[height=0.25\linewidth]{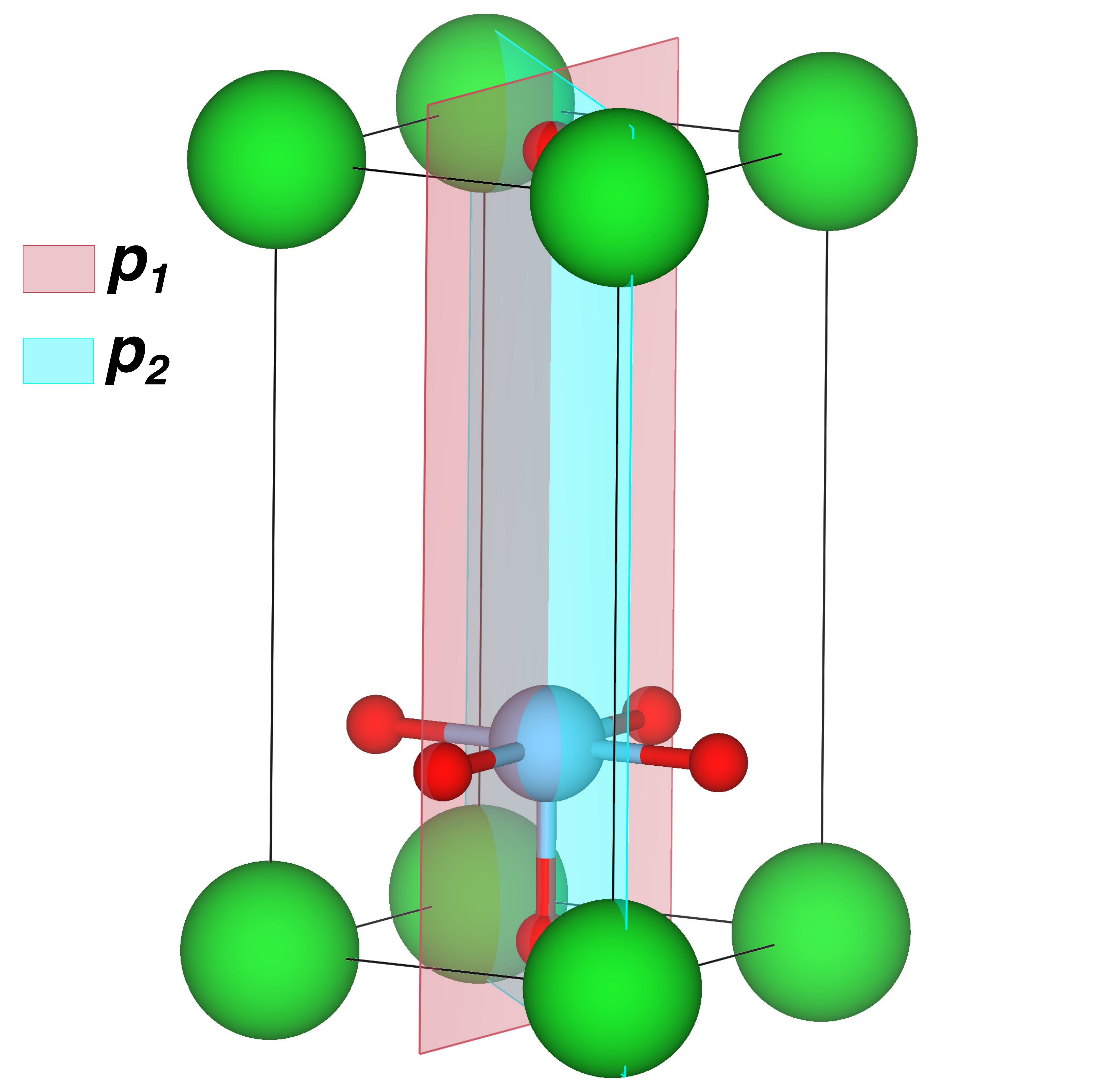}\label{fig:viability:BTO:host}}%
    \hspace{1em}%
    \subfloat[Ba]{\includegraphics[height=0.24\linewidth]{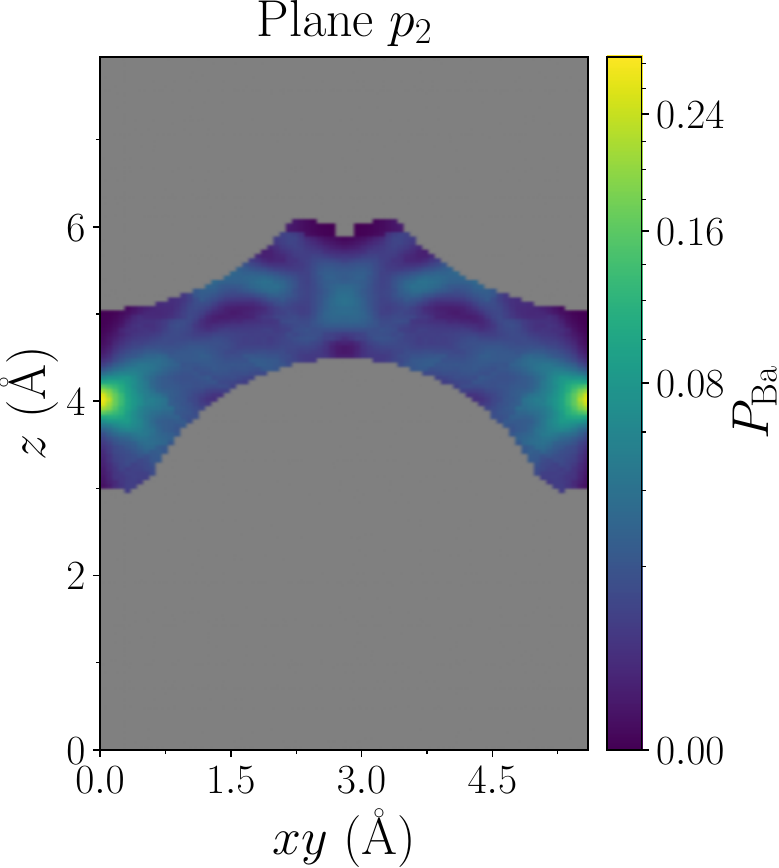}\label{fig:viability:BTO:Ba}}%
    \hspace{1em}%
    \subfloat[Ti]{\includegraphics[height=0.24\linewidth]{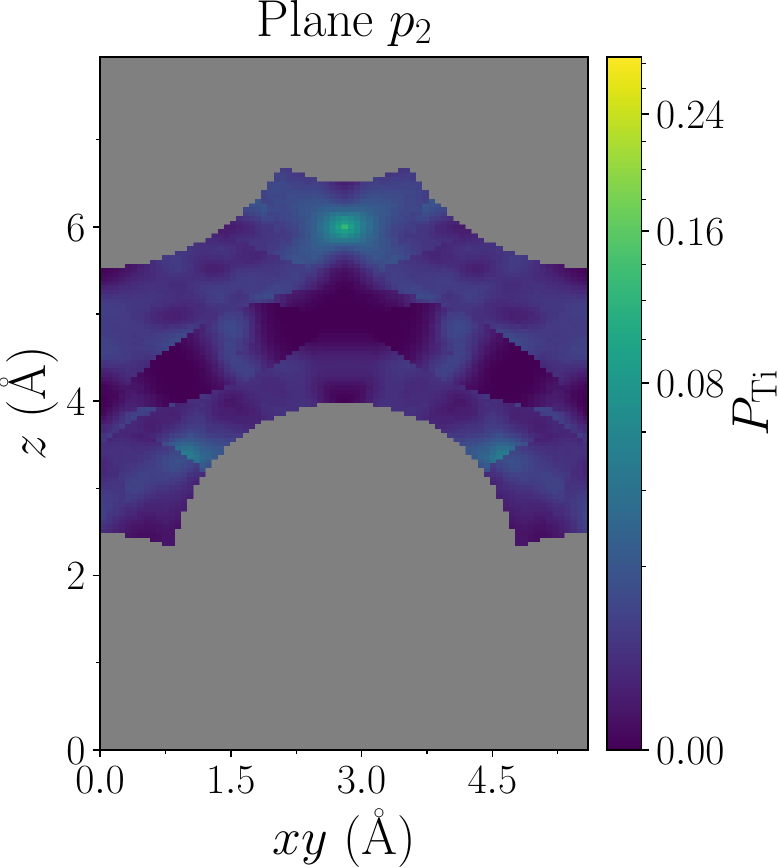}\label{fig:viability:BTO:Ti}}%
    \hspace{1em}%
    \subfloat[O]{\includegraphics[height=0.25\linewidth]{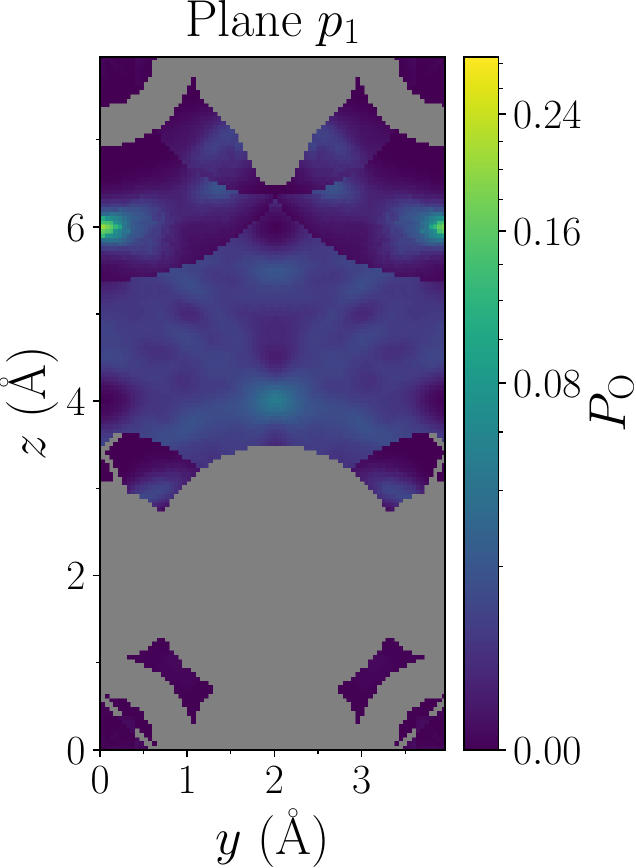}\label{fig:viability:BTO:O}}%
    \caption{
        BaTiO$_{3}$ site viability test.
        \protect\subref{fig:viability:BTO:host} Ball-and-stick representation of the host structure, with planes $p_1$ ((100), intersecting $0.5a$) and $p_2$ ((110), intersecting ($0.5a$, $0.5b$)) represented with translucent planes.
        Spheres represent Ba (green), Ti (blue), and O (red), from largest to smallest.
        Viability heatmaps for \protect\subref{fig:viability:BTO:Ba} Ba and \protect\subref{fig:viability:BTO:Ti} in their most favourable plane, $p_{2}$.
        Viability heatmap for \protect\subref{fig:viability:BTO:O} O in its most favourable plane, $p_{1}$.
        Grey regions indicate sites too close to other atoms to be viable.
    }
    \label{fig:viability:BTO}
\end{figure*}%
}

\subsection{Atom placement methods}
\label{sec:placement_methods}

Preserving the local bonding environment is anticipated to yield more energetically favourable structures.
This expectation is based on several factors:
1) perfect bulk crystals represent the ground state configuration for an infinitely repeating set of atoms, where their highly ordered structure minimises the system’s energy,
2) defects introduce strain, which raises the system’s energy, and
3) maintaining local bonding environments should increase symmetry, often resulting in a lower-energy state.
Therefore, when placing atoms, it is crucial to retain the local bonding configurations associated with known materials, rather than disregarding them, to better capture realistic, low-energy arrangements~\cite{Taylor2020ARTEMISAbInitio}.

Adopting this postulation, one must now sample the lattice to identify sites for potential atom placement.
To perform this, RAFFLE utilises five placement methods, which focus on different aspects such as thoroughness/completeness, computational cost, and entropic maximisation.
The five methods are:
1) a global minimum search,
2) a random walk,
3) a growth method (a variant random walk),
4) a void-finding method, and
5) constrained random placement.
The first method (minimum) focuses on thoroughness and preserving local geometry.
The second two methods (random walk and growth) are used to introduce deviations from the ground state.
The fourth method (void-finding) attempts to simulate the effects of growth patterns in the interface, whilst the final method (random) attempts to recover standard random structure search methods.

The five sampling methods are both similar and contrasting.
The walk, growth, and minimum approaches rely on performing a probability search at each point.
Conversely, the void function is designed to speed up filling large cells and maintain uniform density.
Both the global minimum and void-finding methods involve discretising the space into a fine grid.
The RAFFLE approach employs a ratio of these methods to effectively and efficiently sample the space.
This search is informed by both empirical search functions and purely random search functions, allowing a multitude of structures to be generated whilst accelerating the search towards favourable outcomes.
See
\suppcol{Section SIV}
for an evaluation of the methods and their respective benefits.

\paragraph{Minimum.}%
The first placement method -- \textit{global minimum} -- discretises the full space of the supercell into a grid.
Note, if defining a bounding box, then only the box is discretised into a grid.
At each point, a test atom is postulated, and the suitability of this atom to be formally placed is queried according to the placement functions outlined above.
For all of the points considered, that with the highest probability is selected as the most viable site.
This method is commonly referred to as a greedy global minimum search.
Approaches like GOFEE~\cite{Bisbo2020EfficientGlobalStructure} could be incorporated to reduce its greediness and improve exploration.

\paragraph{Walk.}%
In the second method -- \textit{random walk} -- a random point within the unit cell is selected and its suitability for accommodating a test atom is determined, where this probability is assessed using the aforementioned placement functions (discussed earlier in this section).
The placement probability for the selected point is then assessed using a pass/fail criteria.
If the check fails, nearby points are randomly sampled for improvements in $P_{\alpha}(\vec{r})$, and if a point is found to be better then the process is repeated.
In this manner, a form of pseudo-random walk is conducted until the new $P_{\alpha}(\vec{r})$ passes the check or the number of failed steps exceeds a user-defined limit (default of 10,000).
A net failure to place repeats the procedure, selecting a new random point and initialising a new random walk, until a suitable pseudo-random point can be selected.

\paragraph{Growth.}%
The third method -- \textit{growth} -- is a variant of the \textit{walk} method, but starts its check from the previously placed atom. This mimics a physical growth process.

\paragraph{Void.}%
The fourth method -- the \textit{void-finding} method -- is used to fill voids in the structure.
This method involves discretising the host cell into a set of points and evaluating the distance from each individual point to the nearest atom in the cell. 
This is achieved using the following equation
        
\begin{equation}
  d(\vec{r})=\min_i |\vec{r}-\vec{r}_i|.
\end{equation}

\noindent
where the minimization is taken over all atoms $i$ in the cell. As such, each grid point $d(\vec{r})$ is assigned a value $d(\vec{r})$ representing the distance to its closest atom. The point with the maximum value of $d(\vec{r})$ is then selected as the optimal position for placing a new atom.
This is a simplified approach to identifying the emptiest region of the search space.
The void method closely resembles early void-filling and sphere-packing techniques in crystallography~\cite{Kittel2004IntroductionSolidState}.
The tests show that it efficiently generates many prototypical crystal structures (see
\suppcol{Section SIV}%
).
Additionally, it can serve as both an initial and secondary seeding strategy, creating new nucleation sites that encourage clustering in subsequent placement methods.

\paragraph{Random.}%
For the final placement method -- \textit{random} -- an atom is placed at an arbitrary position within the cell, ensuring it is not too close to existing atoms, as determined by pair covalent radii or user-defined thresholds.
This approach is implemented to replicate existing methods used by existing random structure search algorithms~\cite{Pickard2011AbInitioRandom}.
While random placement alone is unlikely to produce a viable structure, coupling it with relaxation methods allows exploration of various local minima.
Any initial configuration will typically fall within the basin of attraction of a nearby minimum, where relaxation refines it to a stable state.
Given enough iterations, random sampling ensures comprehensive coverage of the configuration space, making it a valuable tool for completeness in structure generation.

Choice of method is randomly selected for each test atom based on a user-defined ratio of the five methods.
If one function is successful then an atom is placed.
If it fails, then a different approach is selected.
As both random walk methods require 10,000 fails and the minimum and void methods are grid dependent, these are normally highly unlikely to fail.
If all functions fail, then the structure generation exits with an error code, outputting the generated structure, which will have a lower stoichiometry than requested.
This criteria being reached suggests that the interface region is fully populated, and would require a larger empty space in the cell to accommodate placement of more atoms.
Once all atoms have been placed, the structure is outputted for energetic evaluation.

In reality, crystal interfaces often contain defects that deviate from perfect lattice matching.
Less symmetric geometries, such as grain boundaries and new interface material phases~\cite{Taylor2019FundamentalMechanismBehind}, naturally occur.
While the perfectly matched interface may be the most stable, the vast number of low-energy alternatives makes them statistically more prevalent.
Therefore, exploring these local minima is essential for a comprehensive understanding of interfaces and their properties.

\subsection{Formation Energy Evaluation}
\label{sec:formation_energy}

Most structure searches focus on identifying structures that are likely to exist on timescales relevant to everyday life.
Therefore, the criteria used in structure searches must fall within the thermodynamic stability limit, and, thus, this work targets the formation energy and resultant relative stability.
A multitude of methods can be applied to a given system to obtain these energies.
Empirical force-field models and machine learned force-field models both require specific structural data, limiting their applicability to structure searching at interfaces.
RAFFLE has been tested using various energetic calculation methods, including density functional theory (DFT) and multiple foundation machine learned potentials (with and without the explicit inclusion of van der Waals corrections).
Example cases detailed in \secref{sec:example_cases} are conducted using the CHGNet~\cite{Deng2023CHGNETPretrainedUniversalNeural} and MACE-MPA-0~\cite{batatia2023FoundationModelAtomistic} machine-learned potentials, whilst additional tests have been performed using DFT and other foundation models and are detailed in the
\suppcol{Section SVII}%
.

It is important to recognise that RAFFLE is agnostic to choice of energetic evaluation method.
However, it is not agnostic to the energetic calculation method being different between training data and new data (i.e. all data fed into a single instance of a RAFFLE generator must use the same calculation method for obtaining energetics to ensure valid direct comparisons between system energetics).

For the empirical weighting method outlined in \secref{sec:weight}, the formation energy is calculated from a system's constituent elements,
        
\begin{equation}
       E_{\mathrm{f},a} = \frac{ E_{\mathrm{tot},a} - \sum_{i} n_i E_{i} }{\sum_{i} n_i },
\label{eq:formation_energy}
\end{equation}

\noindent
where $E_{\mathrm{tot},a}$ is the total energy of system $a$ (as obtained from DFT, machine learned potentials, or any other preferred method), $E_{i}$ is the reference energy of constituent element $i$, and $n_{i}$ is the number of atoms of element $i$ in system $a$.
The constituent energy is usually taken as the energy of an individual atom in the element's bulk form at room temperature, but this can be specified by the user.

To attribute the formation energy to each element pair, we consider the number of 2-body element-pair interactions within the structure.
This is given by  

\begin{equation}
    E_{\mathrm{f},a,\alpha\beta} = E_{\mathrm{f},a} \frac{ \omega_{2,a,\alpha\beta} }{ n_{2,a} },
\end{equation}  

\noindent
where the 2-body element-pair weight, $\omega_{2,a,\alpha\beta}$, and the total number of 2-body interactions, $n_{2,a}$, are system-dependent and defined as  

\begin{equation}
    \omega_{2,a,\alpha\beta} = ( 2 r_{\mathrm{cov},\alpha\beta} )^2 \sum_{i \neq j} \mathrm{c}_{ij} \delta_{s_{i},\alpha} \delta_{s_{j},\beta},
\end{equation}  

\noindent
and  

\begin{equation}
    n_{2,a} = \sum_{i \neq j} \delta_{\mathrm{c}_{ij} \neq 0},
\end{equation}  

\noindent
respectively.
Here, $r_{\mathrm{cov},\alpha\beta}$ is the average covalent radius of elements $\alpha$ and $\beta$, multiplied by 2 to represent the typical interatomic distance.
The function $\mathrm{c}_{ij}$ is the 2-body cutoff function from \eqref{eq:cutoff:2body}, and indices $i$ and $j$ refer to atoms within system $a$.  

For angular descriptors, the formation energy is assigned per element:  

\begin{equation}
    E_{\mathrm{f},a,\alpha} = E_{\mathrm{f},a} \frac{ \omega_{2,a,\alpha} }{ n_{2,a} },
\end{equation}  

\noindent
where the element-specific weight is given by  

\begin{equation}
    \omega_{2,a,\alpha} = \sum_{\beta} \omega_{2,a,\alpha\beta}.
\end{equation} 

Additionally, when considering interface structures generated from previous iterations, we aim to make them more favourable in the generalised descriptors, as they represent the precise chemical environment under investigation.
To achieve this, we remove the formation energy of the host, which accounts for the energetic cost of surface terminations due to bond cleavage.
Thus, the formation energy of the host is subtracted from the generated structures.

The formation energy of the interface atoms can be calculated as

\begin{equation}
    E_{\mathrm{f},a} = \frac{E_{\mathrm{tot}} - n_{\mathrm{host}}E_{\mathrm{f,host}} - \sum_{j}n_{j}E_{j}}{\sum_{j}n_{j}},      
\end{equation}

\noindent
where $E_{\mathrm{f,host}}$ is the formation energy of the host structure, calculated as described in \eqref{eq:formation_energy}, and $n_{\mathrm{host}}$ is the number of host atoms.
The summation index $j$ iterates over all elements of atoms introduced by RAFFLE, with $n_{j}$ denoting the number of atoms of element $j$ and $E_{j}$ representing its reference energy.

\subsection{Example cases}
\label{sec:example_cases}

In this subsection, structure prediction is performed for set of bulk and interface systems using the RAFFLE package.
The resulting search spaces are analysed to demonstrate RAFFLE’s capabilities across different systems.
The RAFFLE parameters used for all the example cases can be found in \tabref{tab:example_cases:RAFFLE:params} and the exact workflow for the example cases is detailed in \secref{sec:methods:raffle_workflow}

Principal Component Analysis (PCA) facilitates the comparison of structural similarity by reducing high-dimensional data to its most significant components~\cite{You2024PrincipalComponentAnalysis}.
In this context, the structural data is projected onto the top $N$ principal dimensions (here, $N=2$), capturing the most relevant structural variations.
For all PCA plots, the left (right) graph in each figure represents the structures before (after) structural relaxation, with a fixed cell.
The PCA models in each figure are trained on the relaxed structures, and the unrelaxed data is subsequently projected into the same space.
This approach ensures direct comparisons.

\subsubsection{Comparison with random structure search}
\label{sec:example_cases:rss}

\begin{figure*}[ht!]
    \centering%
    \subfloat[RAFFLE - Energy sampling]{\includegraphics[height=0.3\linewidth]{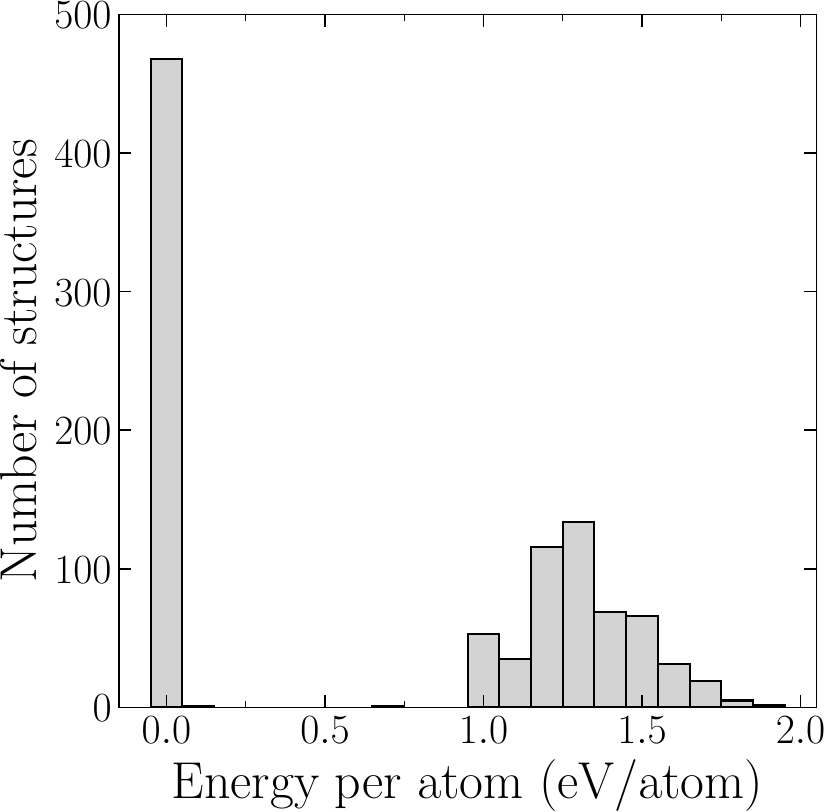}\label{fig:comparison:RAFFLE:struc_count}}%
    \hspace{2em}%
    \subfloat[RAFFLE - PCA]{\includegraphics[height=0.3\linewidth]{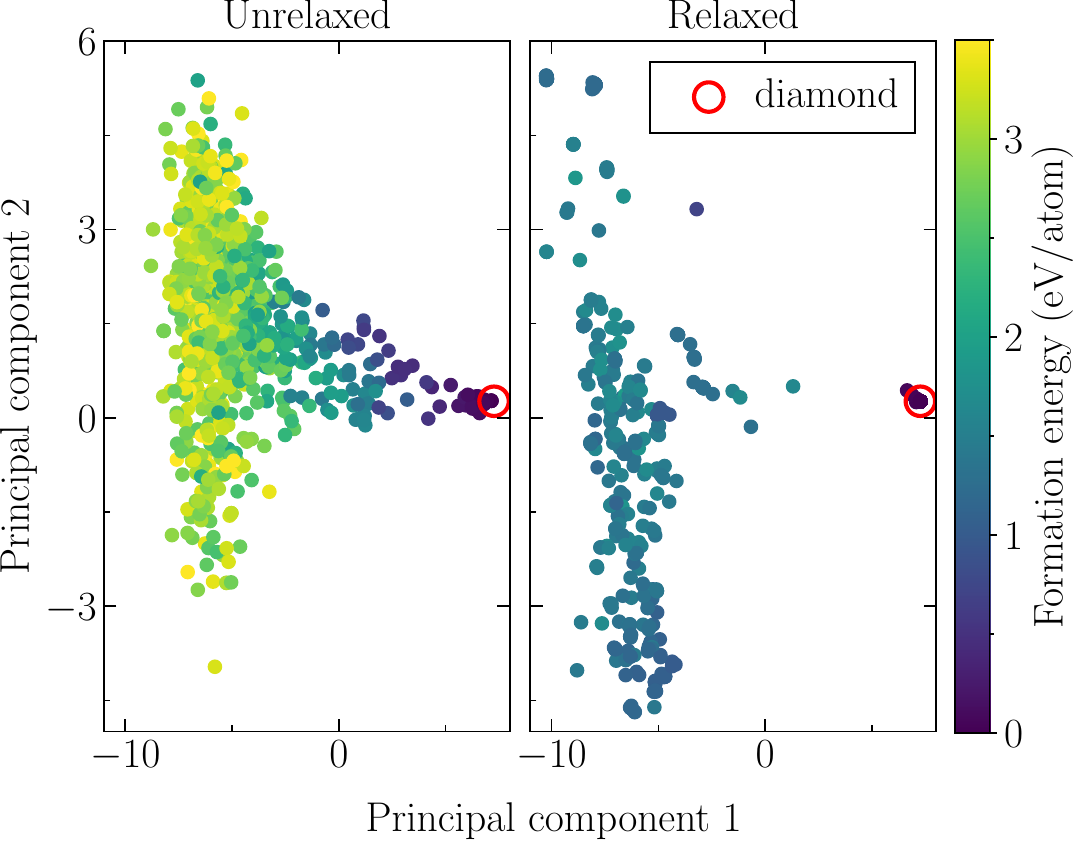}\label{fig:comparison:RAFFLE:pca}}%
    \hspace{2em}%
    \subfloat[RSS - Energy sampling]{\includegraphics[height=0.3\linewidth]{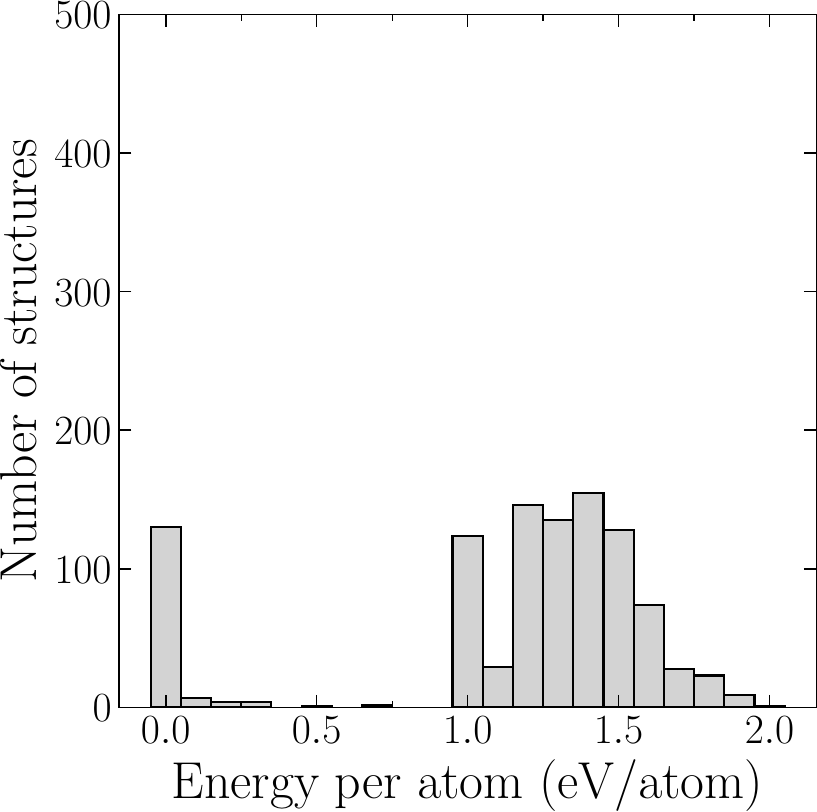}\label{fig:comparison:RSS:struc_count}}%
    \hspace{2em}%
    \subfloat[RSS - PCA]{\includegraphics[height=0.3\linewidth]{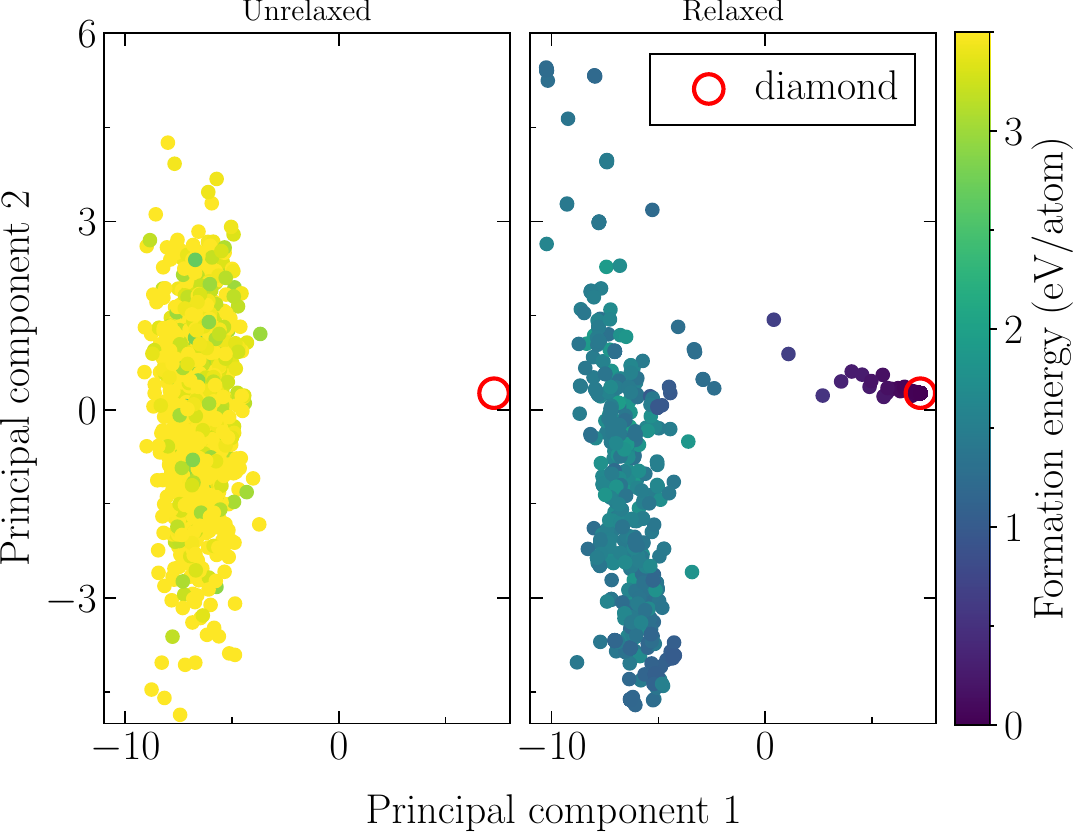}\label{fig:comparison:RSS:pca}}%

    \caption{
        Comparison of RAFFLE and random structure search (RSS).
        RAFFLE and RSS (as implemented in AGOX~\cite{Christiansen2022AtomisticGlobalOptimization}) each generate 1000 structures to search for the diamond phase of carbon.
        Histogram of relaxed structures versus energy above the diamond phase for \protect\subref{fig:comparison:RAFFLE:struc_count} RAFFLE and \protect\subref{fig:comparison:RSS:struc_count} RSS.
        Principal component analysis (PCA) of unrelaxed and relaxed structures for \protect\subref{fig:comparison:RAFFLE:pca} RAFFLE and \protect\subref{fig:comparison:RSS:pca} RSS: method-generated (left) unrelaxed and (right) relaxed structures.
        Graphite is not generated by either method due to the enforced cell and density.
        The search is performed for 8 carbon atoms in a cubic cell ($a = 3.567$~\si{\angstrom}).
        Energetics are computed using CHGNet~\cite{Deng2023CHGNETPretrainedUniversalNeural}, and structural relaxations (fixed cell) are performed using FIRE~\cite{Bitzek2006StructuralRelaxationMade} (\texttt{fmax=0.05}, \texttt{steps=100}).
    }
    \label{fig:comparison}
\end{figure*}

We first apply RAFFLE to identify the diamond phase of carbon, comparing its performance with random structure search (RSS).

For this test, RAFFLE is not provided with an initial database.
Instead, we generate structures by placing eight carbon atoms into a cubic host cell (lattice constant $3.567$~\si{\angstrom}) and repeating this process 200 times, yielding 1000 structures.
The explored phase space, reduced via principal component analysis (PCA), is shown in \figref{fig:comparison:RAFFLE:pca}.
To assess stability, we repeat the test 20 times with different random seeds, consistently obtaining qualitatively similar results (see
Section SVIII%
).
The choice of unit cell and atom count constrains the search to phases within diamond’s density range, preventing the identification of graphite.

For comparison, we perform RSS under identical conditions -- using the same host cell, energy calculator, and relaxation optimiser.
The explored phase space for RSS is shown in \figref{fig:comparison:RSS:pca}, where all data is projected onto the same PCA space derived from the relaxed RAFFLE structures.
Before atomic relaxation, RAFFLE explores a broader phase space than RSS, covering a wider range of principal component values while also sampling more structures near the minimum-energy diamond phase.
Additionally, RAFFLE reveals clusters within the search space that RSS does not access and samples a greater range along $\mathrm{PC2} = 0$.
After relaxation, both methods yield similar distributions, though RSS produces more structures along this line—possibly due to an imposed iteration limit of 100 for computational efficiency.
Notably, RAFFLE generates over 400 structures that relax to diamond, compared to just 150 from RSS.

\subsubsection{Bulk}
\label{sec:example_cases:bulk}

\afterpage{%
\begin{figure*}[hbt!]{}
    \centering
    \subfloat[Carbon]{\includegraphics[width=0.3\linewidth]{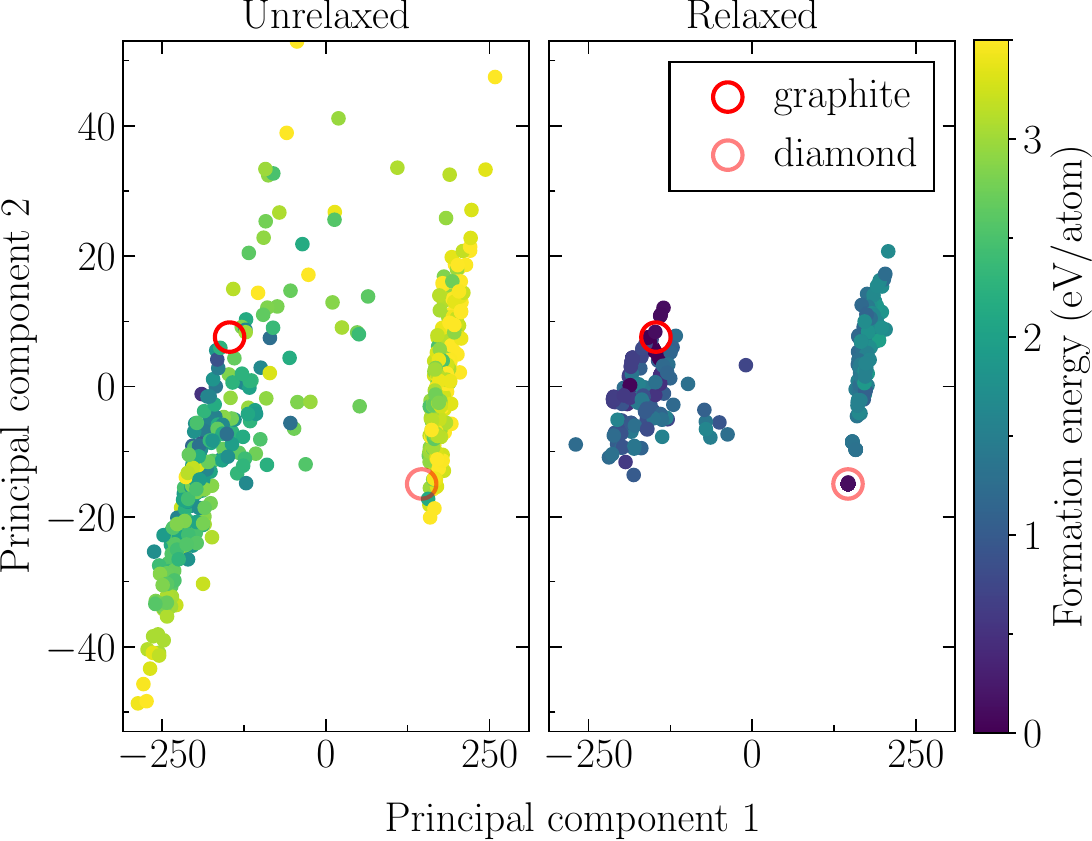}\label{fig:bulk_pcas:diamond_graphite}}
    \hspace{1em}
    \subfloat[Aluminium]{\includegraphics[width=0.3\linewidth]{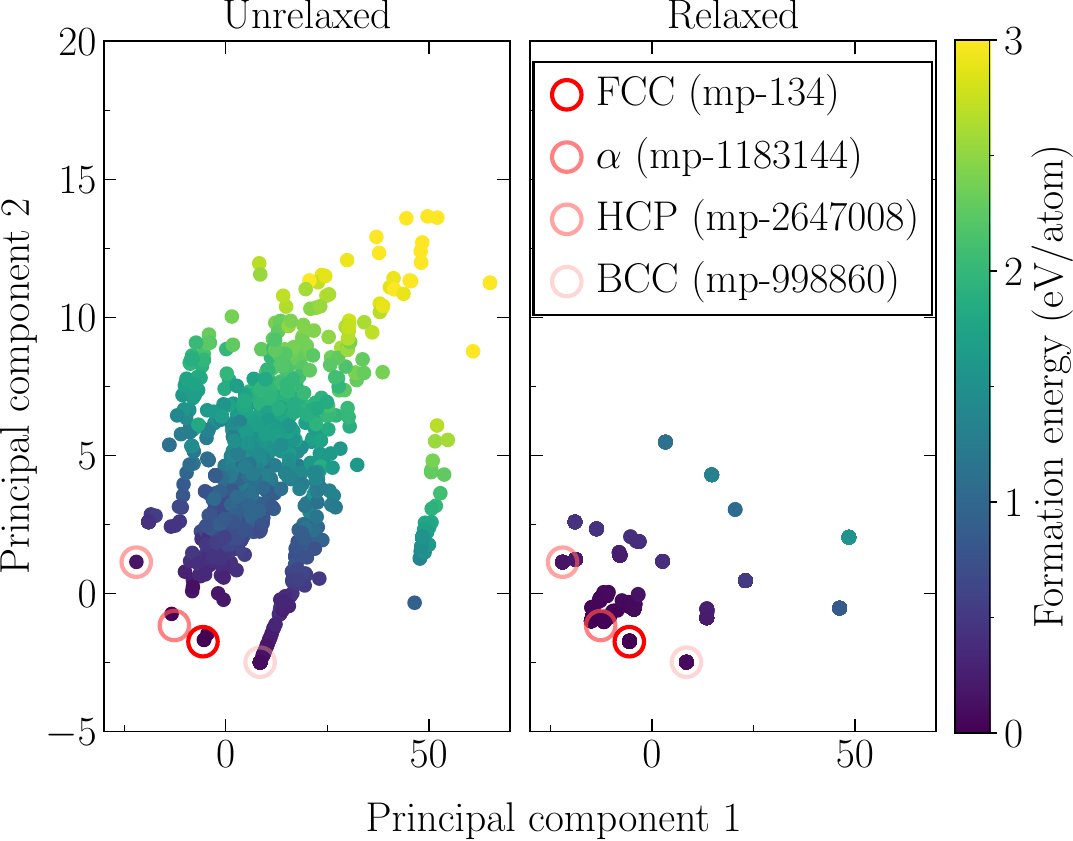}\label{fig:bulk_pcas:Al}}
    \hspace{1em}
    \subfloat[MoS$_{2}
$]{\includegraphics[width=0.3\linewidth]{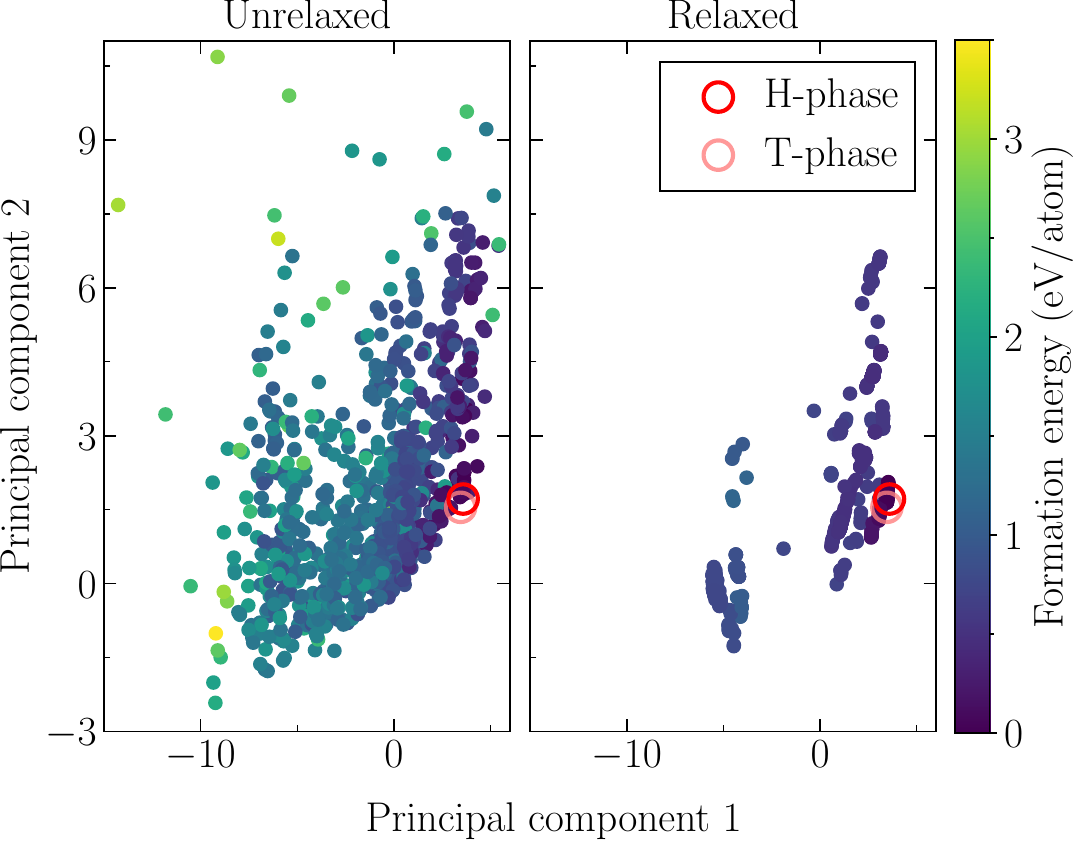}\label{fig:bulk_pcas:MoS2}}

    \caption{
        Principal component analysis (PCA) of RAFFLE-generated bulk structures (pre- and post-relaxation).
        Red circles highlight known experimental or theoretical phases.
        \protect\subref{fig:bulk_pcas:diamond_graphite} Bulk carbon search using cubic and hexagonal Bravais lattices (500 structures).
        \protect\subref{fig:bulk_pcas:Al} Bulk aluminium search using cubic and hexagonal Bravais lattices (1200 structures).
        \protect\subref{fig:bulk_pcas:MoS2} Bulk MoS$_{2}$ search using a single hexagonal Bravais lattice (1000 structures).
        Subplots present RAFFLE-generated (left) unrelaxed and (right) relaxed structures.
        For carbon, the graphite ABC stacking (dark red circle) is not fully recovered due to the cell constraints, instead identifying AB stacking.
        For aluminium, known Materials Project~\cite{Jain2013CommentaryMaterialsProject} phases are rescaled to the closest available lattice constant.
        FCC, HCP, and BCC phases are recovered exactly, while the $\alpha$ lanthanum-like phase is only partially recovered due to stacking constraints in the chosen smaller cells.
        Energetics are computed using CHGNet~\cite{Deng2023CHGNETPretrainedUniversalNeural}, and structural relaxations (fixed cell) are performed using FIRE~\cite{Bitzek2006StructuralRelaxationMade} (\texttt{fmax=0.05}, \texttt{steps=100}).
    }
    \label{fig:bulk_pcas}
\end{figure*}
}

We apply RAFFLE to three bulk systems: carbon, aluminium, and MoS$_{2}$.
Although RAFFLE is not specifically designed for bulk structure searches, these cases demonstrate its broader applicability within a well-established search space.
For practical bulk structure searches, we recommend using specialised tools optimised for this purpose, such as USPEX~\cite{Oganov2006CrystalStructurePrediction}, AIRSS~\cite{Pickard2006HighPressurePhases}, and CALYPSO~\cite{Wang2012CALYPSOMethodCrystal}, which incorporate symmetry handling and other efficiency enhancements that are less relevant for interface searches.
Additionally, the AGOX package~\cite{Christiansen2022AtomisticGlobalOptimization}, designed for surface global optimisation, also includes functionality for bulk random structure searches.

\paragraph{Carbon.}
\label{par:example_cases:bulk:C}

We now explore RAFFLE’s ability to identify multiple existing phases with distinctly different densities within a shared phase space.
To this end, we alternate between two host cells: a cubic cell with a lattice constant of $3.567$~\si{\angstrom} and a hexagonal close-packed (HCP) cell with lattice parameters $a = 2.47$~\si{\angstrom}, $b = 4.94$~\si{\angstrom}, and $c = 7.80$~\si{\angstrom}.
In each case, 8 atoms are placed per cell, and no initial database is provided, meaning the search begins with no prior knowledge.

The search proceeds as follows: 50 iterations are performed using the HCP cell, followed by 50 iterations using the cubic cell, with prior learning retained between the two.
The HCP cell is explored first to demonstrate that, even though the most energetically favourable phase (graphite) is identified first, the search still succeeds in locating the diamond phase.
The resulting PCA search space is shown in \figref{fig:bulk_pcas:diamond_graphite}.
The search successfully recovers graphite with AB stacking and the diamond phase, though ABC-stacked graphite (the lowest energy stacking) is not identified due to cell constraints.

This demonstrates that, while the method is naturally biased towards the most energetically favourable phase -- graphite, in this case -- continuing the search in a cubic cell allows the diamond phase to be consistently identified.
Two different placement method ratios are tested, with results presented in
\suppcol{Section SIX}%
.

\paragraph{Aluminium.}
\label{par:example_cases:bulk:Al}

We now apply RAFFLE to the search for bulk aluminium phases, demonstrating its ability to identify multiple potentially stable phases with similar formation energies (as highlighted by the Materials Project~\cite{Jain2013CommentaryMaterialsProject}).
As before, the search begins with no initial database.
Two Bravais lattices -- cubic and hexagonal -- are used as host structures.
For each lattice type, distinct host cells are generated using six different lattice constants in the range $a = c = 3.1$--$5.4$\si{\angstrom}, in steps of $0.46$\si{\angstrom}.
Atoms are then placed into these host cells to approximate a density of $2.7$~\si{\gram/\cm\cubed}.
The search proceeds by cycling through each host lattice and repeating the process 20 times, generating a total of 1200 structures.
The resulting PCA is shown in \figref{fig:bulk_pcas:Al}.

The search successfully identifies phases equivalent to (or closely approximating) those found in the materials database, subject to the constraints imposed by the small cells used in the search.
This limitation most significantly affects the $\alpha$ lanthanum-like phase, where the small unit cell cannot support the full ABCB stacking observed in the Materials Project entry (mp-1183144).
Instead, a similar structure with AB stacking is obtained.
The search recovers the FCC (mp-134), $\alpha$ lanthanum-like (mp-1183144)~\cite{Murray1985AluminiumCopperSystem}, HCP (mp-2647008)~\cite{Pickard2010AluminiumTerapascalPressures}, and BCC (mp-998860)~\cite{Vailionis2011EvidenceSuperdenseAluminium} phases, with the aforementioned stacking limitation for the $\alpha$ phase.
Since each search occurs at fixed cell dimensions, the presence and relative stability of these phases depend on how closely the lattice constant aligns with the equilibrium value of each structure.
Additionally, a simple cubic phase is observed.
Due to the constraints on the lattice constants sampled, the computed order of formation energies may not be entirely accurate, as some phases experience artificial strain.

The structures located within $40\lesssim \mathrm{PC1} \lesssim 60$ and $-1\lesssim \mathrm{PC2} \lesssim 5$ in PCA space are hexagonal, with each layer forming an in-plane triangular bonding structure.
However, these configurations are energetically unfavourable, with formation energies exceeding $1.0$~\si{\electronvolt}/atom. Overall, RAFFLE’s ability to recover not only the ground-state structure but also high-pressure and metastable phases highlights its versatility.

\paragraph{\texorpdfstring{\textnormal{\textbf{MoS}}$_{2}$}{MoS2}.}
\label{par:example_cases:bulk:MoS2}

We now extend RAFFLE’s application to multi-species, multi-phase systems, specifically van der Waals materials.
The initial database contains only bulk molybdenum and bulk sulphur, meaning the search starts without prior knowledge of Mo–S bonding.
The host cell used is an HCP structure with lattice parameters $a = 3.19$~\si{\angstrom} and $c = 13.1$~\si{\angstrom}.
The choice of $c$ is deliberate, as it lies between the values observed for the H- and T-phases, allowing the unit cell to accommodate two layers and enabling both AB-stacked H-phase and AA-stacked T-phase configurations to form.
Each structure contains 2 molybdenum and 4 sulphur atoms.
A total of 1000 structures are generated and the the resulting PCA is presented in \figref{fig:bulk_pcas:MoS2}.

Both the H- and T-phases are successfully recovered, including their respective lowest-energy stacking arrangements.
These results remain consistent across different computational methods, including CHGNet (both with and without the DFT-D4 correction~\cite{dftd4-1,dftd4-2,dftd4-periodic}) and the MACE-MP-0 calculator.
In all cases, RAFFLE correctly identifies the H- and T-phase stacking configurations (see
\suppcol{Fig. S6}%
).

\subsubsection{Interfaces}
\label{sec:example_cases:interfaces}

\afterpage{%
\begin{figure*}[hbt!]{}
    \centering
    \subfloat[ScS$_{2}$-Li]{\includegraphics[width=0.40\linewidth]{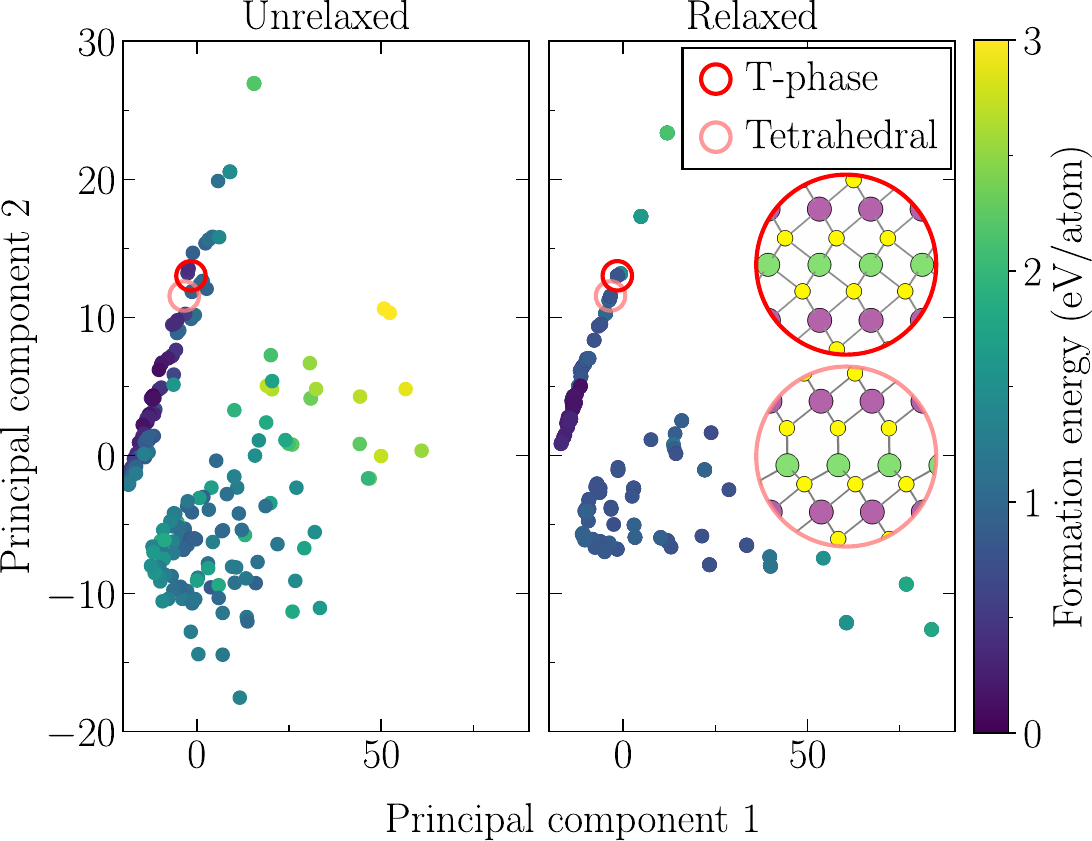}\label{fig:interface_pcas:ScS2-Li}}
    \hspace{1em}
    \subfloat[Si\|Ge]{\includegraphics[width=0.40\linewidth]{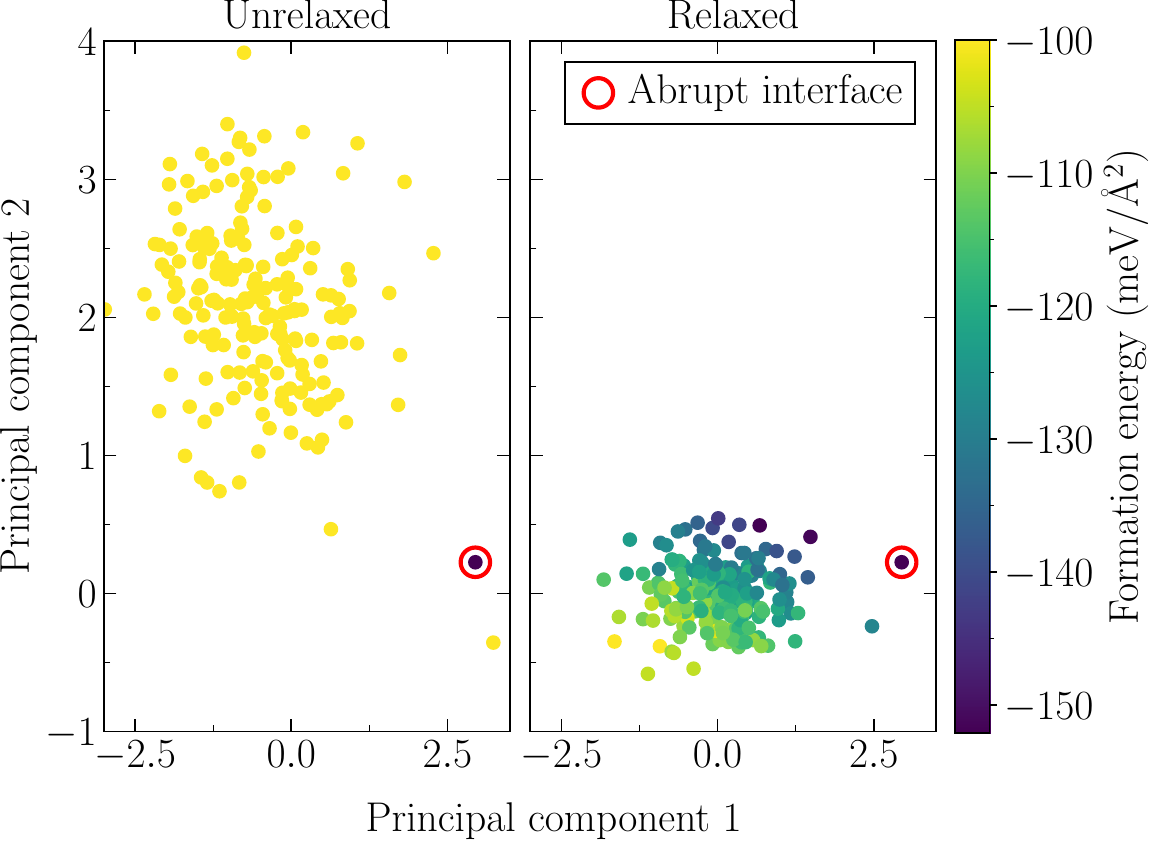}\label{fig:interface_pcas:Si-Ge}}
    \caption{
        Principal component analysis (PCA) of RAFFLE-generated interface structures (pre- and post-relaxation).
        \protect\subref{fig:interface_pcas:ScS2-Li} Li intercalation into ScS$_{2}$ (250 structures).
        The most and second-most favourable forms (dark and light red circles, respectively) are a T-phase structure with sulphur-mediated bonding and a tetrahedrally bonded structure.
        A range of host cell strains and Li insertions (1–2 atoms) are explored.
        Insets show the two most stable structures.
        \protect\subref{fig:interface_pcas:Si-Ge} (001)\|(001) Si\|Ge interface, with 16 Si and 16 Ge atoms inserted into a $2\times2$ host cell (lattice matched to $a = 5.54$~\si{\angstrom}, the midpoint between Si and Ge lattice constants).
        Subplots present RAFFLE-generated (left) unrelaxed and (right) relaxed structures.
        For the ScS$_{2}$-Li case, energetics are computed using CHGNet~\cite{Deng2023CHGNETPretrainedUniversalNeural}, whilst energetics for the Si\|Ge case are computed using MACE-MPA-0~\cite{batatia2023FoundationModelAtomistic}
        Both cases use perform structural relaxations (fixed cell) using FIRE~\cite{Bitzek2006StructuralRelaxationMade} (\texttt{fmax=0.05}, \texttt{steps=100}).
    }
    \label{fig:interface_pcas}
\end{figure*}
}

\paragraph{\texorpdfstring{\textnormal{\textbf{ScS}}$_{2}$ \textnormal{\textbf{Li}}}{ScS2 Li}-intercalation.}
\label{par:example_cases:interfaces:ScS2-Li}

We now demonstrate RAFFLE’s ability to explore intercalation by providing a fixed host structure, ScS$_{2}$, while allowing lithium atoms to be placed freely.  This structure is typical of intercalation into electrodes for batteries.
The initial database includes only bulk ScS$_{2}$ and bulk lithium, meaning no prior knowledge of Li-S or Li-Sc interactions is provided.
The host structure is systematically stretched and compressed along both the $a$-$b$ plane and the $c$-axis, generating 25 unique distortions (five variations in the $a$-$b$ plane and five along $c$).
RAFFLE then inserts between one and two lithium atoms into each distorted host.
For each host, the Li intercalation count is iterated over, then repeated for the next host.
This cycle continues until all host structures have been explored, resulting in 250 generated structures.
The resulting PCA is shown in \figref{fig:interface_pcas:ScS2-Li}, highlighting key intercalation sites within ScS$_{2}$.
The two most stable intercalation phases are depicted in the insets of \figref{fig:interface_pcas:ScS2-Li}.
As seen, lithium adopts two distinct configurations: a T-phase-like layering and a tetrahedral-like coordination.

The results are generated using the CHGNet calculator, and strongly agree with previous results~\cite{Price2023FirstPrinciplesStudy} carried out with GGA-PBE~\cite{perdewGeneralizedGradientApproximation1996} both applying the RAFFLE methodology.

\paragraph{\texorpdfstring{\textnormal{\textbf{Si}}\|\textnormal{\textbf{Ge}}}{Si|Ge} interface.}
\label{par:example_cases:interfaces:Si-Ge}

RAFFLE is now applied to the study of an Si\|Ge (001)\|(001) interface.
The initial database includes only bulk Si and bulk Ge, meaning no prior knowledge of Si-Ge bonding is provided.
The starting structure consists of a silicon slab, constructed as a $2\times2\times2$ supercell expansion of the 8-atom cubic diamond-phase cell.
An equivalent slab is generated for germanium, and the two are combined into a single unit cell with in-plane lattice constants $a=b=5.54$~\si{\angstrom} -- the midpoint between Si and Ge lattice constants.
A vacuum region of $5.54$~\si{\angstrom} is introduced between the slabs, ensuring that atoms are placed only within this bounding region.
Due to periodic boundary conditions, the system naturally forms two interfaces: an abrupt Si\|Ge interface and a second interface that is free to evolve during RAFFLE’s search.

During structure generation, 32 atoms (16 Si, 16 Ge) are placed within the vacuum region.
A total of 200 structures are generated and the resulting PCA is shown in \figref{fig:interface_pcas:Si-Ge}. It is important to note that the system contains two interfaces due to periodic conditions: one abrupt and one generated by RAFFLE.
Additionally, the fully abrupt The abrupt interface is not part of RAFFLE’s search but is included in the PCA as a point of comparison with both unrelaxed and relaxed intermixed structures.

Energetic comparisons are made (using the MACE-MPA-0 calculator) against an abrupt Si\|Ge interface with $2.5$ layers of Si and Ge on either side, which relaxes to a formation energy of $-152.46$~\si{\milli\electronvolt/\angstrom\squared}.
RAFFLE identifies multiple intermixed interfaces with similar formation energies, including two stable structures at $-152.08$ and $-151.46$~\si{\milli\electronvolt/\angstrom\squared}.
GGA-PBE calculations with GPAW~\cite{Mortensen2024GPAWOpenPythonPackage} further show that these RAFFLE-predicted structures are $2.03$ and $1.27$~\si{\milli\electronvolt/\angstrom\squared} more stable than the abrupt interface.
The most stable interfaces retain a diamond-like structure with some Si/Ge intermixing across the boundary.
Further analysis of the structures of these interfaces can be found in
\suppcol{Section SXI}%
.

\paragraph{Graphene-encapsulated \textnormal{\textbf{MgO}}}
\label{par:example_cases:interfaces:C-MgO}

The intercalation of materials between layered structures has become an important approach for studying interface materials.
A detailed investigation of MgO intercalated into graphite has revealed the formation of distinct phases that differ from the bulk~\cite{Vasu2016VanMolecules}.
Using RAFFLE with GGA-PBE, we explored these intercalation phases, identifying the formation of rocksalt, hexagonal, and mixed phases, the latter featuring sub-interfaces between rocksalt and hexagonal regions.
The stability of these phases is strongly influenced by encapsulation effects and layer thickness~\cite{Pitfield2024PredictingPhaseStability}.
Our results indicate that the monolayer rocksalt phase is significantly stabilised by graphene, with a reduced surface energy compared to an isolated surface, making it more favourable than the graphene-like hexagonal monolayer.
Additionally, the search identified rotated and defective variants close in formation energy to these phases, further validating RAFFLE’s ability to capture the C-MgO energy landscape accurately.

\section{Benchmarks}
\label{sec:benchmarks}

\begin{figure*}
    \subfloat[Bond length cutoff]{\includegraphics[width=0.31\linewidth]{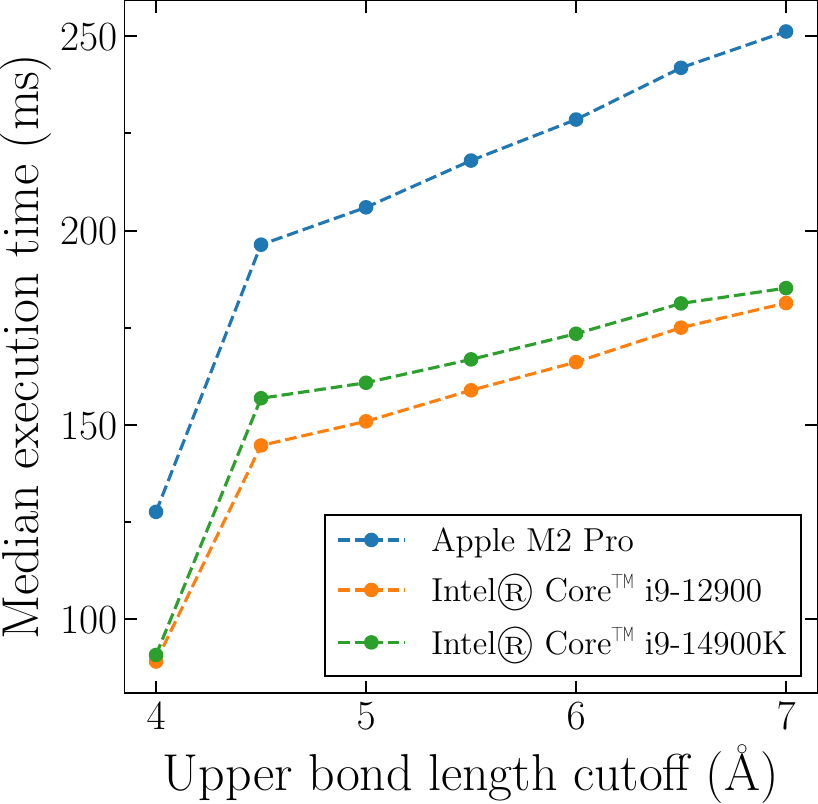}\label{fig:benchmarks:cutoff}}
    \hspace{1em}
    \subfloat[Dataset size]{\includegraphics[width=0.31\linewidth]{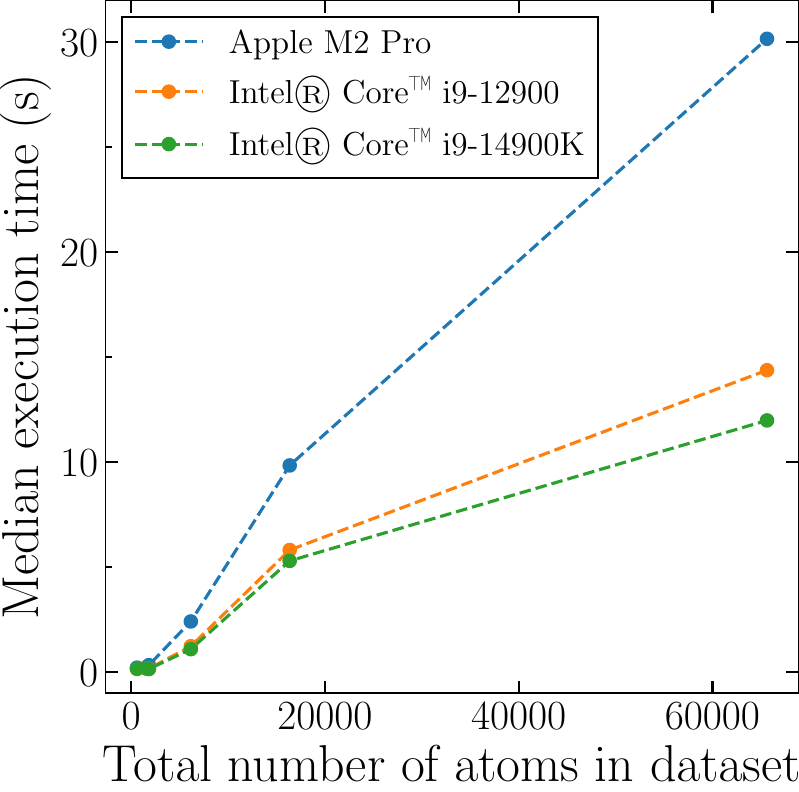}\label{fig:benchmarks:num_atoms}}
    \hspace{1em}
    \subfloat[Grid spacing]{\includegraphics[width=0.31\linewidth]{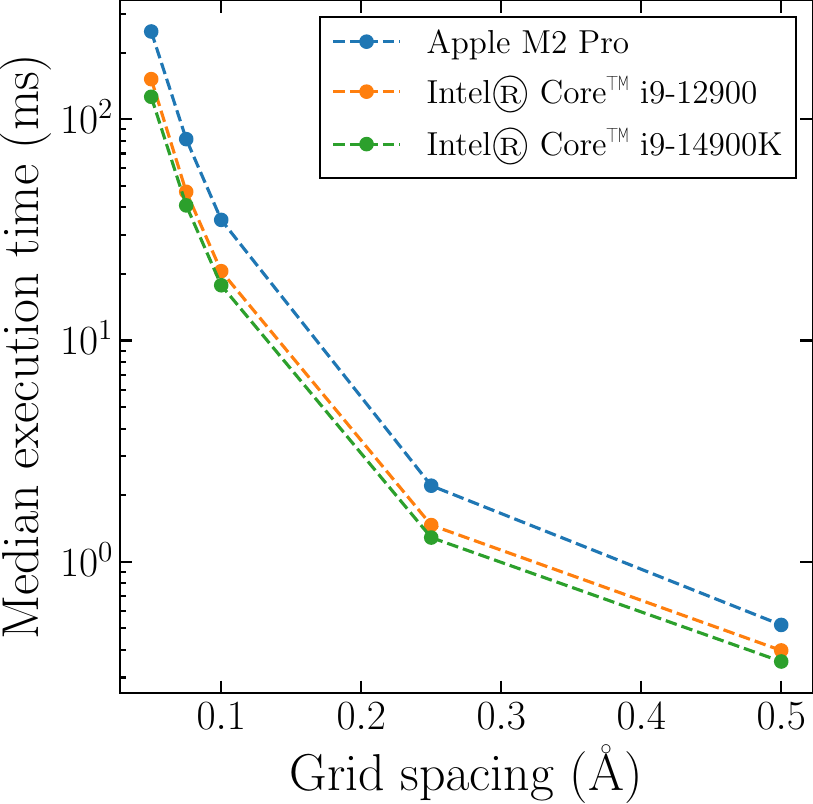}\label{fig:benchmarks:placement_method:min}}
    \caption{
        Benchmarks across multiple architectures.
        \protect\subref{fig:benchmarks:cutoff} Upper bond length cutoff versus time required to learn RAFFLE descriptors for a database of 198 eight-atom carbon materials.
        \protect\subref{fig:benchmarks:num_atoms} Total number of atoms in a dataset versus time to learn RAFFLE descriptors.
        \protect\subref{fig:benchmarks:placement_method:min} Grid spacing versus time to place one carbon atom in a cubic host cell (lattice constant $3.567$\si{\angstrom}) containing a single carbon atom, using the global minimum placement method.
        For all benchmarks, the following parameters are used: $k_{\mathrm{B}}T = 0.4$~\si{\electronvolt}, bin widths of $0.025$~\si{\angstrom}, $\pi/200$~\si{\radian}, and $\pi/200$~\si{\radian} for 2-, 3-, and 4-body descriptors, respectively.
    }
    \label{fig:benchmarks}
\end{figure*}

\Figref{fig:benchmarks} shows a set of benchmarks performed on three architectures (one MacOS and two Linux).
These benchmarks highlight the cost and scalability of varying some of the parameters involved in RAFFLE descriptor learning and structure generation.
It can be seen that learning time is directly dependent on bond length and number of atoms in the dataset.
The exact form of the relations are heavily system dependent (density being a main contributor as this determines the number of $n$-body contributions that are included in evaluation of the generalised descriptors.
Meanwhile, execution time is inversely dependent on grid spacing, $s$, roughly proportional to $1/s^{3}$.
Additional benchmarks with further parameter variations are provided in
\suppcol{Section SV}%
.

\section{Discussions}
\label{sec:discussions}

RAFFLE has been applied to three categories of materials: bulk, intercalation structures, and 3D bulk\|3D bulk interfaces.
The bulk searches using CHGNet align well with known results, reinforcing the reliability of machine-learned potentials for structure prediction~\cite{batatia2023FoundationModelAtomistic}.
For lithium intercalation into transition metal dichalcogenides, RAFFLE successfully reproduces the set of known phases experimentally observed~\cite{vanDijk1980CrystalStructureLiScS2} using either CHGNet or GGA-PBE, providing strong validation of the methodologies applied.
This success extends to MgO intercalation, which involves multiple intercalants and a significantly larger search space.

Si\|Ge interfaces have been extensively studied for decades~\cite{Shiraki2011SiliconGermaniumSiGe,Virgilio2011OpticalGainShort,Chen2014StructuralOpticalCharacteristics,Hu2021DirectObservationPhonon}.
These structures lack van der Waals interactions and are instead strongly influenced by the host lattice.
The results reveal a prevalence of metastable and stable intermixed states, suggesting that an abrupt interface is unlikely to form.
The identified structures fall within the energy range associated with room-temperature thermal fluctuations ($26$~\si{\milli\electronvolt}), making them thermally accessible.
Furthermore, statistical entropy considerations favour disordered interfaces over the abrupt configuration, even if the latter is energetically lower.
This highlights strain relief as a key driver in interface formation~\cite{PaqueletWuetz2023ReducingChargeNoise}.

When the Si\|Ge interfaces are further evaluated using GGA-PBE DFT, the two best RAFFLE-generated structures exhibit lower formation energies than the abrupt interface by $2.03$ and $1.27$~\si{\milli\electronvolt/\angstrom\squared}, reinforcing the idea that intermixing reduces interfacial strain.
While MACE-MPA-0 correctly captures the energy landscape, it struggles with precise energetic ordering, though RAFFLE's sampling effectively navigates this limitation.
Critically, this demonstrates that RAFFLE mitigates against human bias towards abrupt interfaces, which have traditionally been the default approach~\cite{Park2018QuickGuideFirst,Taylor2020ARTEMISAbInitio,Stengel2007AbInitioTheory,Okamoto2006LatticeRelaxationOxide,Mondal2021AbInitioModeling,Hepplestone2014EffectMetalIntermixing,Hocker2021AbInitioInvestigation,Cheng2023AbInitioStudy}.

These studies highlight the importance of choosing an appropriate energy calculator for structure search as has also noted for surfaces~\cite{Focassio2024PerformanceAssessmentUniversal,Pitfield2025AugmentationUniversalPotentials,christiansen2025deltamodelcorrectionfoundationmodel, radova2025finetuningfoundationmodelsmaterials}.
A calculator must accurately represent the relevant chemical energy landscape; otherwise, RAFFLE cannot effectively predict stable structures.
A suitable calculator must, at least, identify the correct local minima, even if their energetic ordering is imperfect.
Current foundation models appear insufficient to accurately capture the properties of interface systems.
While high-accuracy methods such as DFT improve predictions, they are computationally expensive.
If cheaper methods sufficiently approximate the energy landscape, they can significantly accelerate searches, but they must be rigorously validated.

Structure search has advanced considerably, and this work introduces a new capability by integrating physical insights with genetic algorithms.
This allows efficient exploration of complicated systems while avoiding human biases.
However, the approach also has inherent limitations: symmetries commonly exploited in bulk searches to accelerate convergence are often absent in interfaces and surfaces.
This underscores the need for multiple distinct search strategies for highly disordered structures~\cite{Bisbo2022GlobalOptimizationAtomic,Kaappa2021GlobalOptimizationAtomic}.
By leveraging dynamically evolving 2-, 3-, and 4-body distribution functions as structural descriptors, RAFFLE enables vast configurational searches at significantly reduced computational cost compared to traditional random searches.
This opens the door to larger-scale interface studies and could be integrated with methods for exploring charge defects~\cite{MosqueraLois2023IdentifyingGroundState}.
The combination of physical models~\cite{Senftle2016ReaxFFReactiveForce} and recent advances brought about by deep learning (e.g. as foundation models~\cite{PyzerKnapp2025FoundationModelsMaterials,Deng2023CHGNETPretrainedUniversalNeural,batatia2023FoundationModelAtomistic} and novel relaxation techniques~\cite{Hessmann2025AcceleratingCrystalStructure,Yang2024ScalableCrystalStructure}) offers promising avenues for scaling interface structure prediction and accelerating the discovery of novel material phases.

\section{Methods}
\label{sec:methods}

\subsection{Installation and documentation of RAFFLE}
\label{sec:methods:install}

The RAFFLE software package is written in Fortran with a Python wrapper and can be used in three ways: (1) as a stand-alone Fortran executable, (2) as a Fortran library, or (3) as a Python library for direct integration into workflows.
This article focuses on the Python library, as it offers the greatest applicability and interoperability with widely used materials science libraries\cite{ase-paper,Ganose2025Atomate2ModularWorkflows,Deng2023CHGNETPretrainedUniversalNeural,batatia2023FoundationModelAtomistic,Christiansen2022AtomisticGlobalOptimization}.
The recommended method of installation for the Python library is via the \code{pip} package installer.
For further details on installation, refer to the \code{README} or the \code{ReadTheDocs}~\cite{RAFFLEDocs2025}.

\code{pip} is the recommended method of installation for the Python library.
For the Fortran versions of the software (executable and library), \code{fpm} (Fortran Package Manager)~\cite{fpm2024} is the recommended method of installation.
Alternatively, all versions support installation via \code{CMake}.

The RAFFLE library relies on a set of external libraries; these include (where versions tested with are bracketed)
\texttt{f90wrap} (0.2.14--0.2.16),
\texttt{numpy} (1.26.4--2.2),
\texttt{meson} (1.6.0),
\texttt{cython} (3.0.11), and 
\texttt{scikit-build-core} (0.10.7).
Whilst not a requirement, it is also recommended to use ASE; RAFFLE has been tested with
\texttt{ase} (3.23.0).
The Fortran backend of the package has been successfully compiled and the using \texttt{gcc} 13.1.0--14.2.0.
The Python wrapper has been tested using versions 3.11, 3.12, and 3.13.
Earlier compilers and Python versions may encounter problems.

\textbf{Note:} ASE is not a requirement to build RAFFLE and RAFFLE has its own built-in atomic structure object, but the ASE Atoms object is recommended for ease of use and compatibility with other Python-based structure manipulation and evaluation libraries.
As such, ASE is set as a project dependency.

\subsection{Example cases: Parameters and workflow}
\label{sec:methods:raffle_workflow}

\begin{table}
    \centering
    \caption{
        RAFFLE parameters used for the example cases.
        Details regarding how to set the parameters are outlined in
        \suppcol{Section SII.1}%
        .
        The $k_{\mathrm{B}}T$ parameter is set using \texttt{set\_kBT()}, whilst the placement method ratio is defined during generation with the \texttt{method\_ratio} dictionary.
        The ratios are renormalised after being read in.
    }
    \label{tab:example_cases:RAFFLE:params}
    \begin{tabular*}{\textwidth}{@{\extracolsep\fill}c c c c c c c}
        \toprule%
        & \multicolumn{6}{@{}c@{}}{Parameter} \\\cmidrule{2-7}%
        & \multirow{2}{*}{$k_{\mathrm{B}}T$} & \multicolumn{5}{@{}c@{}}{Placement method ratio} \\\cmidrule{3-7}%
        Example & & min & walk & grow & void & rand \\\midrule%
        C         &     0.4 & 1.0 & 0.4 &  0.2 & 0.4 & 0.001 \\%
        Al        &     0.4 & 1.0 & 0.5 &  0.0 &  0.5 & 0.001 \\%
        MoS$_{2}$ &     0.4 & 1.0 & 0.5 &  0.0 &  0.5 &  0.05 \\%
        ScS$_{2}$-Li &  0.4 & 1.0 & 1.0 &  0.0 &  1.0 &   0.5 \\%
        Si\|Ge       &  0.2 & 1.0 & 0.25 &  0.25 &  0.1 &  0.01 \\%
        C-MgO~\cite{Pitfield2024PredictingPhaseStability} & 1.0 & 1.0 & 1.0 & 0.0 & 1.0 & 0.0 \\%
        \botrule%
    \end{tabular*}
\end{table}

The exact workflow for the example cases follows this general pattern:
\begin{enumerate}
    \item RAFFLE descriptors are initialised on a starting dataset
    \item structures are generated using RAFFLE in batches of 5,
    \item energies are evaluated and atomic positions relaxed,
    \item structures and energies are added to the RAFFLE database, which RAFFLE learns from, and
    \item repeat the process a for a set of host cells an number of times each.
\end{enumerate}
The starting dataset depends on the example case.
For bulk cases, no data regarding the known phases for the desired stoichiometry is included.

For the following systems, we use a set of methods for building structural descriptors.
For the carbon-diamond and MoS$_{2}$ searches (i.e. single-Bravais lattice searches), the AGOX Fingerprint descriptor (with default parameters) is used.
For systems with varying cell sizes and number of atoms (and the Si\|Ge system), the local descriptor known as SOAP (smooth overlap of atomic positions)~\cite{Bartok2013RepresentingChemicalEnvironments} is employed (with radial cutoff of $5$~\si{\angstrom}, and then averaged over the atoms in the system to obtain a fingerprint for a \textit{super-atom}, which describes the whole system, similar to the techniques outlined in other studies~\cite{Rosenbrock2017DiscoveringBuildingBlocks,Himanen2020DScribeLibraryDescriptors}.

\subsection{Energetic and structural calculations}
\label{sec:methods:calculations}

For all example cases in the main article, energetics are evaluated using the CHGNet calculator~\cite{Deng2023CHGNETPretrainedUniversalNeural}, except for the Si\|Ge study, which utilises the MACE-MPA-0 calculator~\cite{batatia2023FoundationModelAtomistic}.
In all cases, structural relaxations (fixed cell) are performed using FIRE~\cite{Bitzek2006StructuralRelaxationMade} (with relaxation parameters \texttt{fmax=0.05} and \texttt{steps=100}).
These relaxation parameters mean that atomic positions relaxation is performed for 100 steps or until forces are converged to within $0.05$~\si{\electronvolt/\angstrom}, whichever is reached first.
Parity plots detailing the comparison between energies calculated using the GGA-PBE DFT functional~\cite{perdewGeneralizedGradientApproximation1996} and CHGNet predictions for a variety of interface structure searches can be seen in
\suppcol{Section SVI}%
, which, in general, provide strong qualitative incentive towards application of CHGNet in the context of the example cases.
Some cases have been repeated with CHGNet+DFT-D4~\cite{dftd4-1,dftd4-2,dftd4-periodic}, MACE-MP-0, VASP~\cite{kresseEfficientIterativeSchemes1996,kresseEfficiencyAbinitioTotal1996} (GGA-PBE)~\cite{perdewGeneralizedGradientApproximation1996}, and GPAW~\cite{Mortensen2024GPAWOpenPythonPackage}  (GGA-PBE), which are presented in
\suppcol{Section SVII}%
.
All variations result in qualitatively equivalent results (with slight quantitative differences due to different energetic values).
These supplementary tests highlight the method’s general applicability separate from of the choice of energy calculator, as long as it is valid or consistent within the search space defined.
Note, for any single example case, the entire process must be conducted with the same calculator throughout so as to allow correct learning.
For the Si\|Ge example, the CHGNet was found to incorrectly model the potential energy surface of the chemical environment.
Instead, the MACE-MPA-0 calculator was used as it showed much greater agreement with GGA-PBE results.

For the Si\|Ge GGA-PBE energies calculated using GPAW, a planewave cutoff of $500$~\si{\electronvolt} was used with a $\Gamma$-centred $3\times3\times3$ Monkhorst-Pack grid~\cite{Monkhorst1976SpecialPointsBrillouin}, with no spin polarisation being considered.
The Si and Ge pseudopotential valence orbitals are $3s^2 3p^2$ and $4s^2 4p^2$, respectively.
GPAW energies are obtained for structures taken directly from the example case without performing further atomic relaxation using the GPAW calculator.

\backmatter

\bmhead{Supplementary information}
Supplementary information provided with this study includes figures and text corresponding to the following information:
guide for setting parameters and implementing RAFFLE into workflow,
limitations of the code,
further benchmarking,
tests and analysis of placement methods, 
validation of energetic calculators used, and
further analysis of example cases.

\bmhead{Author contributions}
NTT contributed conceptualisation, software development, methodology, data curation, investigation, validation, supervision, and visualisation.
JP contributed conceptualisation, software development, methodology, data curation, investigation, validation, and visualisation.
FHD contributed investigation, validation, and visualisation.
SPH was the project supervisor, contributing conceptualisation and methodology.
NTT, JP, and SPH contributed to funding acquisition.
All authors contributed to analysis, and article writing and editing, and read and approved the final manuscript.

\bmhead{Acknowledgements}

The authors thank C. J. Price and E. Baker for discussions.
The work provided by N. T. Taylor was supported in part by the Government Office for Science and the Royal Academy of Engineering under the UK Intelligence Community Postdoctoral Research Fellowships scheme.
J. Pitfield was supported by VILLUM FONDEN through Investigator Grant, 
Project No. 16562, and by the Danish National Research Foundation
through the Center of Excellence “InterCat” (Grant Agreement No:
DNRF150). Additionally, we thank the EPSRC for funding J. Pitfield (EP/L015331/1) and F. H. Davies (EP/X013375/1) via the EPSRC Centre for Doctoral Training in Metamaterials, and the Leverhulme for funding S. P. Hepplestone and N. T. Taylor (RPG-2021-086).
Via our membership of the UK's HEC Materials Chemistry Consortium, which is funded by EPSRC (EP/R029431), this work used the ARCHER2 UK National Supercomputing Service within the framework of a Grand Challenge project.
The authors acknowledge the use of the University of Exeter High-Performance Computing (HPC) facility ISCA for this work.

\section*{Declarations}

\section{Data availability}
Initial bulk structures were obtained form the Materials Project~\cite{Jain2013CommentaryMaterialsProject} at \url{https://next-gen.materialsproject.org/}.
The data for reproducing the results presented here can be found in the RAFFLE GitHub repository at \url{https://github.com/ExeQuantCode/RAFFLE}.
The datasets generated and analysed during the current study are available in the RAFFLE repository, \url{https://github.com/ExeQuantCode/RAFFLE}.

\section{Code availability}
The RAFFLE software used in this study is available from GitHub and can be accessed via this link \url{https://github.com/ExeQuantCode/RAFFLE}.
The scripts used to generate the data and plots for the systems discussed in \secref{sec:example_cases} are available in the \code{example/python\_pkg} directory of the code repository, enabling full reproducibility of the presented results.

\bibliography{main}

\end{document}


\title[Supplementary Information Title]{
\centering
Supplementary Information\\%
RAFFLE: Active learning accelerated interface structure prediction
}

\author*[1]{\fnm{Ned Thaddeus} \sur{Taylor}}\email{n.t.taylor@exeter.ac.uk}

\author[2]{\fnm{Joe} \sur{Pitfield}}

\author[1]{\fnm{Francis Huw} \sur{Davies}}

\author*[1]{\fnm{Steven Paul} \sur{Hepplestone}}\email{s.p.hepplestone@exeter.ac.uk}

\affil*[1]{\orgdiv{Department of Physics and Astronomy}, \orgname{University of Exeter}, \orgaddress{\street{Stocker Road}, \city{Exeter}, \postcode{EX4 4QL} \country{United Kingdom}}}

\affil[2]{\orgdiv{Center for Interstellar Catalysis, Department of Physics and Astronomy}, \orgname{Aarhus University}, \orgaddress{\postcode{DK-8000}, \city{Aarhus C}, \country{Denmark}}}

\maketitle

\section{Brief overview of code}
\label{sec:overview}

This section provides a brief overview of the motivation behind RAFFLE’s development.

RAFFLE is an open-source software package (GPLv3) designed to construct and identify new material phases at the interface between two crystals.
It can be used it identify intermixing and novel interface phases that emerge due to strain compensation and high-energy growth conditions.
By integrating physical principles with genetic algorithms, RAFFLE efficiently explores the vast configuration space of interfaces.

The source code and examples are available at \url{https://github.com/ExeQuantCode/RAFFLE}.
The software has been tested and developed using the GNU 14.1.0 Fortran compiler on macOS and Unix/Linux, alongside Python 3.12 and 3.13.
The Python library can be installed via pip using:

\begin{lstlisting}
    pip install raffle
\end{lstlisting}

\noindent
or

\begin{lstlisting}
    pip install 'raffle[ase]'
\end{lstlisting}

The Fortran library and executable can be installed using the Fortran Package Manager.
All three options (Python library, Fortran library, and Fortran executable) can be installed using CMake.

\section{How to run RAFFLE}
\label{sec:guide}

This section provides guided examples of the RAFFLE Python library, covering parameter setup, expected inputs and outputs, and a practical example of integrating RAFFLE into existing Python workflows.

Whilst RAFFLE is packaged with three implementations (Fortran executable, Fortran library, and Python library), the authors anticipate that the Python library will be of greatest interest to the community and therefore provide guidance on this implementation here.

\subsection{User-definable parameters}
\label{sec:guide:parameters}

Parameters within the RAFFLE workflow have physics-grounded definitions.
However, these can be modified by the user when calling the library procedures and initialising types.

The following guide explains how to set user-definable parameters and follows the workflow of generating a Si-Ge structure.
\paragraph{Initialisation:} First, we import the RAFFLE library and initialise our structure generator:

\begin{lstlisting}
# Initialise RAFFLE generator
from raffle.generator import raffle_generator

generator = raffle_generator()
\end{lstlisting}

This is the only object that needs to be directly imported from RAFFLE.
It is recommended to use the Atomic Simulation Environment (ASE)~\cite{ase-paper} for handling structure data; however, RAFFLE also includes its own atomic structure object if needs be.
All variables associated with the descriptors (distribution functions) must be handled appropriately.

\noindent
\textbf{Reference energies:}
Energy references do not have built-in default values, as they depend on specific calculation conditions (e.g. calculator choice, DFT functional choice, pseudopotential selection).
To avoid misleading or inaccurate results, users must define reference energies explicitly:

\begin{lstlisting}[firstnumber=5]
# Set reference energies
generator.distributions.set_element_energies( {
    'Si': -5.31218, # energy per atom of Si bulk obtained from DFT etc.
    'Ge': -4.44257  # energy per atom fo Ge bulk obtained from DFT etc.
} )
\end{lstlisting}

\noindent
\textbf{Gaussian parameters and cutoffs:}
Energy scaling, Gaussian smearing, width, and cutoff tolerances can be set as follows:

\begin{lstlisting}[firstnumber=11]
# Set Gaussian parameters
generator.distributions.set_kBT(0.2)
generator.distributions.set_sigma(
    [0.1, 0.2, 0.3] # sigma for 2-, 3-, 4-body distribution functions
)
generator.distributions.set_width(
    [0.1, 0.2, 0.3] # Gaussian width for 2-, 3-, 4-body distribution functions
)
generator.distributions.cutoff_min(
    [0.0, 0.0, 0.0] # minimum value for 2-, 3-, 4-body distribution functions
)
generator.distributions.cutoff_max(
    [6.0, 3.14159, 3.14159] # maximum value for 2-, 3-, 4-body distribution functions
)
generator.distributions.set_radius_distance_tol(
    [1.5, 2.5, 3.0, 6.0] 
    #lower/upper multipliers for element pair radii used in 3- & 4-body distribution functions
)
\end{lstlisting}

\noindent
The distance tolerance is a multiple of the element-pair covalent radius, similar to that used in AGOX~\cite{Christiansen2022AtomisticGlobalOptimization}.
By default, for any element pair, the average of their covalent radii is used.
However, this can be customised:

\begin{lstlisting}[firstnumber=28]
# Set reference element-pair covalent radii
generator.distributions.set_bond_radii( {
    ('Si', 'Ge'): 1.165 # average bond length
} )
\end{lstlisting}

The user can define the grid on which the generator operates.
Note that the \code{grid} and \code{grid\_spacing} arguments are mutually exclusive.
The grid offset determines the displacement of grid points from the spatial origin $(0,0,0)$.

\begin{lstlisting}[firstnumber=34]
# Define grid for placement methods
generator.set_grid(
    grid = [1,2,3],
    grid_spacing = 0.1,
    grid_offset = [0.1, 0.1, 0.1]
)
\end{lstlisting}

\subsection{Inputs and outputs}
\label{sec:guide:inputs_outputs}

Once the desired parameters have been set, the host structure must be defined \green{(see main article for more details on host)}:

\begin{lstlisting}[firstnumber=40]
# Set the host structure
host = Atoms(...)
generator.set_host(host)
\end{lstlisting}

\noindent
An optional bounding box can restrict atom placement to a specified region, with limits expressed in fractional coordinates relative to the lattice vectors $(\vec{a}, \vec{b}, \vec{c})$:

\begin{lstlisting}[firstnumber=40]
# Set the fractional limits of atom position placement
a_min = 0.0; b_max = 0.0; c_min = 0.3
a_max = 1.0; b_max = 1.0; c_max = 0.8
generator.set_bounds( [
    [a_min, b_min, c_max],
    [a_max, b_max, c_max]
]  )
\end{lstlisting}

\noindent
\textbf{Initial database:}
An initial database should be provided from which RAFFLE learns descriptors:

\begin{lstlisting}[firstnumber=47]
# Set the database
database = [Atoms(...)]
generator.distributions.create(database)
\end{lstlisting}

\noindent
A common approach to obtaining an initial bulk database is to download the required data from the Materials Project~\cite{Jain2013CommentaryMaterialsProject}.
A step-by-step tutorial on this process is available in the software's Read\textit{the}Docs~\cite{RAFFLEDocs2025} under ``Databases Tutorial''.

\noindent
\textbf{Structure generation:}
Once all parameters and inputs have been set, structure generation can proceed:

\begin{lstlisting}[firstnumber=50]
# Generate structures
generator.generate(
    seed = 0,
    num_structures = 5,
    stoichiometry = { 'Si': 2, 'Ge': 3 },
    method_ratio = { # define ratio of placement methods
        "void": 5.0,
        "rand": 1.0,
        "walk": 2.0,
        "grow" 3.0,
        "min": 8.0
   }
)
\end{lstlisting}

\noindent
\textbf{Retrieving generated structures:}
The generated structures can be retrieved as follows:

\begin{lstlisting}[firstnumber=63]
# Retrieve structures
structures = generator.get_structures()
\end{lstlisting}

\noindent
This returns a list of ASE \code{Atoms} objects, with length equal to the number of structures generated during this RAFFLE generator instance.

\noindent
\textbf{Convergence considerations:}
If RAFFLE is used with the same random seed across multiple iterations, it should eventually converge in cases of bulk materials.
This occurs because the generalised descriptors remain unchanged unless new features (within an energetically favourable range) are introduced that significantly modify them.
Specifying the seed reinitialises the random seed to the value for that call of \code{generate()}.

\subsection{Worked example}
\label{sec:guide:example}

Below is an example script for a single iteration of the RAFFLE generator for structure search.

\begin{lstlisting}
# Single iteration of RAFFLE structure search
from ase.io import read, write
from raffle.generator import raffle_generator

generator = raffle_generator()

host = read("host.xyz")
generator.set_host(host)
generator.set_grid(grid_spacing=0.1)
generator.set_bounds([[0, 0, 0.5], [1, 1, 0.75]])
generator.distributions.set_element_energies(
    { 'C': -9.063733 }
)

database = read("database.xyz", index=":")
generator.distributions.create(database)

structures, status = generator.generate(
    num_structures = 1,
    stoichiometry = { 'C': 2 },
)

write("output.xyz", structures)
\end{lstlisting}

The following script provides an example of iterating the RAFFLE generator to update the descriptors and improve placement.

\begin{lstlisting}
# Iterative RAFFLE structure search
from chgnet.model.dynamics import CHGNetCalculator
from ase.optimize import FIRE
calculator = CHGNetCalculator()

for i in range(10):
    structures, status = generator.generate(
        num_structures = 2,
        stoichiometry = { 'C': 2 },
        calc = calculator
    )

    for structure in structures:
        optimiser = FIRE(structure)
        optimiser.run(fmax=0.05)
    
    generator.update(structures)
\end{lstlisting}

In this quick guide, the CHGNet machine-learned potential~\cite{Deng2023CHGNETPretrainedUniversalNeural} is employed for fast and accurate evaluation of energetic values.
It should be noted that such machine-learned potentials are currently limited in capabilities for non-bulk structures, so extreme care should be taken if utilising them for evaluation of interface energetics.
However, recent works have shown a pathway towards augmenting such potentials to improve their qualitative accuracy and ordering of energetics for surface structures~\cite{Pitfield2025AugmentationUniversalPotentials, christiansen2025deltamodelcorrectionfoundationmodel, radova2025finetuningfoundationmodelsmaterials}.

\section{Limitations of the code}
\label{sec:limitations}

The authors acknowledge that the distribution functions are not continuous, particularly at $R_{\mathrm{cut}}$.
This can introduce discontinuities in the viability grid.
However, since the viability grid is only used to determine atom placement at specific points, continuity is not a requirement.

As discussed in the main article, while the RAFFLE software package is agnostic to the choice of energy calculator, it does not support changes in the energy calculator within a single RAFFLE generator.
In other words, all data used to construct the descriptors must be generated using the same energy calculator.
If different calculators are used, inconsistencies in energy trends between them will likely result in meaningless descriptors learned by RAFFLE.
If one wishes to combine data from different sources, various approaches can be employed~\cite{Pitfield2025AugmentationUniversalPotentials,Kingsbury2022FlexibleScalableScheme}.

Finally, the void placement method places atoms in sites furthest from another atom.
A future consideration would be to convert this to a density map and, instead, placing atoms at points of lowest atomic density.
This requires considering atoms as non-point-like objects with an associated distribution in space, which is already utilised in the distribution functions.

\section{Showcase of placement methods}
\label{sec:placement_showcase}

\begin{figure*}[t]
    \centering
    \subfloat[min]{\includegraphics[height=0.25\linewidth]{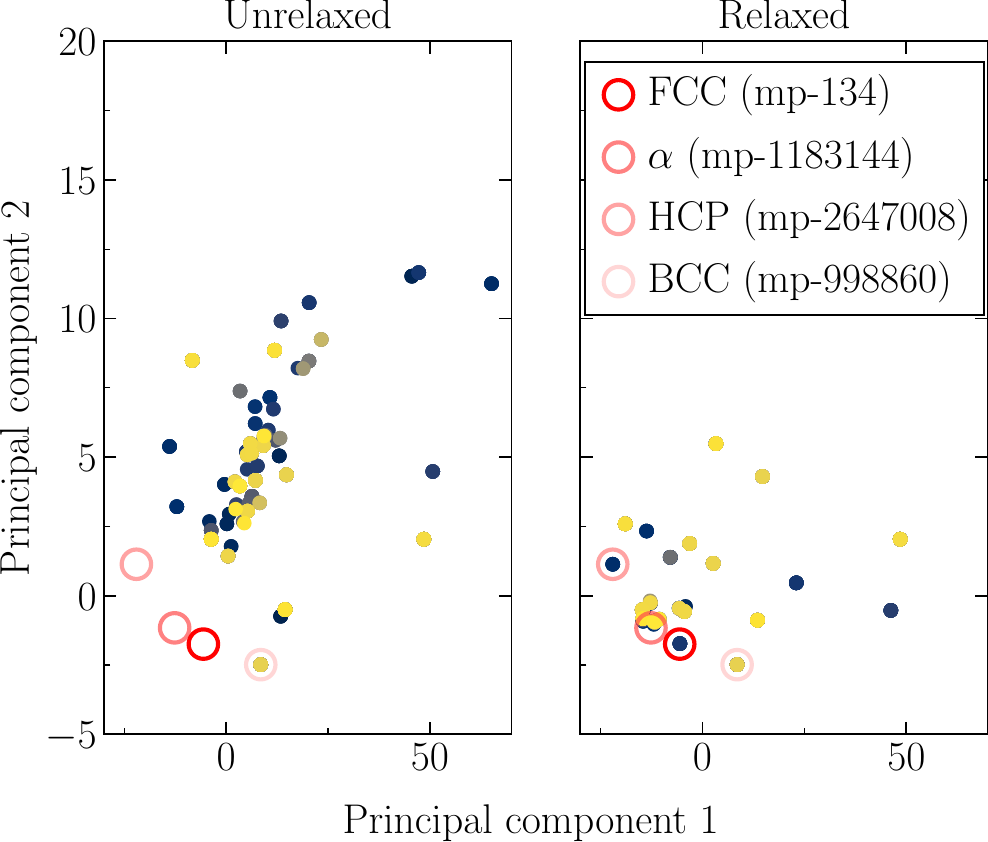}\label{fig:method:min}}%
    \hspace{1em}%
    \subfloat[17:3 min:rand]{\includegraphics[height=0.25\linewidth]{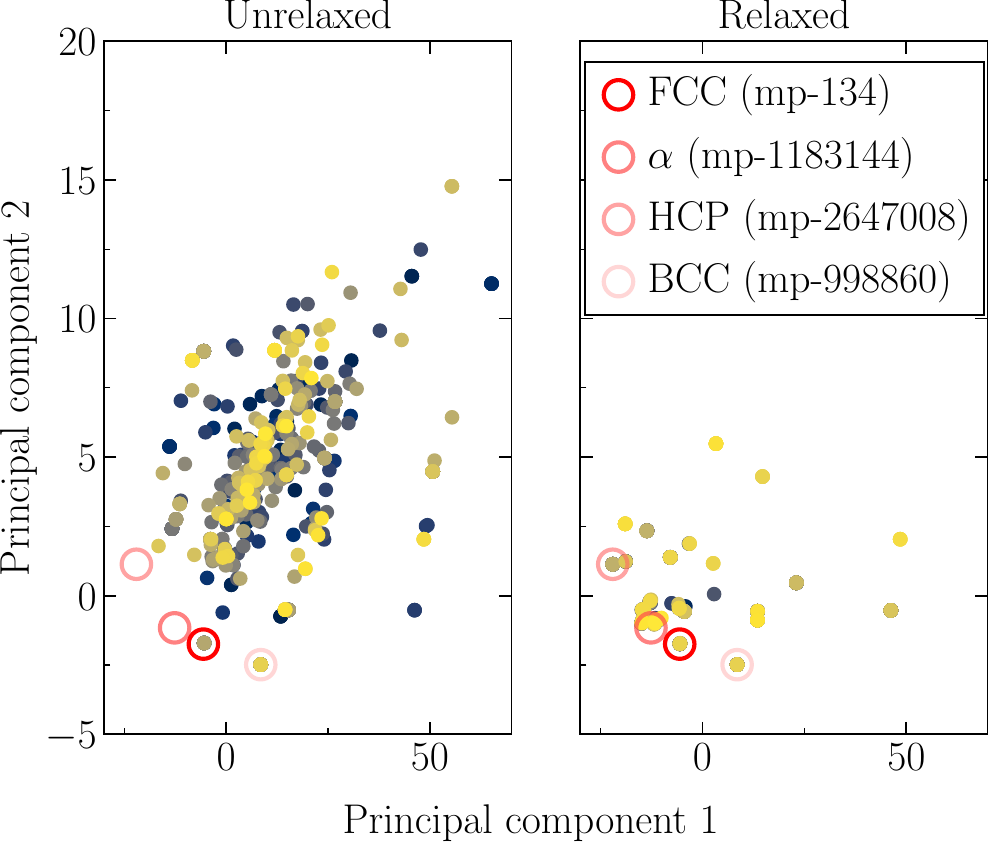}\label{fig:method:min+rand}}%
    \hspace{1em}%
    \subfloat[rand]{\includegraphics[height=0.25\linewidth]{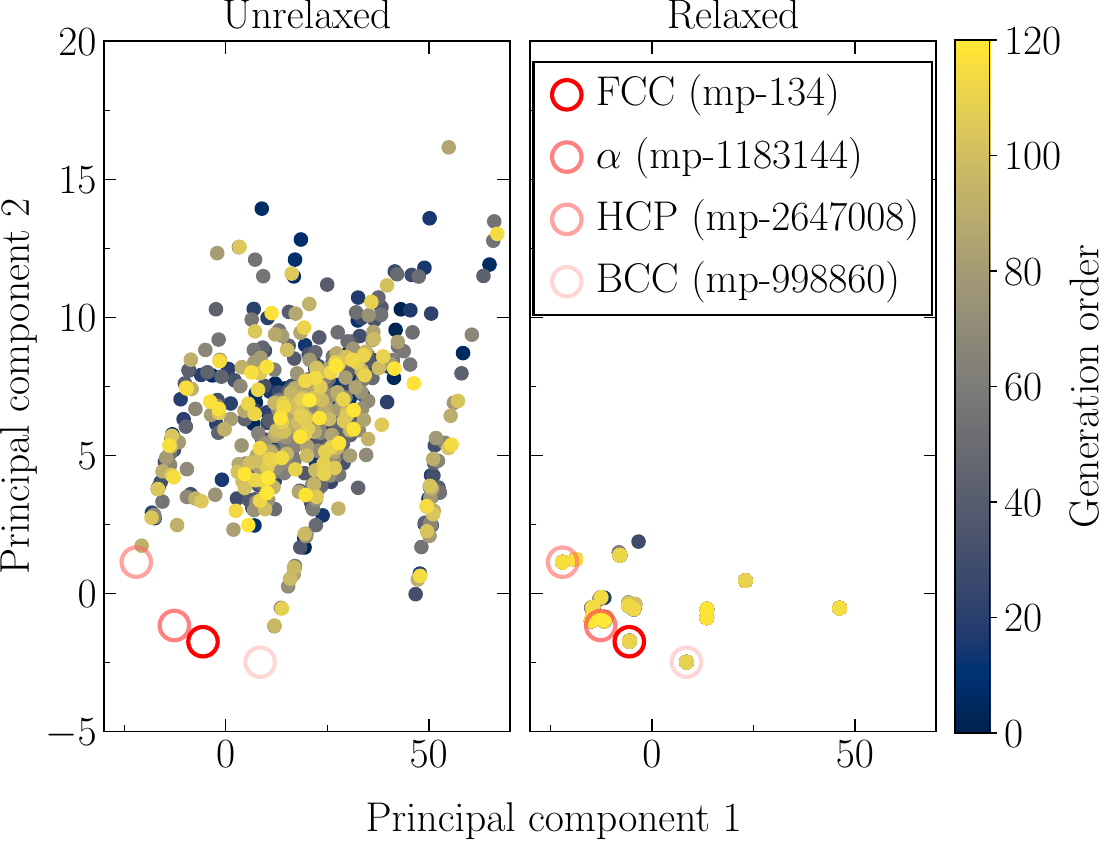}\label{fig:method:rand}}%
    \newline
    \subfloat[walk]{\includegraphics[height=0.25\linewidth]{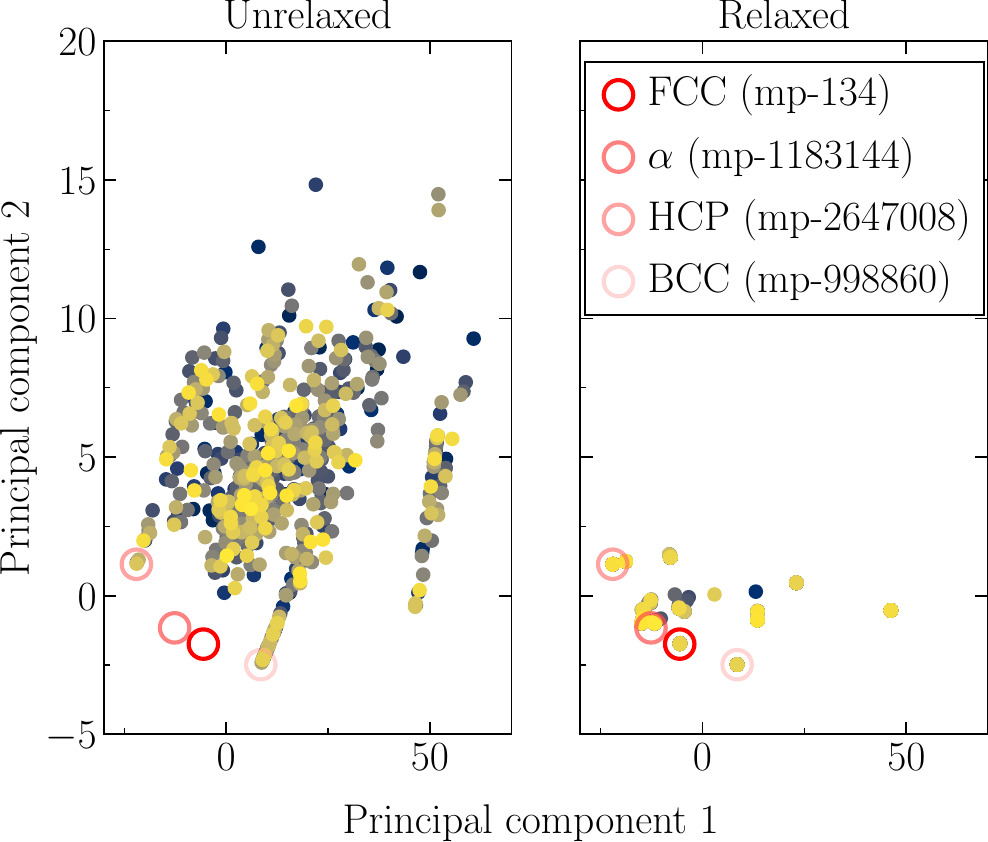}\label{fig:method:walk}}%
    \hspace{1em}%
    \subfloat[17:3 walk:rand]{\includegraphics[height=0.25\linewidth]{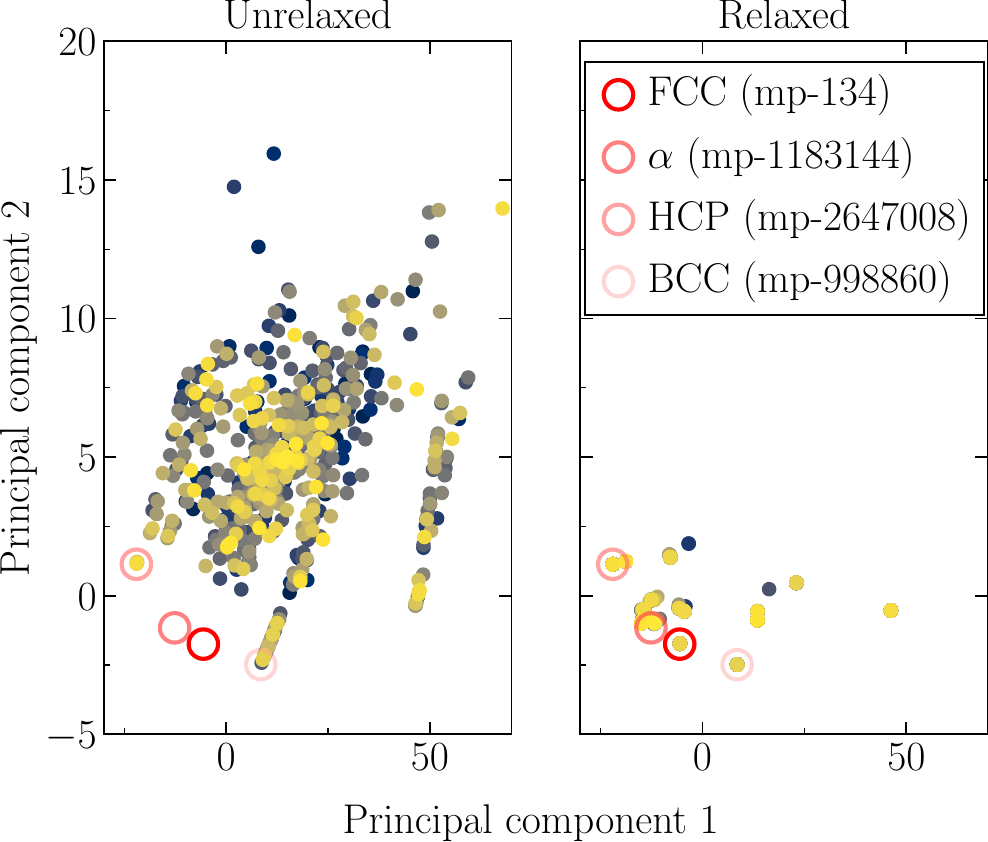}\label{fig:method:walk+rand}}%
    \hspace{1em}%
    \subfloat[grow]{\includegraphics[height=0.25\linewidth]{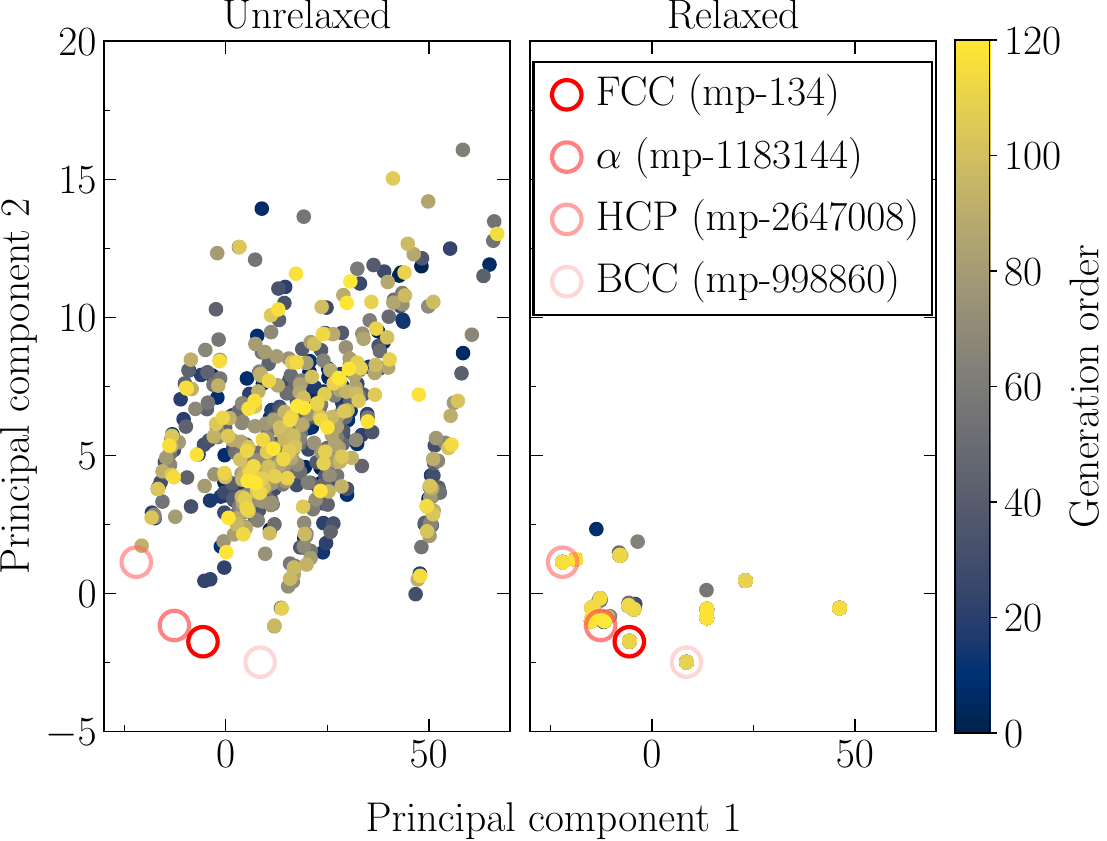}\label{fig:method:grow}}%
    \newline
    \subfloat[17:3 grow:rand]{\includegraphics[height=0.25\linewidth]{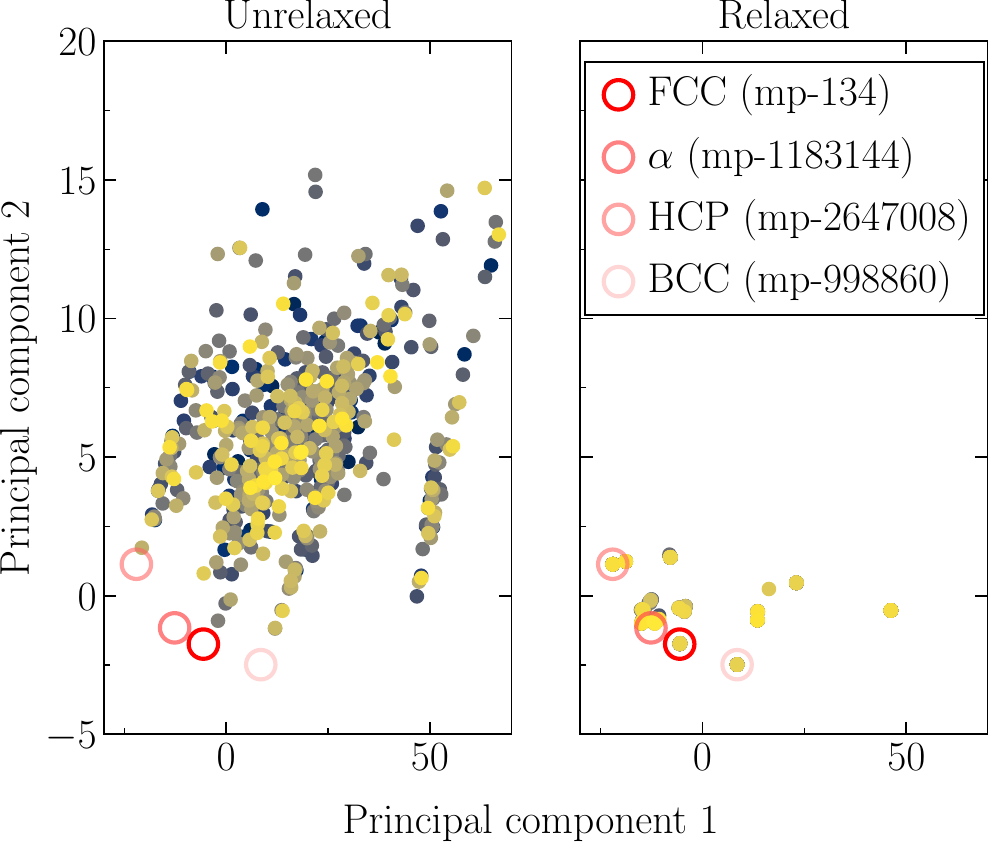}\label{fig:method:grow+rand}}%
    \hspace{1em}%
    \subfloat[void]{\includegraphics[height=0.25\linewidth]{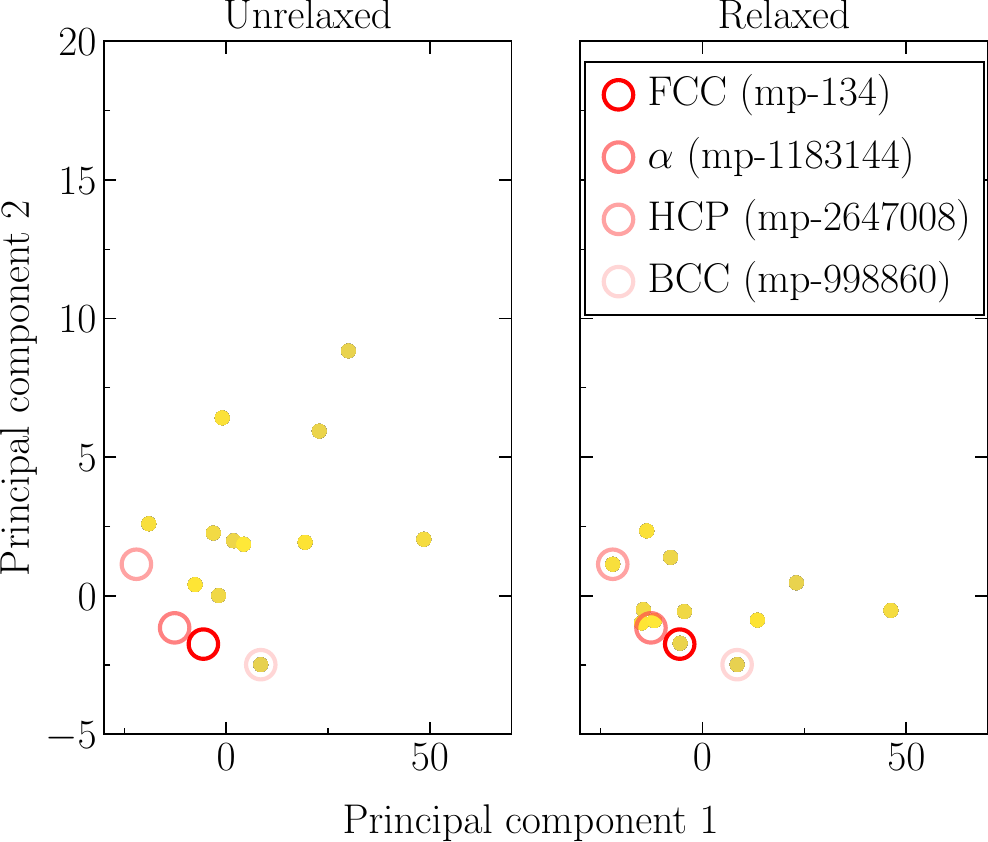}\label{fig:method:void}}%
    \hspace{1em}%
    \subfloat[17:3 void:rand]{\includegraphics[height=0.25\linewidth]{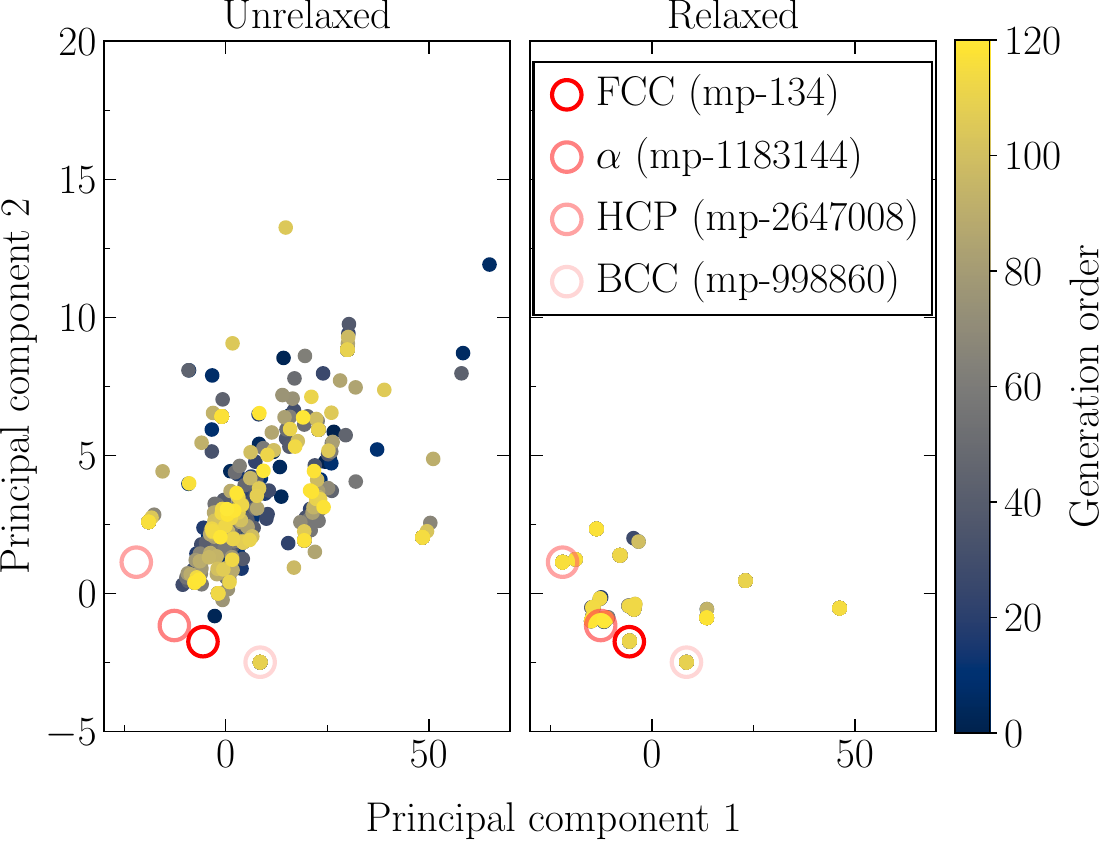}\label{fig:method:void+rand}}%
    \caption{
        Principal component analysis of bulk aluminium structures with cubic and hexagonal Bravais lattices, showing pre- and post-relaxation structures (cell fixed) generated using different RAFFLE placement methods and combinations.
        Red circles highlight known aluminium phases (FCC and BCC are experimentally verified).
        Colour of points represent order of generation (higher on colour bar represents later-generated structures).
        \protect\subref{fig:method:min} \emph{Minimum} placement,
        \protect\subref{fig:method:min+rand} 17:3 \emph{Minimum}:\emph{Random},
        \protect\subref{fig:method:rand} \emph{Random},
        \protect\subref{fig:method:walk} \emph{Walk},
        \protect\subref{fig:method:walk+rand} 17:3 \emph{Walk}:\emph{Random},
        \protect\subref{fig:method:grow} \emph{Growth},
        \protect\subref{fig:method:grow+rand} 17:3 \emph{Growth}:\emph{Random},
        \protect\subref{fig:method:void} \emph{Void}, and
        \protect\subref{fig:method:void+rand} 17:3 \emph{Void}:\emph{Random}.
        Energetics are calculated using CHGNet~\cite{Deng2023CHGNETPretrainedUniversalNeural}, with atomic relaxations performed using FIRE~\cite{Bitzek2006StructuralRelaxationMade} (cells are fixed).
        }
    \label{fig:method}
\end{figure*}%

We compare the five sampling methods introduced in the main article—\emph{Void}, \emph{Random}, \emph{Walk}, \emph{Growth}, and \emph{Minimum}—by applying each exclusively (100\%) and with a small fraction (15\%) of purely random placements.
\Figref{fig:method} presents representative structures for bulk aluminium, projected via principal component analysis, illustrating the configurations generated under each condition.

The \emph{Minimum} method (\figref{fig:method:min}) systematically searches the entire cell, placing atoms at the most energetically favourable sites.
This leads to rapid convergence and can even precisely reproduce the BCC phase.
However, it struggles to explore alternative configurations due to its tendency to reinforce known structures, akin to being trapped in local minima.
Introducing 15\% random placements (\figref{fig:method:min+rand}) broadens the search space, mitigating this issue and enabling the identification of the FCC phase -- the most stable aluminium structure.
Notably, both the pure \emph{Minimum} and the 17:3 \emph{Minimum}:\emph{Random} methods yield a higher number of metastable relaxed structures, making them particularly useful for studying highly disordered interfaces.

The purely \emph{Random} method (\figref{fig:method:rand}) explores the configuration space indiscriminately, ensuring broad coverage and potentially discovering rare structures.
However, it is highly inefficient, requiring a large number of iterations to reliably sample all possible metastable states.
This inefficiency is analogous to the Coupon Collector’s Problem~\cite{Boneh1997CouponCollectorProblem}: if each meta-stable state has an equal chance of being sampled post-relaxation, the iterations required to find all states scale as $N\mathrm{log}(N)$.
For interfaces with hundreds of meta-stable states, the most stable ones, having larger basins of attraction, are more easily discovered.
While this method eventually identifies known phases in the simple bulk aluminium test case, a significant portion of its search consists of rejected placements, limiting its practical efficiency.

The \emph{Walk} method (\figref{fig:method:walk}) starts from a random point and makes local moves towards higher placement probabilities, resembling a steepest-descent search.
It successfully identifies the BCC and HCP phases after multiple generations.
Adding 15\% random placements (\figref{fig:method:walk+rand}) does not significantly alter the sampled space, as the method already incorporates a level of randomness.
This suggests that \emph{Walk} could serve as a structured alternative to pure \emph{Random} placements in hybrid approaches.

A similar trend is observed with the \emph{Growth} method (\figref{fig:method:grow}), which places atoms near previously positioned ones, mimicking physical nucleation.
The addition of 15\% random placements (\figref{fig:method:grow+rand}) does not meaningfully change the generated structures, reinforcing the idea that the method inherently explores a diverse range of configurations.

The \emph{Void} method (\figref{fig:method:void}) prioritises placement in the largest available spaces, leading to a highly uniform filling of the cell.
Introducing 15\% randomness (\figref{fig:method:void+rand}) diversifies placements, broadening the range of generated structures.

Overall, these results highlight the distinct biases of each method.
Deterministic approaches such as \emph{Minimum} and \emph{Void} efficiently locate stable configurations but undersample atypical arrangements.
In contrast, methods incorporating randomness -- \emph{Random}, \emph{Walk}, and \emph{Growth} -- can produce more diverse structures but at the cost of computational inefficiency.
Introducing a small fraction of random placements balances these trade-offs, improving the exploration of configuration space and increasing the likelihood of discovering new structures.

\subsection{Basins of Attraction}
\label{sec:placement_showcase:boa}

\begin{figure}
    \centering
    \includegraphics[width=0.5\linewidth]{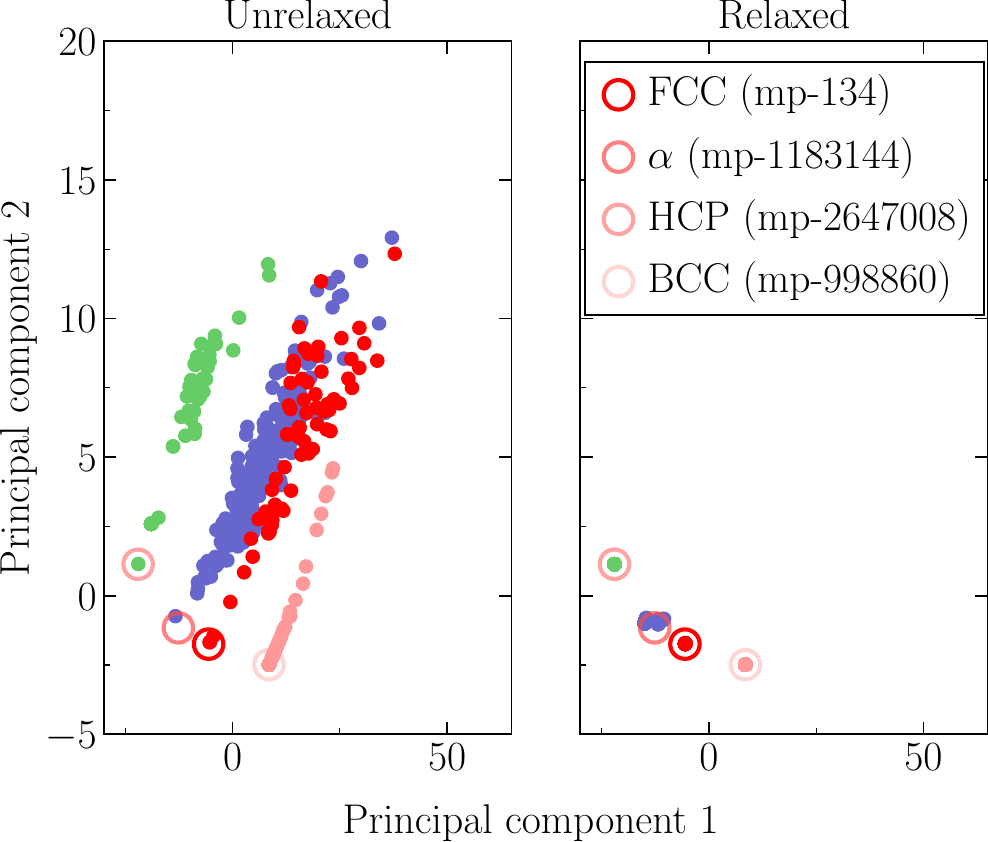}
    \caption{
    Principal component analysis of bulk aluminium showing only the structures that relax into one of the four listed phases.
    The pre- and post-relaxation geometries are shown and were generated using RAFFLE with fixed cells.
    Energetics are calculated using CHGNet~\cite{Deng2023CHGNETPretrainedUniversalNeural}, with atomic relaxations performed using FIRE~\cite{Bitzek2006StructuralRelaxationMade} (cells are fixed).
    }
    \label{fig:boa}
\end{figure}
In atomic structure generation, multiple distinct initial configurations converge to the same final relaxed structure upon geometry optimization. The relaxation trajectory of each initial configuration is driven by the potential energy surface toward a local minimum. The collection of all such initial configurations that will relax to the same minimum defines the basin of attraction for that structure. A proper understanding of these basins is necessary for comprehending the convergence behaviour and for evaluating the robustness of the optimization algorithms employed.

In figure~\ref{fig:boa} we show the PCA of only the structures that relax into one of the four identified known phases.
The initial structures that converge to the same minimum are coloured the same.
This allows us to provide a visual representation of the basin of attraction for each known phase.
Notably, the PCA graphs show that the most energetically favourable phases exhibit the largest basins of attraction.
This suggests that these phases are both more stable energetically and are accessible by broader set of initial configurations. 

In complex interface structures, the energy landscape becomes highly complex due to additional degrees of freedom and interfacial interactions, leading to a multitude of local minima and metastable configurations.
Because of this complex potential energy landscape, we can expect any real interface structure to end up trapped at local minima.
We still expect that the most energetically favourable structures will have the largest basins of attraction.
This, however, is less helpful in interface prediction where we will need to properly evaluate the metastable configurations as well.
This requires the kind of guided search methodology described in the main text.

\section{Additional benchmarks}
\label{sec:benchmarks:addit}

\begin{figure*}[ht!]
    \centering
    \subfloat[Void]{\includegraphics[width=0.35\linewidth]{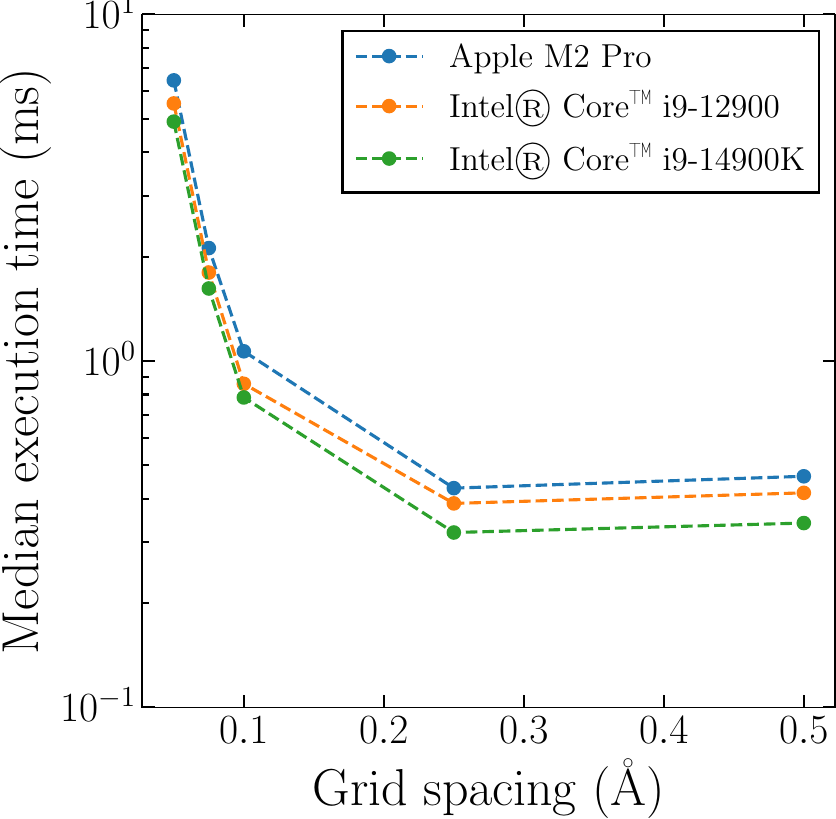}\label{fig:benchmarks:placement:void}}
    \hspace{2em}
    \subfloat[Random]{\includegraphics[width=0.35\linewidth]{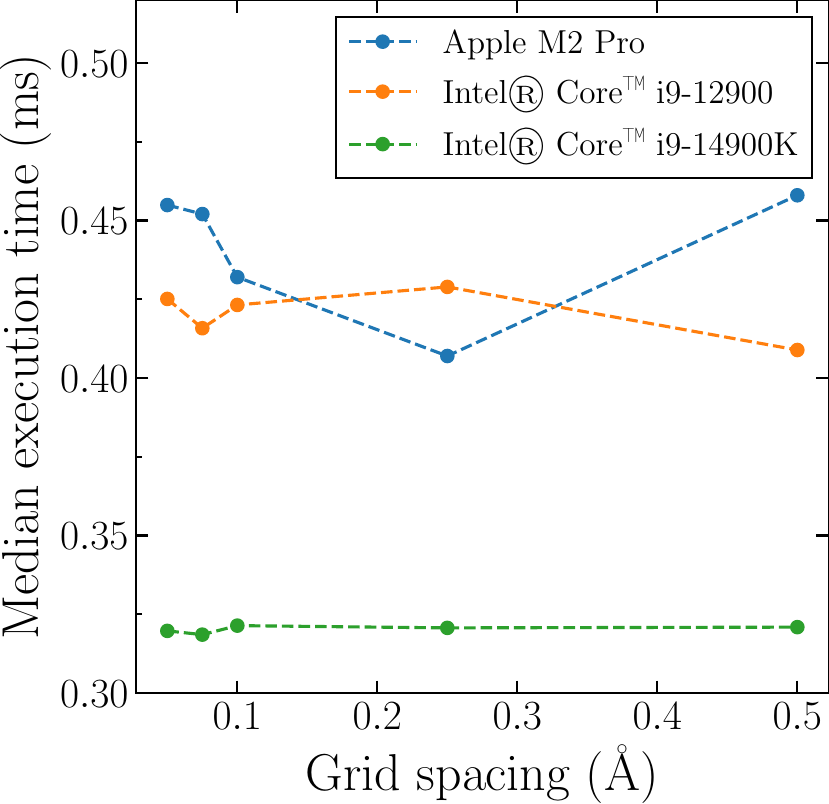}\label{fig:benchmarks:placement:rand}}
    \hspace{2em}
    \subfloat[Random walk]{\includegraphics[width=0.35\linewidth]{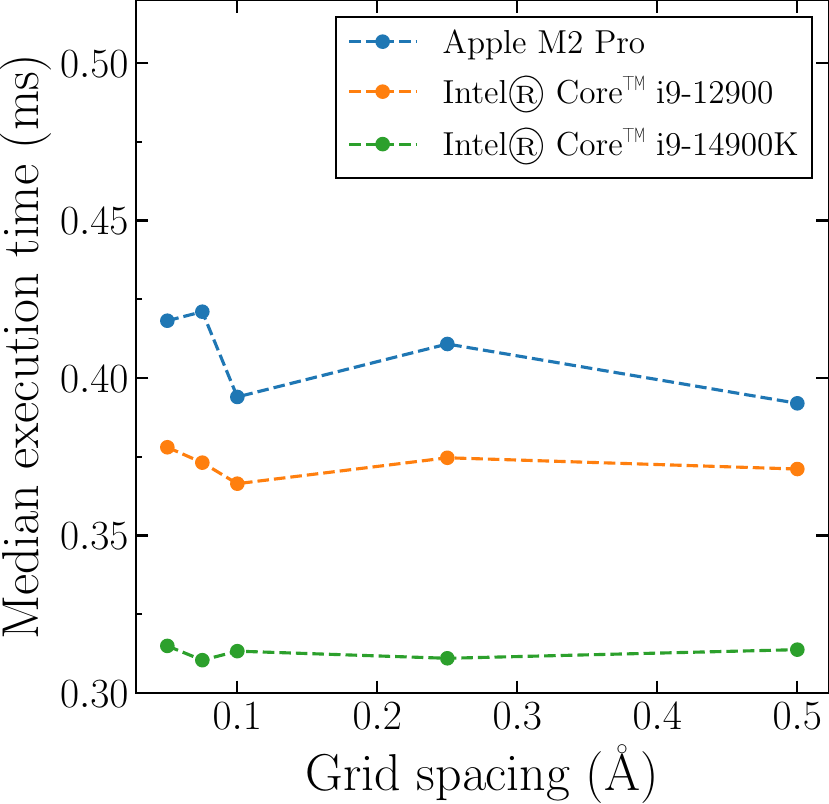}\label{fig:benchmarks:placement:walk}}
    \hspace{2em}
    \subfloat[Growth]{\includegraphics[width=0.35\linewidth]{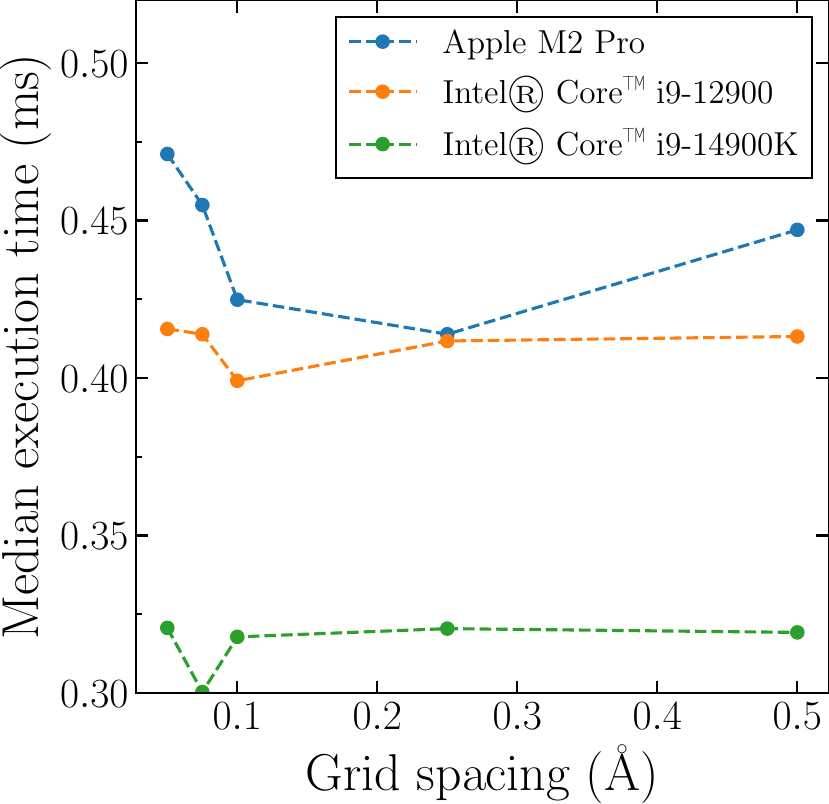}\label{fig:benchmarks:placement:grow}}
    
    \caption{
        Benchmarks of atom placement methods across multiple architectures.
        Grid spacing versus time to place a single carbon atom in a cubic host cell (lattice constant $3.567$\si{\angstrom}) containing one pre-existing carbon atom, using the \protect\subref{fig:benchmarks:placement:void} \textit{void}, \protect\subref{fig:benchmarks:placement:rand} \textit{random}, \protect\subref{fig:benchmarks:placement:walk} \textit{random walk}, and \protect\subref{fig:benchmarks:placement:grow} \textit{growth} placement methods.
        For all benchmarks presented, the following parameters are used: $k_{\mathrm{B}}T = 0.4$~\si{\electronvolt}, bin widths of $0.025$~\si{\angstrom}, and angular bin widths of $\pi/200$~\si{\radian} for 2-, 3-, and 4-body terms, respectively.
    }
    \label{fig:benchmarks:placement}
\end{figure*}

Additional benchmarks are presented in \figref{fig:benchmarks:placement} for further insight into the four placement methods not benchmarked in the main article -- void, random, walk, and growth.
As expected, the random, walk, and growth methods are independent of grid size, as they sample points in continuous space using a random number generator rather than operating on a fixed grid.
In contrast, the void method depends on grid size, as it identifies the point furthest from any atom by evaluating discrete grid points.
Compared to the min placement method, these four approaches operate on significantly shorter timescales.
The associated scripts and notebook can be found in \texttt{./example/python\_pkg/benchmarks/}.

\section{Energy calculator comparison}
\label{sec:en_check}

\begin{figure*}[ht!]
    \centering
    \subfloat[Carbon]{\includegraphics[width=0.3\linewidth]{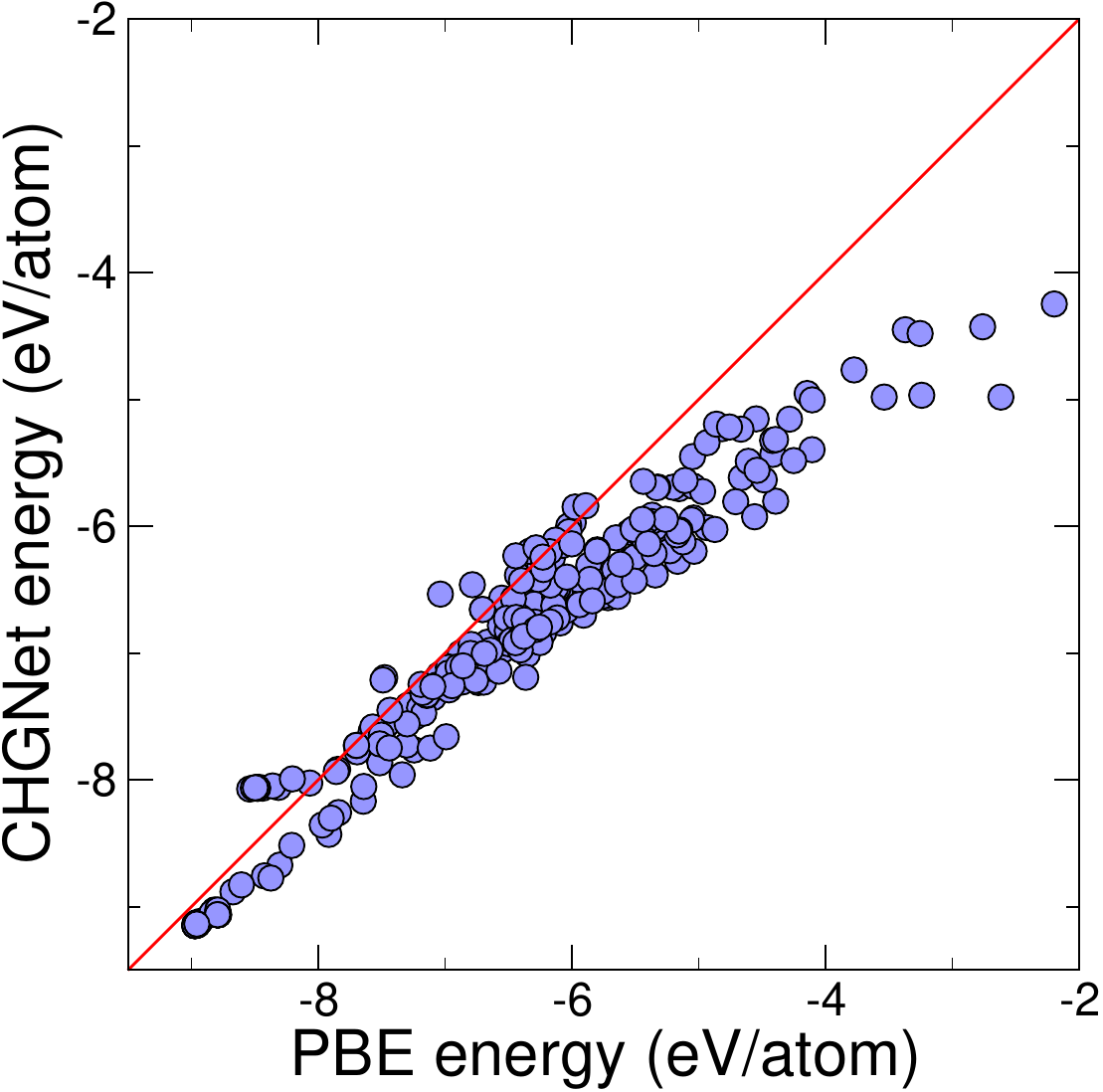}\label{fig:comparison:C}}
    \hspace{1em}
    \subfloat[Aluminium]{\includegraphics[width=0.3\linewidth]{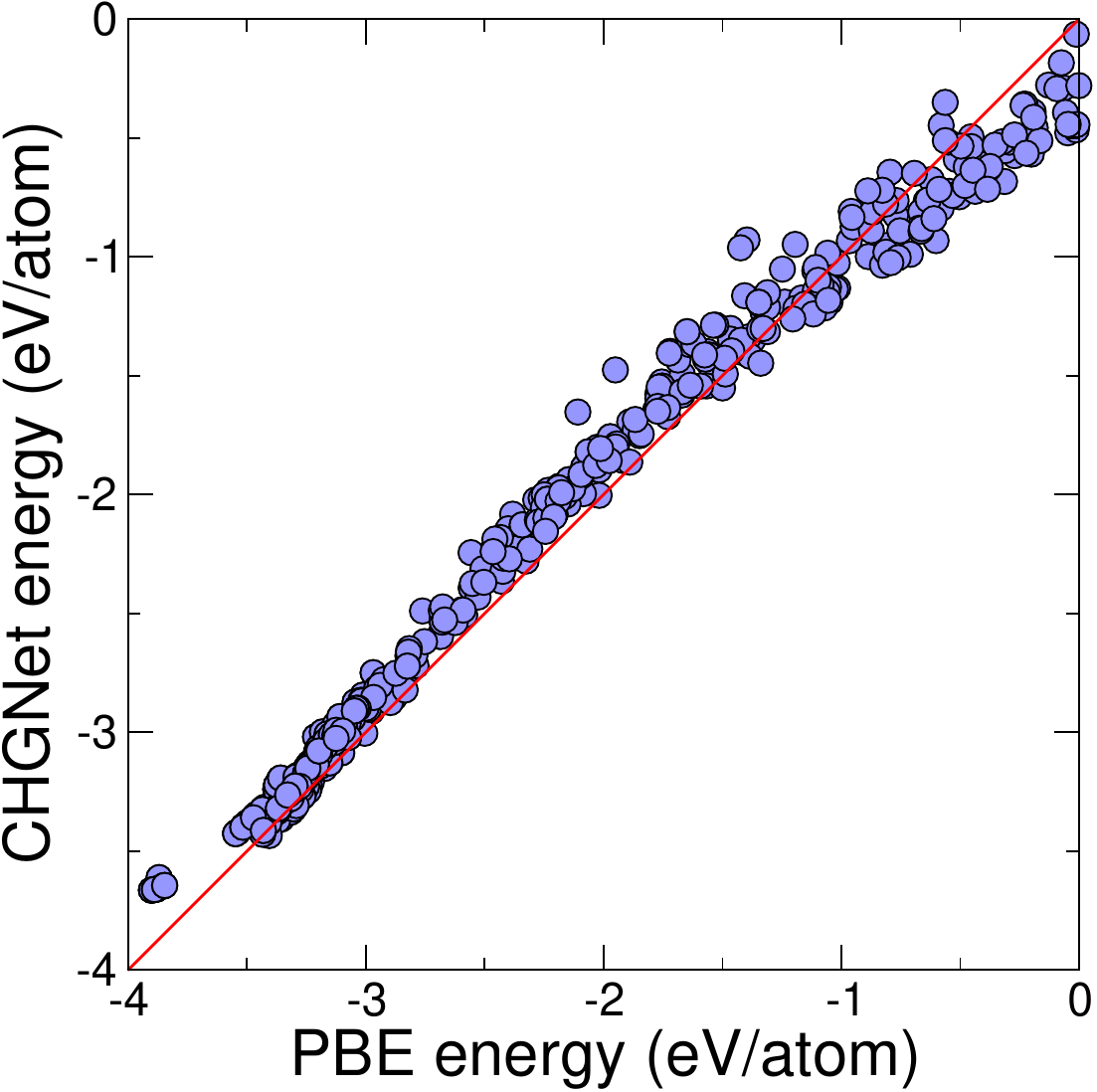}\label{fig:comparison:Al}}
    \hspace{1em}
    \subfloat[ScS$_{2}$-Li intercalation]{\includegraphics[width=0.3\linewidth]{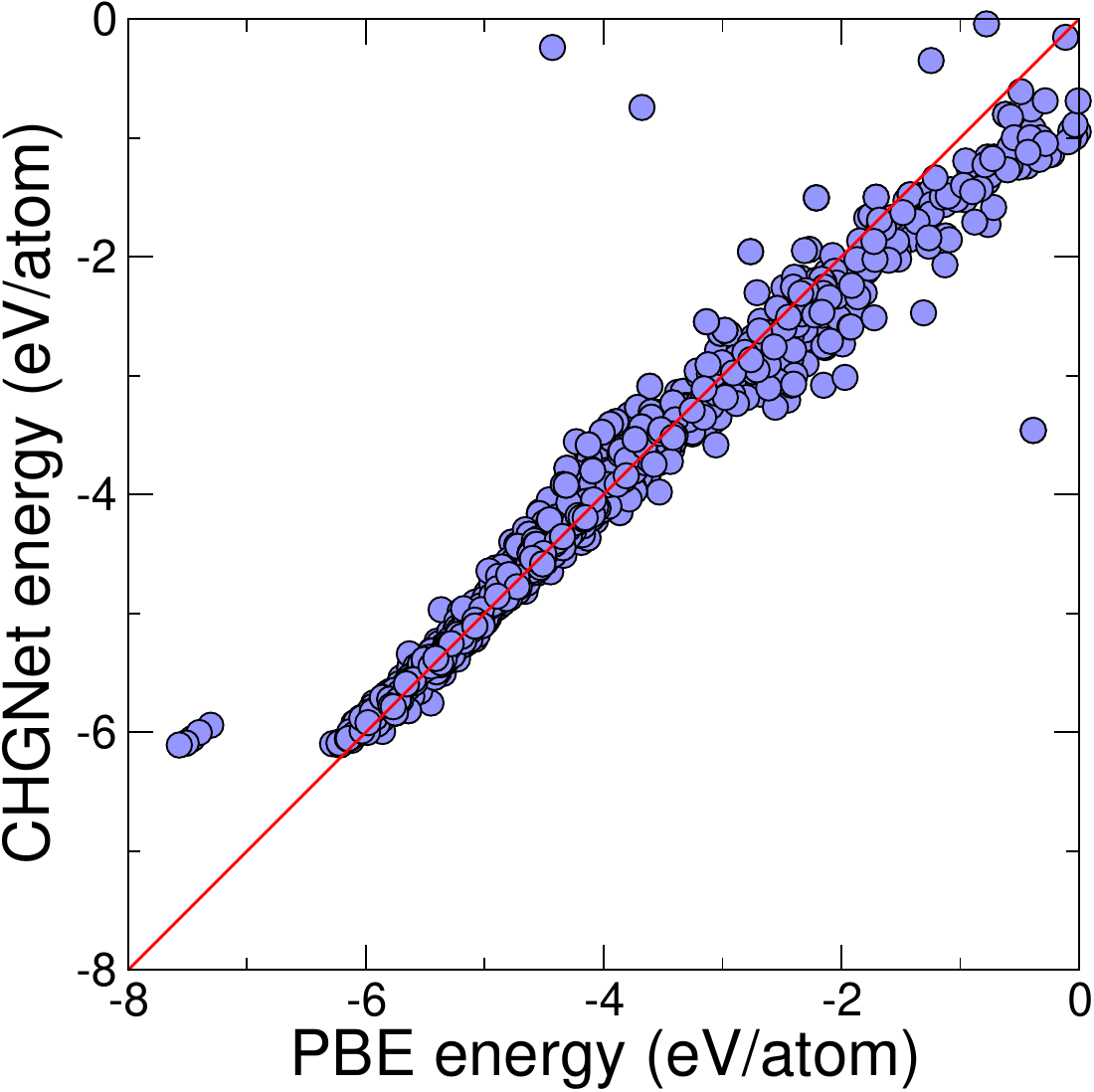}\label{fig:comparison:ScS2-Li}}
    \hspace{1em}
    \subfloat[C-MgO]{\includegraphics[height=0.3\linewidth]{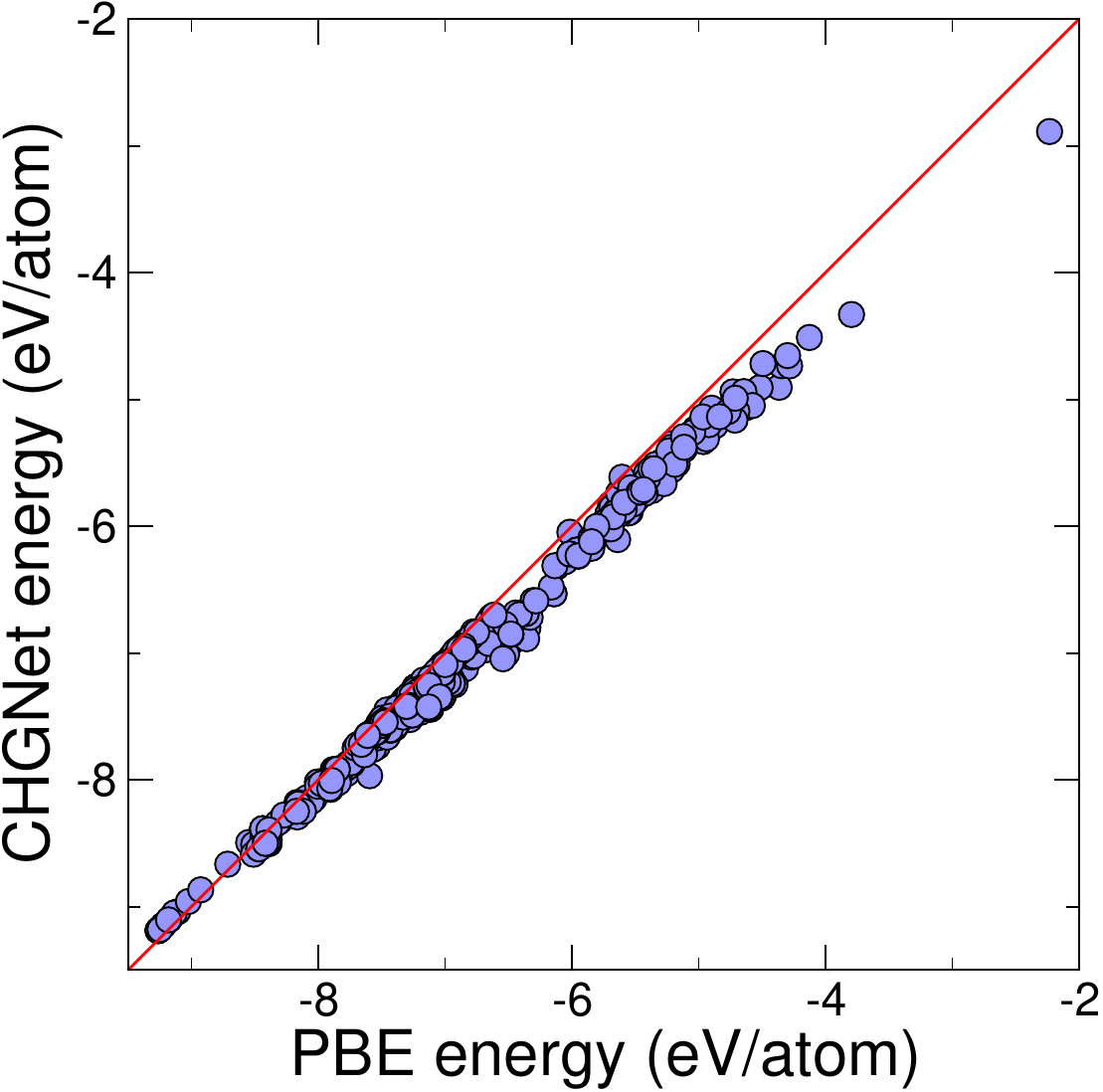}\label{fig:comparison:C-MgO}}
    \hspace{1em}
    \subfloat[Si\|Ge interface]{\includegraphics[height=0.3\linewidth]{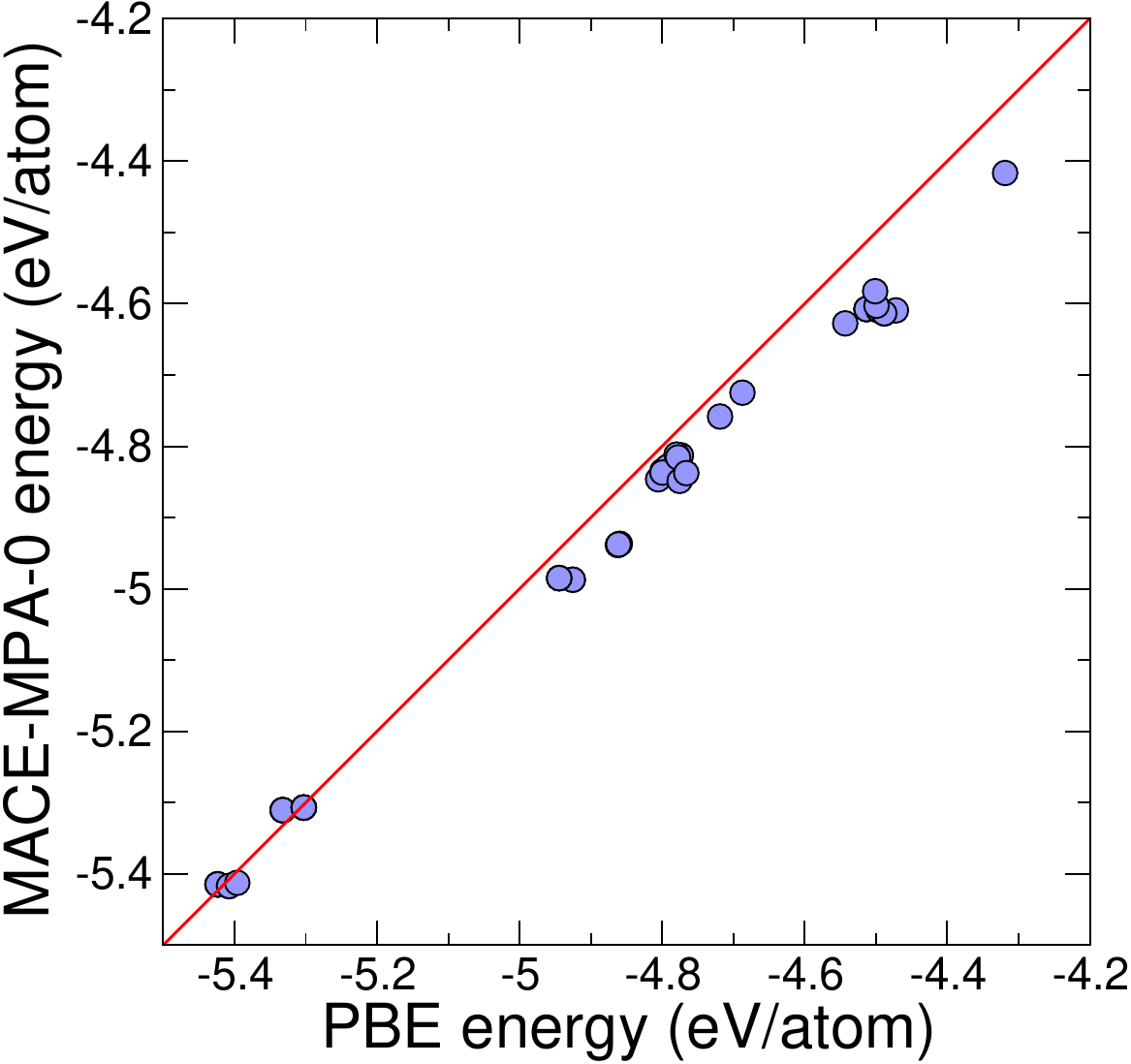}\label{fig:comparison:Si-Ge}}
    
    \caption{
        Parity plot of CHGNet and GGA-PBE energies for various material phases.
        Energy calculations are compared for structures representative of those typically generated by random structure search methods: \protect\subref{fig:comparison:C} bulk carbon, \protect\subref{fig:comparison:Al} bulk aluminium, \protect\subref{fig:comparison:C-MgO} graphene-encapsulated MgO, and \protect\subref{fig:comparison:Si-Ge} a set of Si|Ge interfaces.
        For the aluminium dataset, structures with energies greater than $0$~\si{\electronvolt}/atom are excluded as such energetically unfavourable systems are unlikely to be explored in effective structure search methods.
    }
    \label{fig:comparison}
\end{figure*}

We assess the validity of using machine-learned potentials to explore chemical spaces.
Foundation models have proven effective as surrogates for GGA-PBE~\cite{perdewGeneralizedGradientApproximation1996} DFT in known bulk materials and have shown promise in random structure search (RSS).
To ensure the suitability of the respective foundation model for this search space, we compare its energy predictions against DFT for a range of explored systems, including bulk aluminium, diamond carbon, and interfaces such as Li-intercalated ScS$_{2}$ structures, graphene-encapsulated MgO (C-MgO), and Si\|Ge.
These parity plots are presented in \figref{fig:comparison}.

While energy per atom is a valid unit for comparing single-species bulk materials, it is less appropriate for intercalation and interface systems.
The choice of reference states affects formation energy, so the context of each system must be considered. For intercalation, energy per intercalant unit relative to the host is more suitable, while for interfaces, energy per unit area is preferable.
However, due to the complexity of these datasets and the diversity of structures studied, energy per atom has been used for all parity plots.
This provides a reasonable estimate of search space accuracy, even if it does not fully capture the context of each system.

\section{Sensitivity of RAFFLE to choice of calculator}
\label{sec:calculators}

\begin{figure*}[ht!]
    \centering
    \subfloat[]{\includegraphics[width=0.45\linewidth]{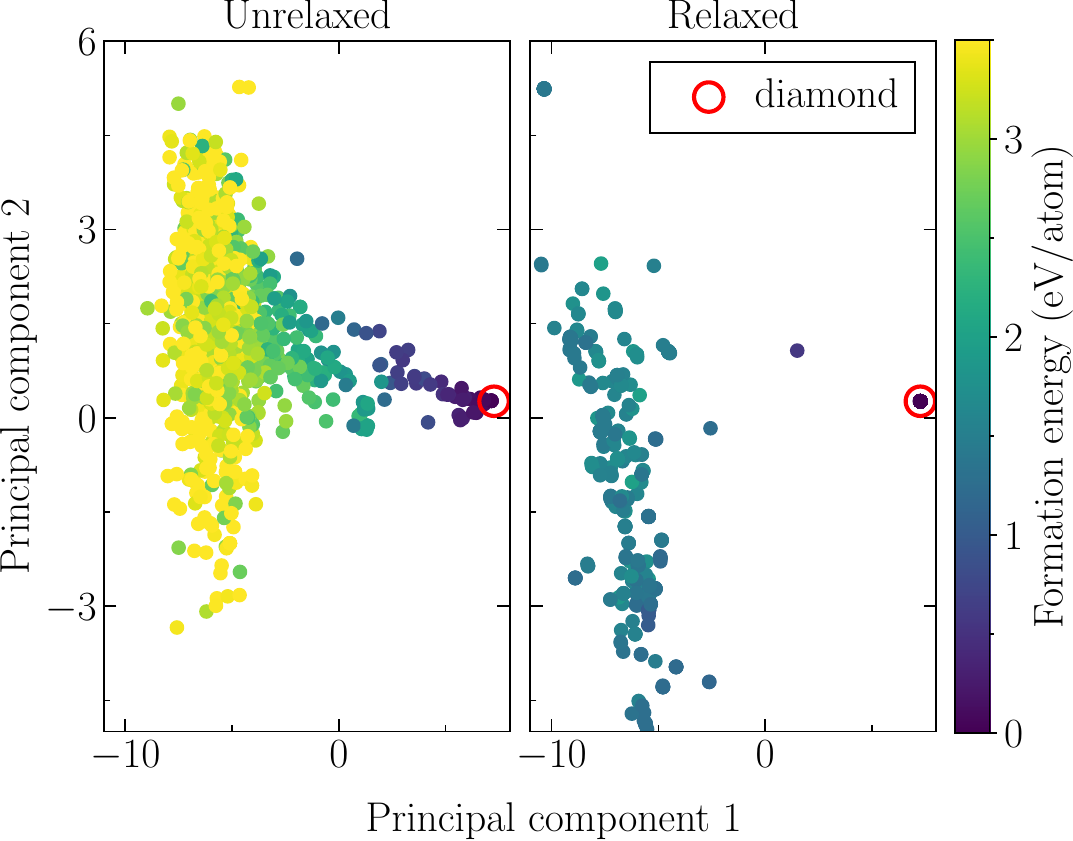}\label{fig:C:VASP:RAFFLE:pca}}
    \hspace{1em}
    \subfloat[]{\includegraphics[width=0.45\linewidth]{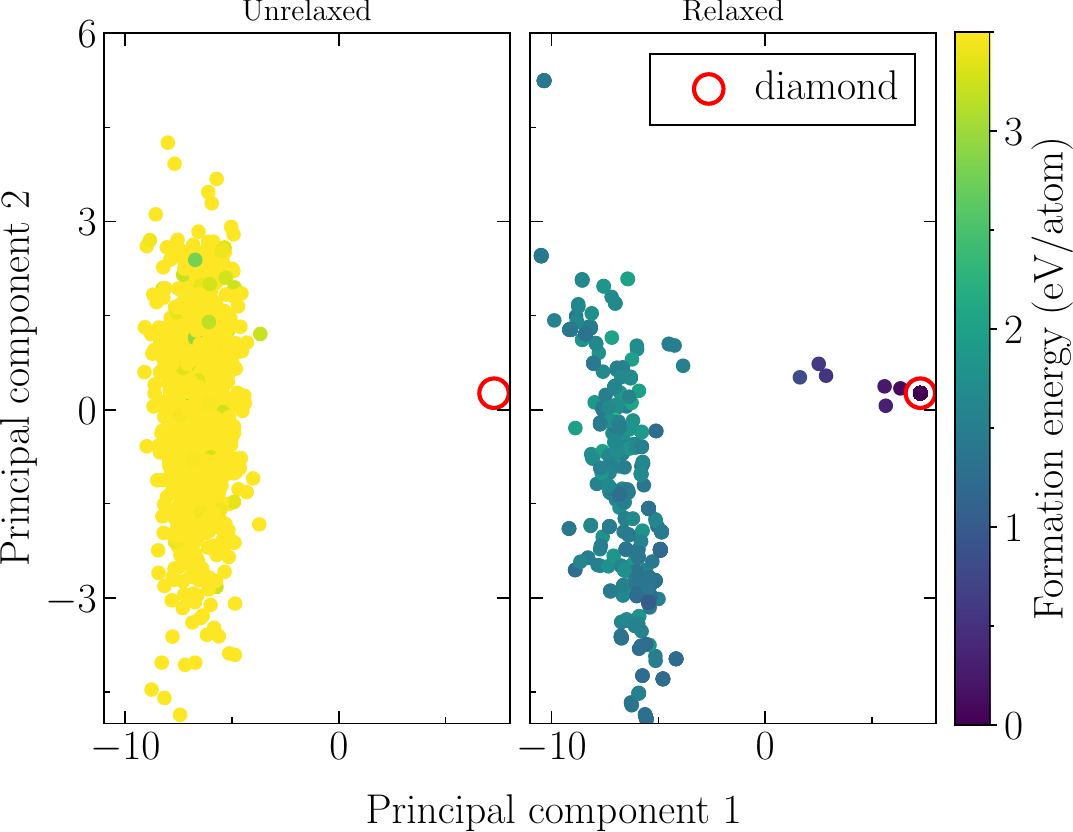}\label{fig:C:VASP:RSS:pca}}
    
    \caption{
        Comparison of RAFFLE and random structure search (RSS) as implemented in AGOX~\cite{Christiansen2022AtomisticGlobalOptimization} for generating 1000 structures in a search for the diamond phase of carbon.
        Principal component analysis (PCA) of unrelaxed and relaxed structures generated using \protect\subref{fig:C:VASP:RAFFLE:pca} RAFFLE and \protect\subref{fig:C:VASP:RSS:pca} RSS, respectively.
        Graphite is not generated by either method due to the enforced cell and density
        The search is performed for 8 carbon atoms in a cubic cell ($a = 3.567$~\si{\angstrom}).
        Energetics are computed using the VASP calculator~\cite{kresseEfficiencyAbinitioTotal1996,kresseEfficientIterativeSchemes1996} using the DFT GGA-PBE functional~\cite{perdewGeneralizedGradientApproximation1996}, and structural relaxations (fixed cell) are performed using FIRE~\cite{Bitzek2006StructuralRelaxationMade} (\texttt{fmax=0.05}, \texttt{steps=100}).
    }
    \label{fig:C:VASP}
\end{figure*}

In \figref{fig:C:VASP}, we present the carbon diamond search using RAFFLE and RSS with the VASP~\cite{kresseEfficiencyAbinitioTotal1996,kresseEfficientIterativeSchemes1996} density functional theory (DFT) calculator.
Here, the PBE form of the generalised gradient approximation (GGA)~\cite{perdewGeneralizedGradientApproximation1996} is used.
The basic C pseudopotential is used, which contains orbitals  $2s^2 2p^2$.
Energy cutoff is set to $400$~\si{\electronvolt}, whilst the momentum space is sampled using a $\Gamma$-centred $3\times3\times3$ Monkhorst-Pack grid~\cite{Monkhorst1976SpecialPointsBrillouin}.

For best comparison with the results presented in the main article, the principal component analysis (PCA) is fit using the CHGNet data, with the VASP data being transformed/mapped into this 2D principal component space.
When comparing the results to those presented in \green{Figure 6 in the main article}, it can be seen that the same qualitative data is retrieved for both the RAFFLE and RSS methods.
Slightly higher energetic values are calculated for structures between $-10<\mathrm{PC1}<-3$.

\begin{figure*}[ht!]
    \centering
    \subfloat[]{\includegraphics[width=0.45\linewidth]{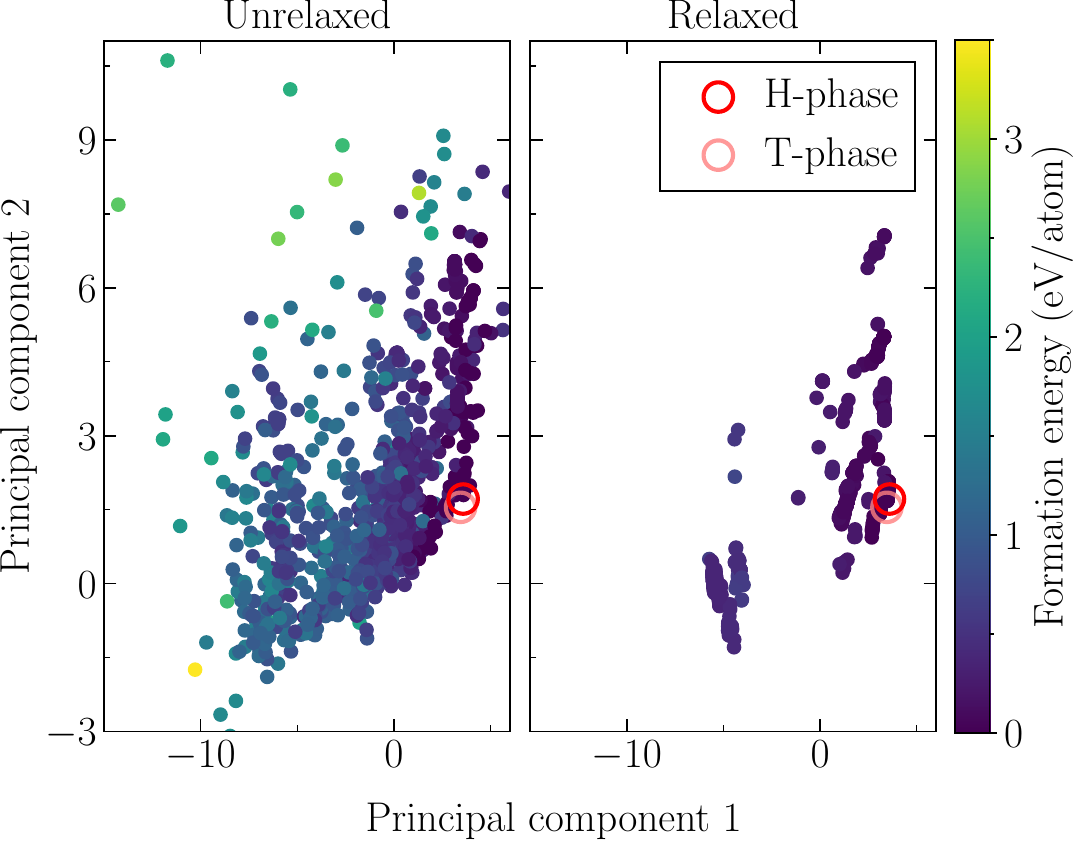}\label{fig:MoS2:vdW:d4:pca}}
    \hspace{1em}
    \subfloat[]{\includegraphics[width=0.45\linewidth]{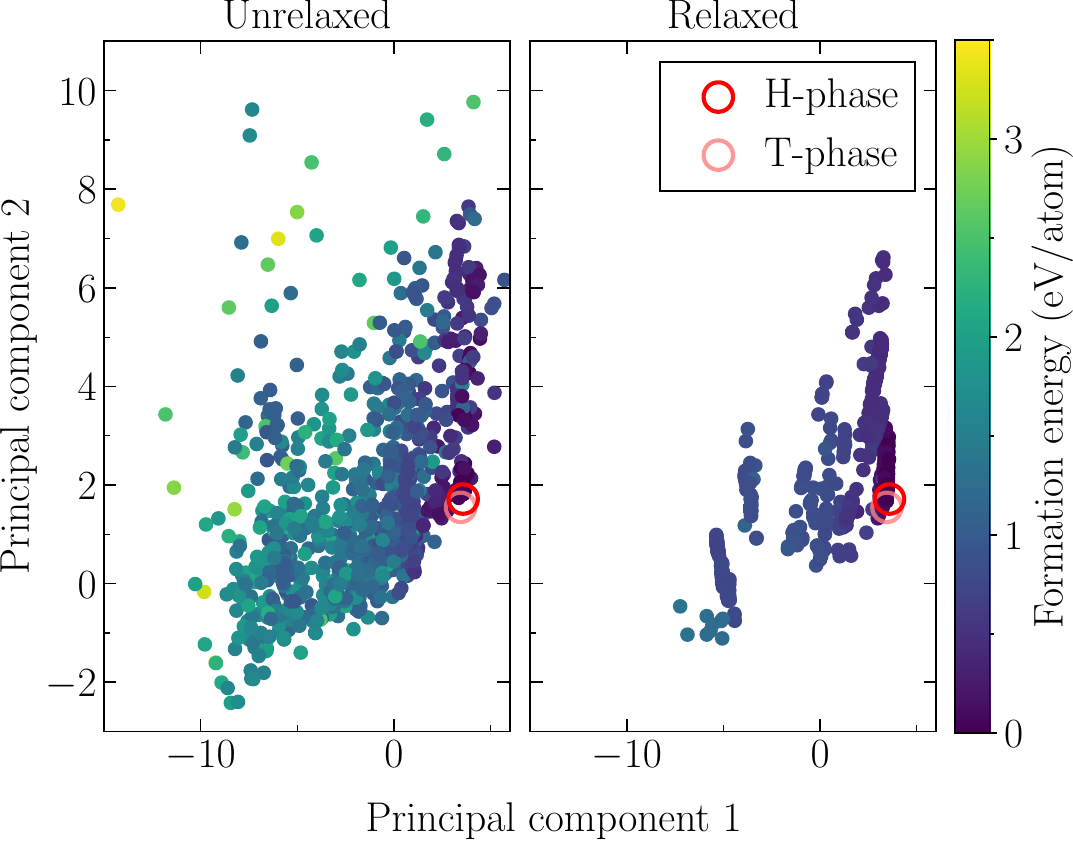}\label{fig:MoS2:vdW:MACE:pca}}
    
    \caption{
        Principal component analyses of RAFFLE-generated MoS$_{2}$ structures (pre- and post-atomic relaxation).  
        Red circles highlight the known H- and T-phases. 
        Bulk MoS$_{2}$ search using a single hexagonal Bravais lattice (1000 structures).
        Cell choice is the same as in the main article (lattice constants $a = 3.19$~\si{\angstrom} and $c = 13.1$~\si{\angstrom}.
        Search performed using the \protect\subref{fig:MoS2:vdW:d4:pca} CHGNet+DFT-D4 and \protect\subref{fig:MoS2:vdW:MACE:pca} MACE-MP-0 energetic calculators (\texttt{mace-torch} version 0.3.9), respectively.
        In both, atomic relaxations area performed using FIRE~\cite{Bitzek2006StructuralRelaxationMade} (cells are fixed).
    }
    \label{fig:MoS2:vdW}
\end{figure*}

Next, the comparison of choice of foundation model is made.
For the search for the bulk phase of MoS$_{2}$, the main article uses the CHGNet calculator.
In \figref{fig:MoS2:vdW}, the same search is presented using a sum of the CHGNet and DFT-D4 calculators, and using the MACE-MP-0 calculator (with dispersion turned on).
It can be seen that the same qualitative data is retrieved in all.
The same search space is explored.
Whilst the two CHGNet results show a similar search space explored for the relaxed structures, the MACE calculator show a bit more variation in the set of relaxed structures.
Furthermore, it is seen that the structures calculated using MACE identified within $-15<\mathrm{PC1}<0$ have higher formation energies as compared to the two CHGNet methods.

\section{Random number seed}
\label{sec:random_seed}

\begin{figure*}[h!]
    \centering
    \subfloat[]{\includegraphics[width=0.45\linewidth]{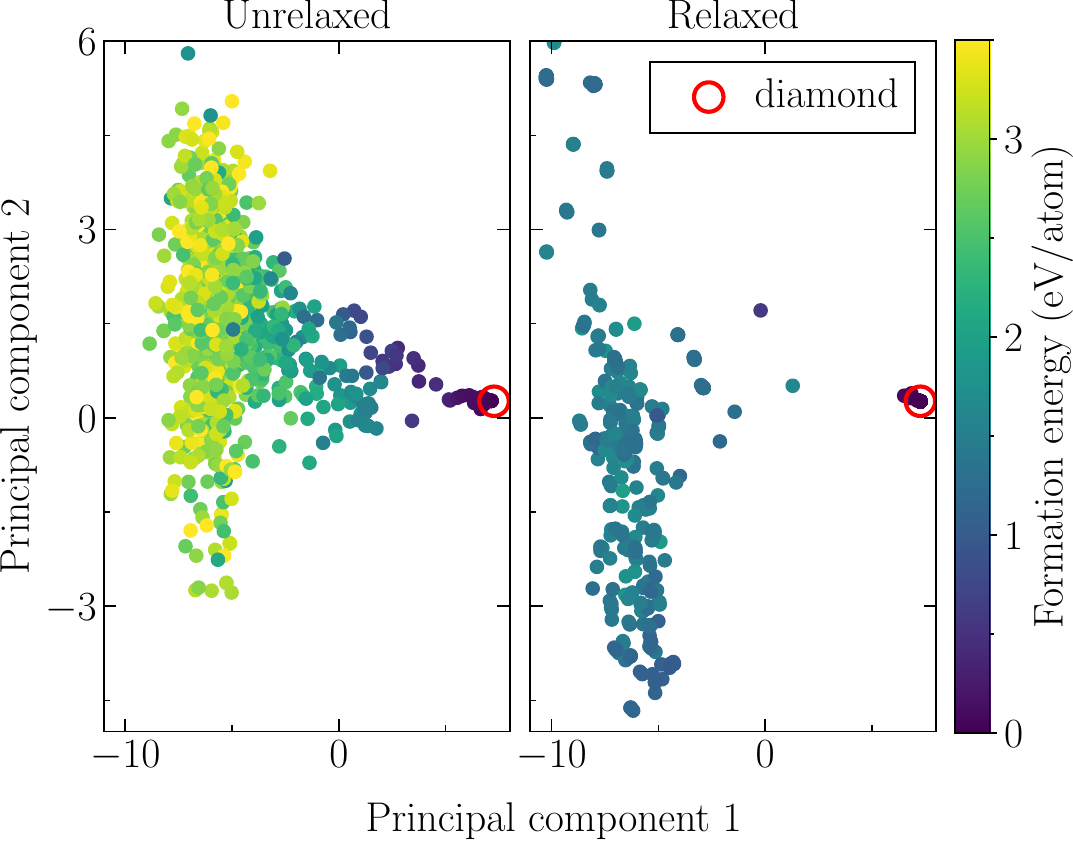}\label{fig:seed:1}}
    \hspace{1em}
    \subfloat[]{\includegraphics[width=0.45\linewidth]{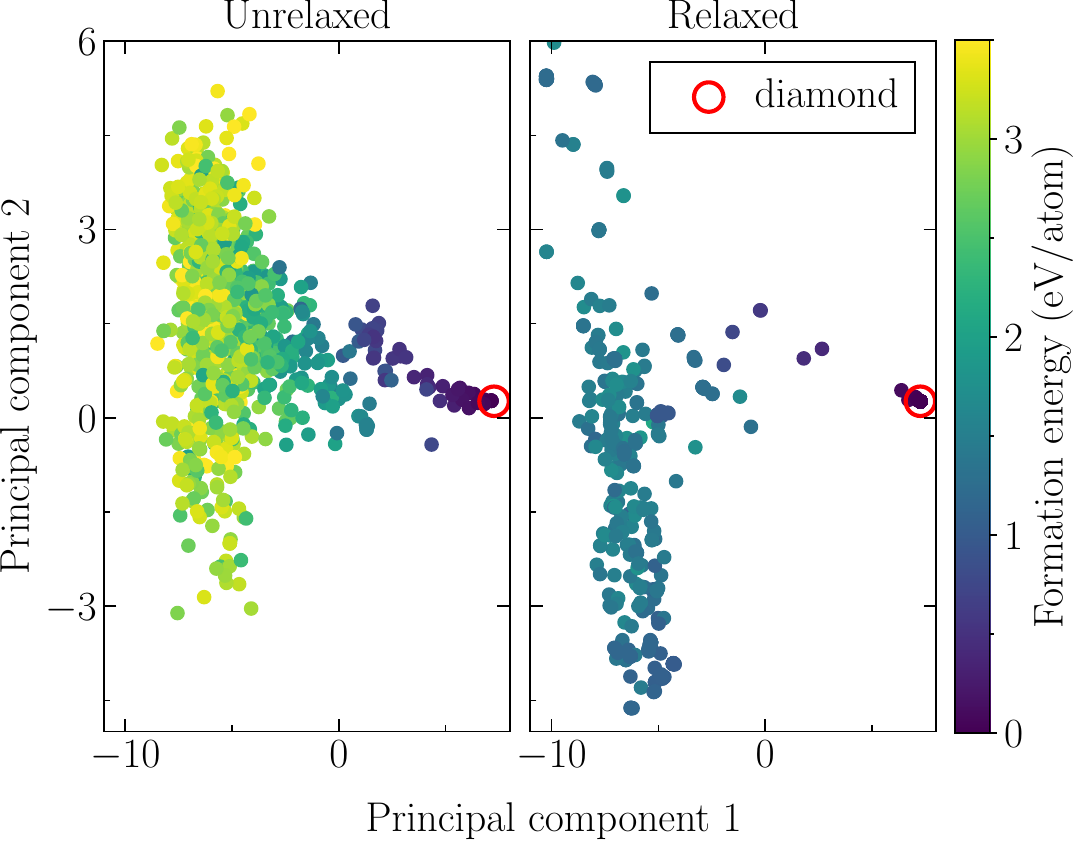}\label{fig:seed:2}}
    \hspace{1em}
    \subfloat[]{\includegraphics[width=0.45\linewidth]{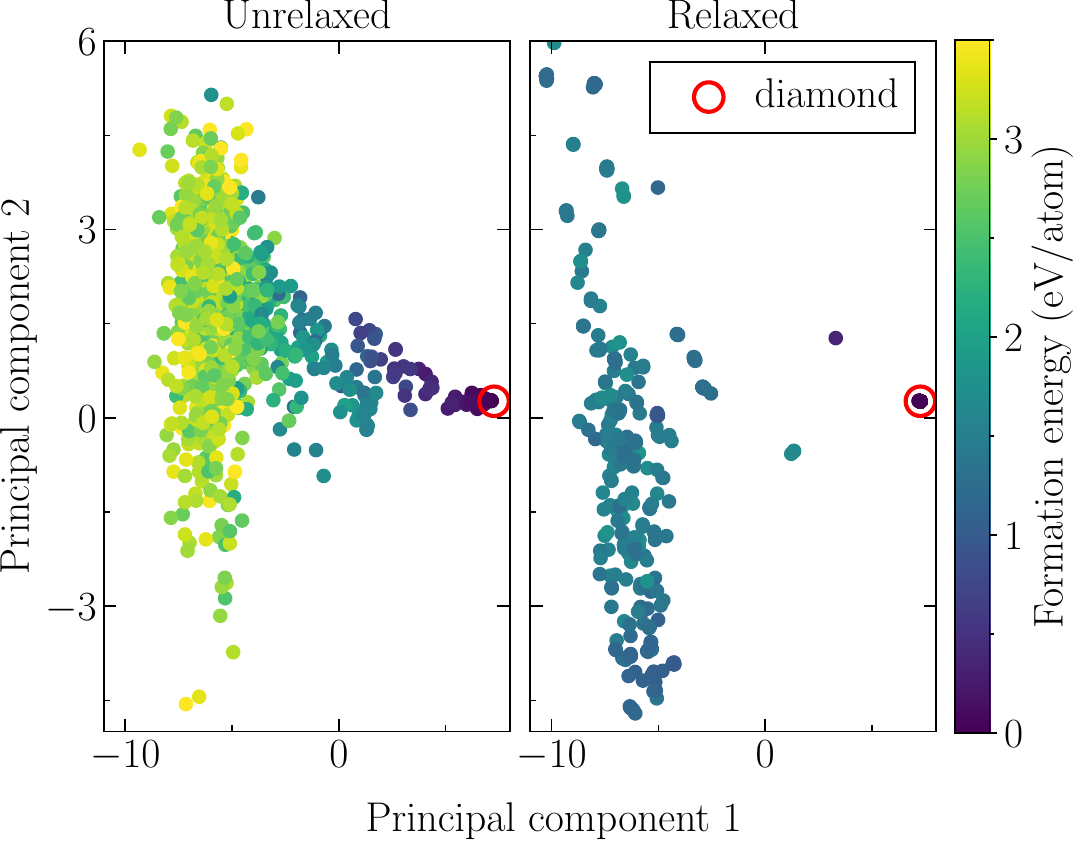}\label{fig:seed:3}}
    \hspace{1em}
    \subfloat[]{\includegraphics[width=0.45\linewidth]{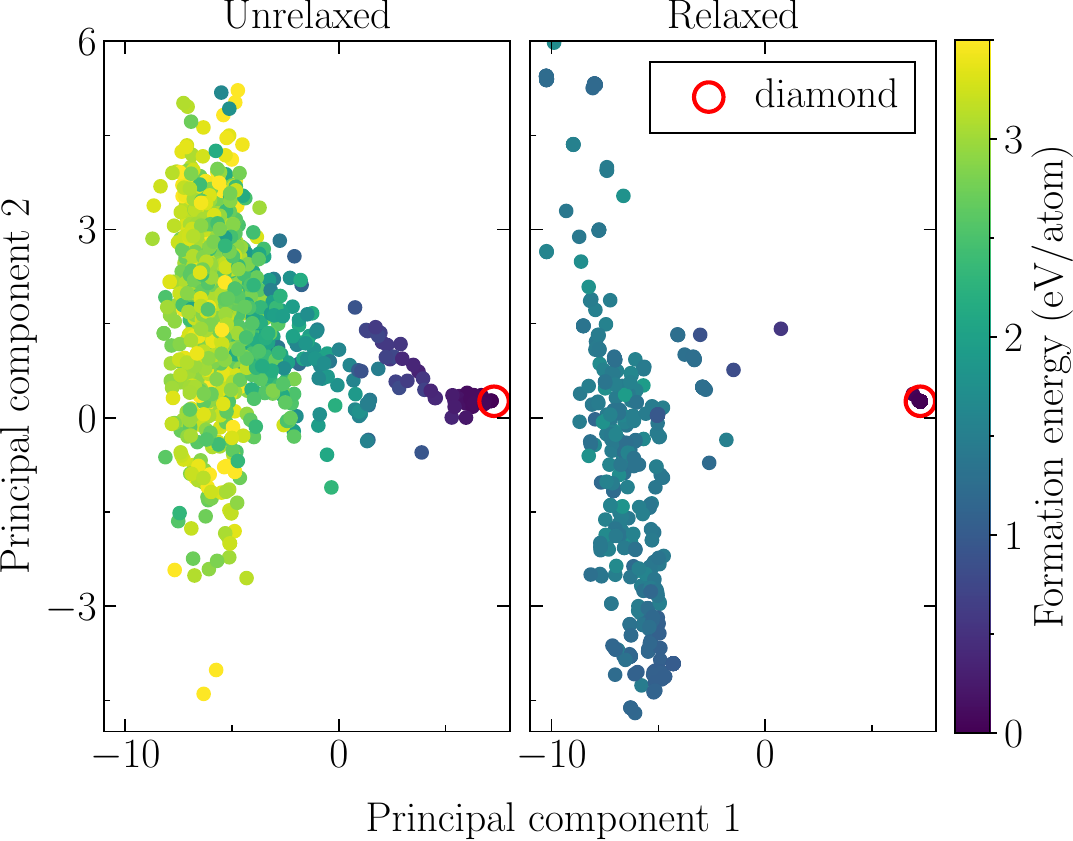}\label{fig:seed:4}}
    
    \caption{
        Principal component analyses of RAFFLE applied to the search for the carbon diamond phase using a set of random seeds.
        For each subfigure, 1000 structures were generated from a single cubic cell with lattice constant $a = 3.567$\si{\angstrom}, with RAFFLE placing 8 atoms.
        The left and right graphs in each subfigure show the principal component analysis (PCA) of unrelaxed and relaxed structures, respectively.
        The random seeds used are \protect\subref{fig:seed:1} 1, \protect\subref{fig:seed:2} 2, \protect\subref{fig:seed:3} 3, and \protect\subref{fig:seed:4} 4, while the corresponding data in the main article used seed 0.
        Energetics were obtained using the CHGNet calculator~\cite{Deng2023CHGNETPretrainedUniversalNeural}, and structural relaxations (fixed cell) are performed using FIRE~\cite{Bitzek2006StructuralRelaxationMade} (\texttt{fmax=0.05}, \texttt{steps=100}).
    }
    \label{fig:seed}
\end{figure*}

We assess the repeatability of RAFFLE’s results using a set of random seeds.
The carbon diamond bulk search serves as a test case, with the same procedure repeated for 20 different seeds.
Four sample results are shown in \figref{fig:seed}, and one in \green{Figure 6 in the main article}; the remaining 15 exhibit similar trends with no clear outliers, so are not presented here.
In each case, the principal component model is fit to the main article’s data, and the new results are transformed accordingly to improve clarity and comparison.
Qualitative trends are reproduced for all tested random seeds.

\section{Comparison of placement method ratio}
\label{sec:C:compare_placement_method_ratio}

\begin{figure*}[h!]
    \centering
    \subfloat[]{\includegraphics[width=0.45\linewidth]{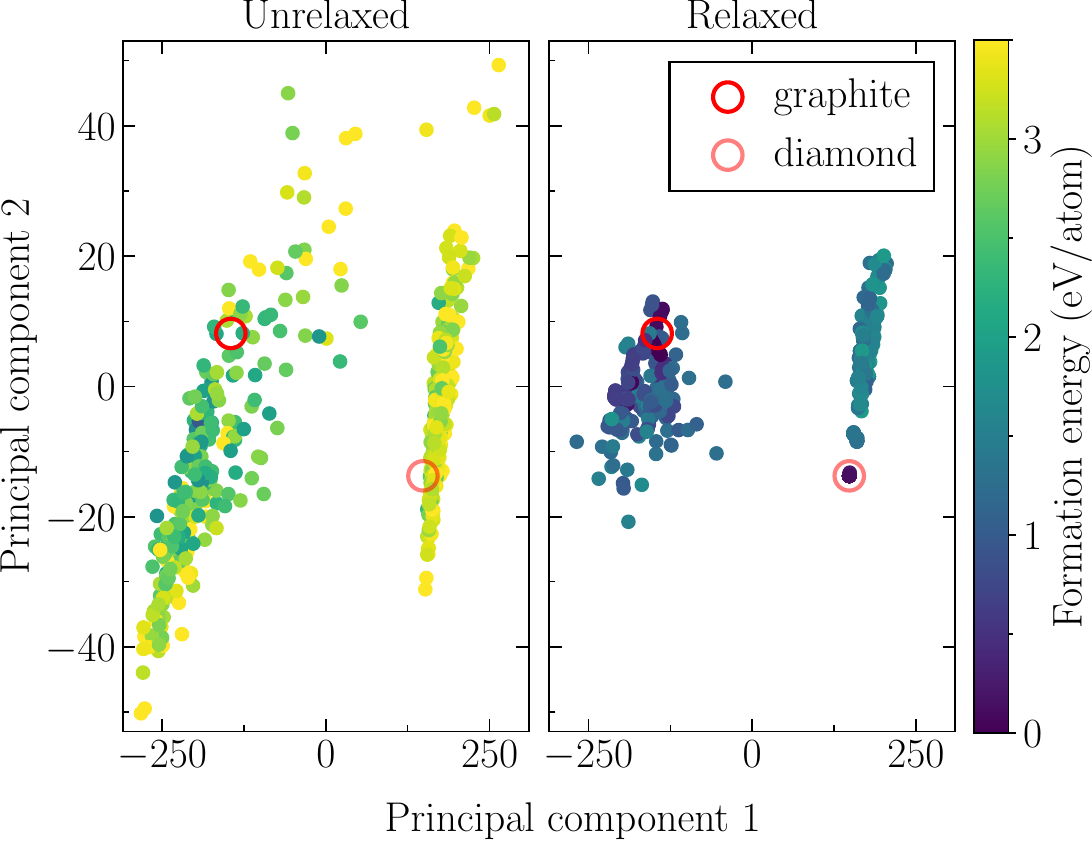}\label{fig:C:ratio:1}}
    \hspace{1em}
    \subfloat[]{\includegraphics[width=0.45\linewidth]{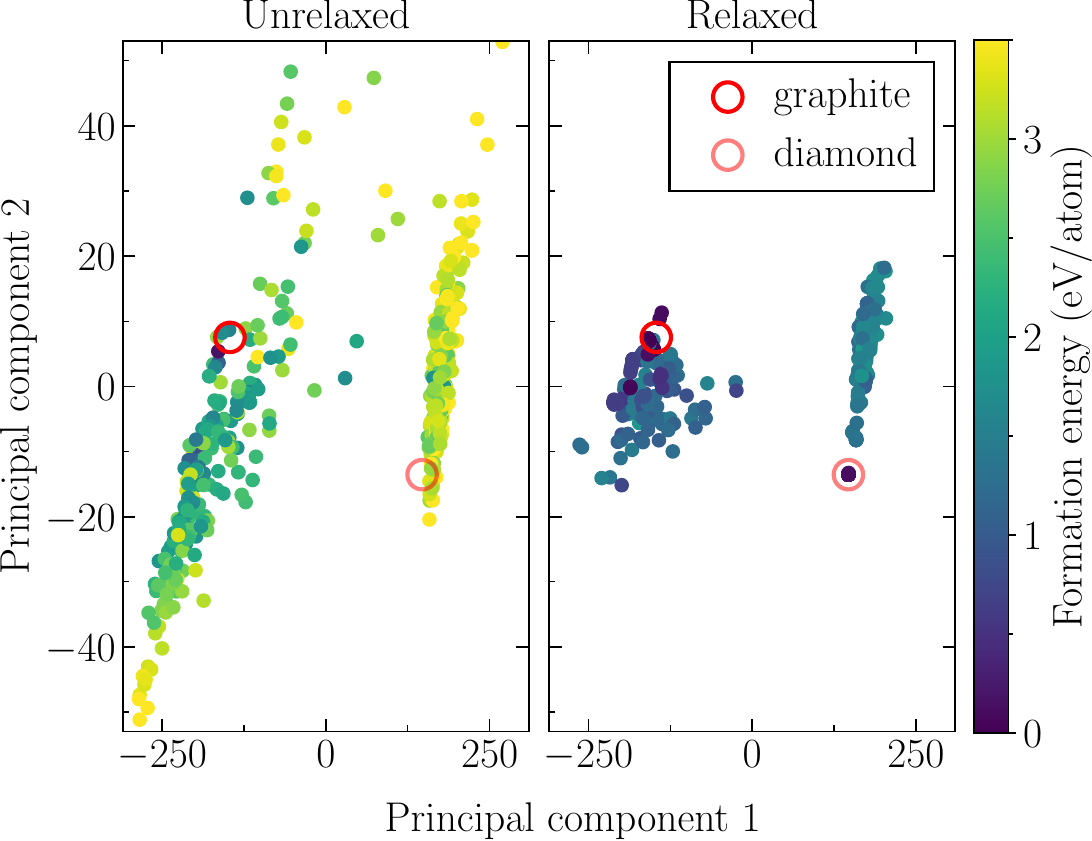}\label{fig:C:ratio:2}}
    
    \caption{
        Principal component analyses of RAFFLE applied to the search for the carbon diamond phase using different placement method ratios.
        For each subfigure, 500 structures were generated following the workflow outlined in the main article.
        The left and right graphs in each subfigure show the principal component analysis (PCA) of unrelaxed and relaxed structures, respectively.
        The placement method ratios used are \protect\subref{fig:C:ratio:1} \texttt{void: 0.5, rand: 0.001, walk: 0.5, grow: 0.0, min: 1.0} and \protect\subref{fig:C:ratio:2} \texttt{void: 0.5, rand: 0.001, walk: 0.25, grow: 0.25, min: 1.0}.
        Energetics were obtained using the CHGNet calculator~\cite{Deng2023CHGNETPretrainedUniversalNeural}, and structural relaxations (fixed cell) are performed using FIRE~\cite{Bitzek2006StructuralRelaxationMade} (\texttt{fmax=0.05}, \texttt{steps=100}).
    }
    \label{fig:C:ratio}
\end{figure*}

A comparison of placement method ratios on the bulk carbon search is provided.
The bulk carbon structure search is conducted on cubic and hexagonal lattices using two additional placement method ratios, as shown in \figref{fig:C:ratio}.
In both cases, the qualitative trends are consistent with each other and with the results presented in \green{Figure 7a in the main article}.

\section{Additional comparisons to random structure search}
\label{sec:addit_rss}

\begin{figure*}[h!]
    \centering
    \subfloat[Al]{\includegraphics[width=0.45\linewidth]{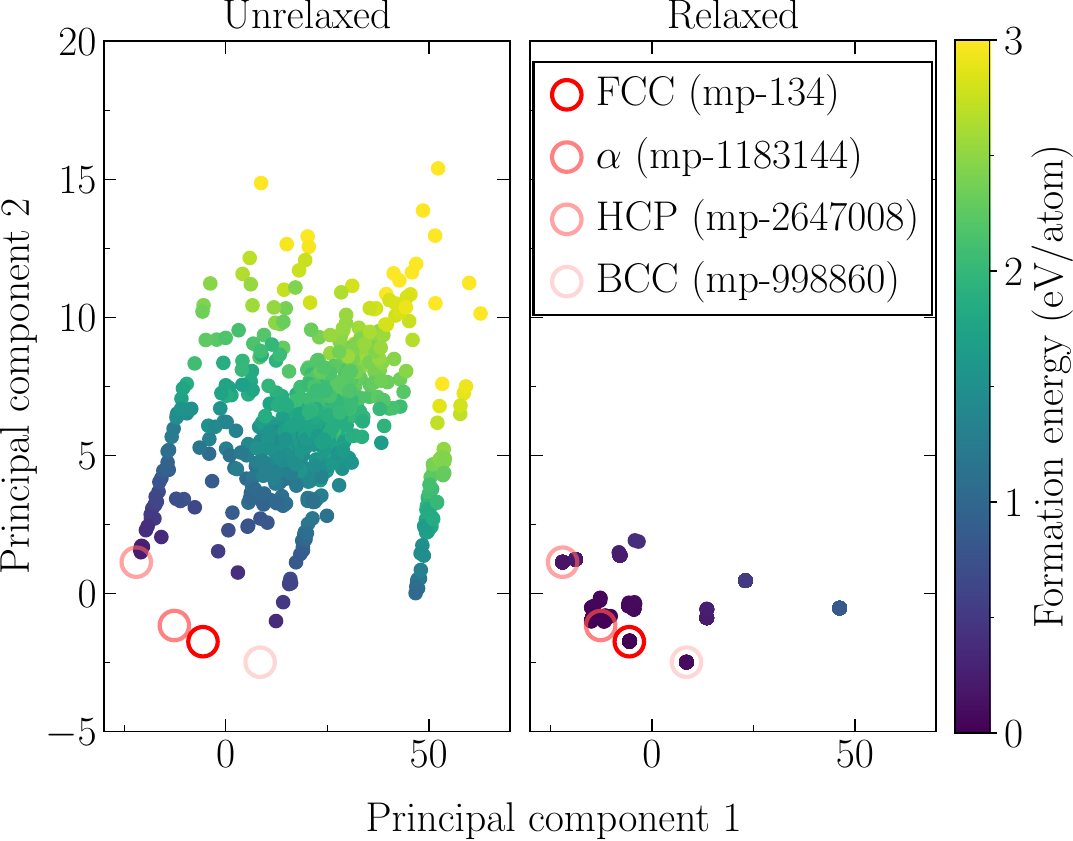}\label{fig:rss:Al}}
    \hspace{1em}
    \subfloat[MoS$_{2}$]{\includegraphics[width=0.45\linewidth]{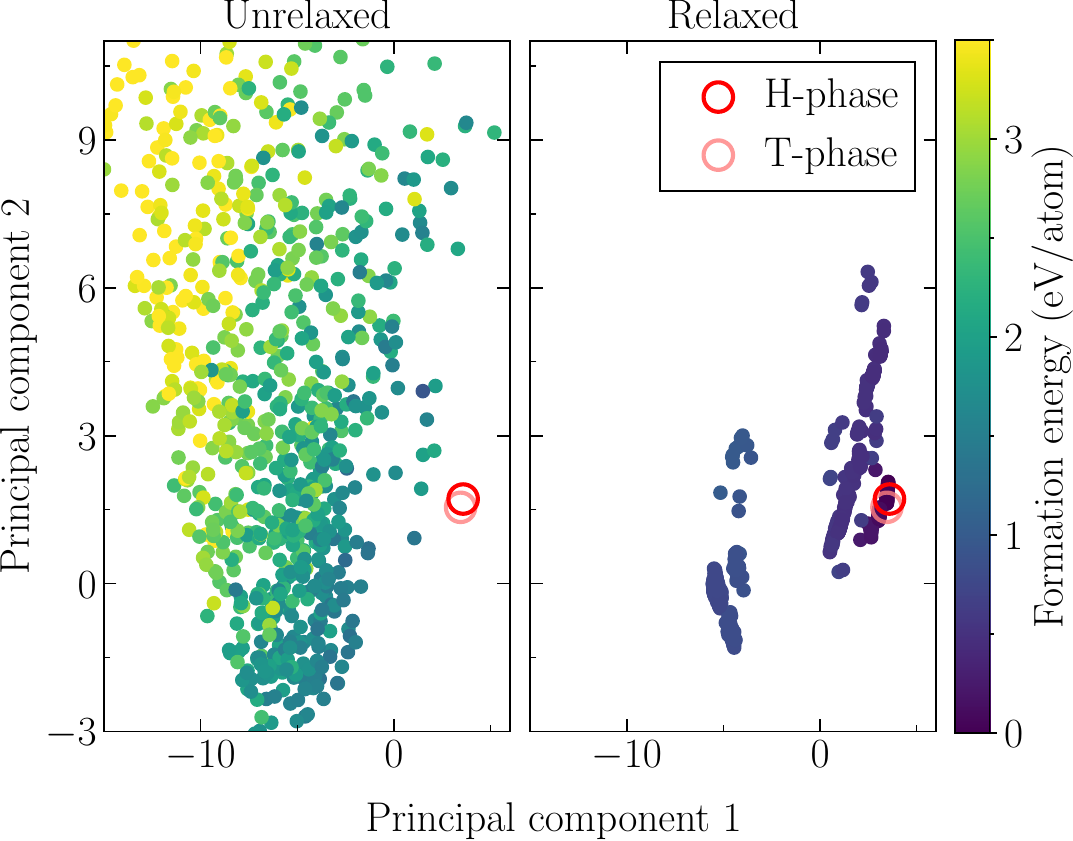}\label{fig:rss:MoS2}}
    
    \caption{
        Principal component analyses of random structure search applied to bulk materials.
        Random structure search, as implemented in AGOX~\cite{Bisbo2022GlobalOptimizationAtomic} applied to \protect\subref{fig:rss:Al} aluminium and \protect\subref{fig:rss:MoS2} MoS$_{2}$, respectively.
        Both searches are applied in the same form as the RAFFLE structure searches presented in \green{Figures 7b and c in the main article}.
        For aluminium, a total of 600 structures are generated using cubic and hexagonal lattices for a range of lattice constants (between $a=c=3.1$--$5.4$~\si{\angstrom}.
        For MoS$_{2}$, a total of 1000 structures are generated using a hexagonal cell with lattice constants $a = 3.19$~\si{\angstrom} and $c = 13.1$~\si{\angstrom}.
        The data presented in the PCAs have been transformed to the principal components fit to their respective RAFFLE searches, to allow for direct comparison between the two methods.
        Energetics are obtained using the CHGNet calculator~\cite{Deng2023CHGNETPretrainedUniversalNeural} and structures are relaxed from their initial generated structures using the FIRE optimiser~\cite{Bitzek2006StructuralRelaxationMade}.
    }
    \label{fig:rss}
\end{figure*}

\Figref{fig:rss} shows the results of RSS applied to the bulk phases of aluminium and MoS$_{2}$.
The RSS and RAFFLE approaches (with RAFFLE data presented in the main article) yield qualitatively similar outcomes.
However, three key differences emerge:
1) RAFFLE generates a greater number of unrelaxed structures with lower formation energies,
2) RAFFLE structures are biased towards PCA values closer to the ground-state structures, and
3) RAFFLE produces a more diverse set of unique structures post-relaxation.

\section{S\lowercase{i}|G\lowercase{e} interface analysis}
\label{sec:Si-Ge:analysis}

\begin{figure*}[ht!]
    \centering
    \subfloat[Abrupt: $-152.46$~\si{\milli\electronvolt/\angstrom\squared}]{\includegraphics[width=0.3\linewidth]{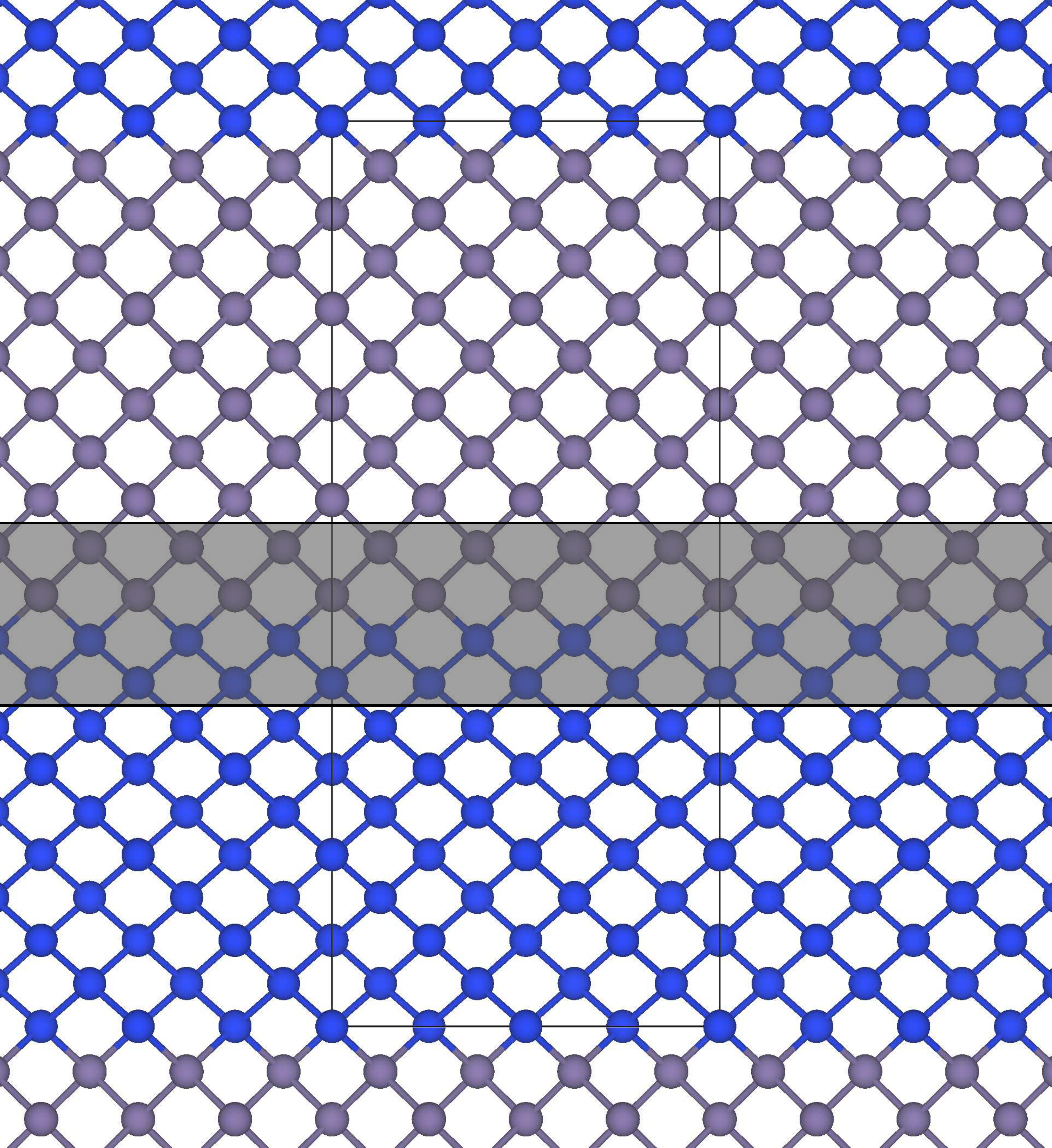}\label{fig:Si-Ge:MACE:structures:abrupt}}
    \hspace{1em}
    \subfloat[Lowest: $-152.08$~\si{\milli\electronvolt/\angstrom\squared}]{\includegraphics[width=0.3\linewidth]{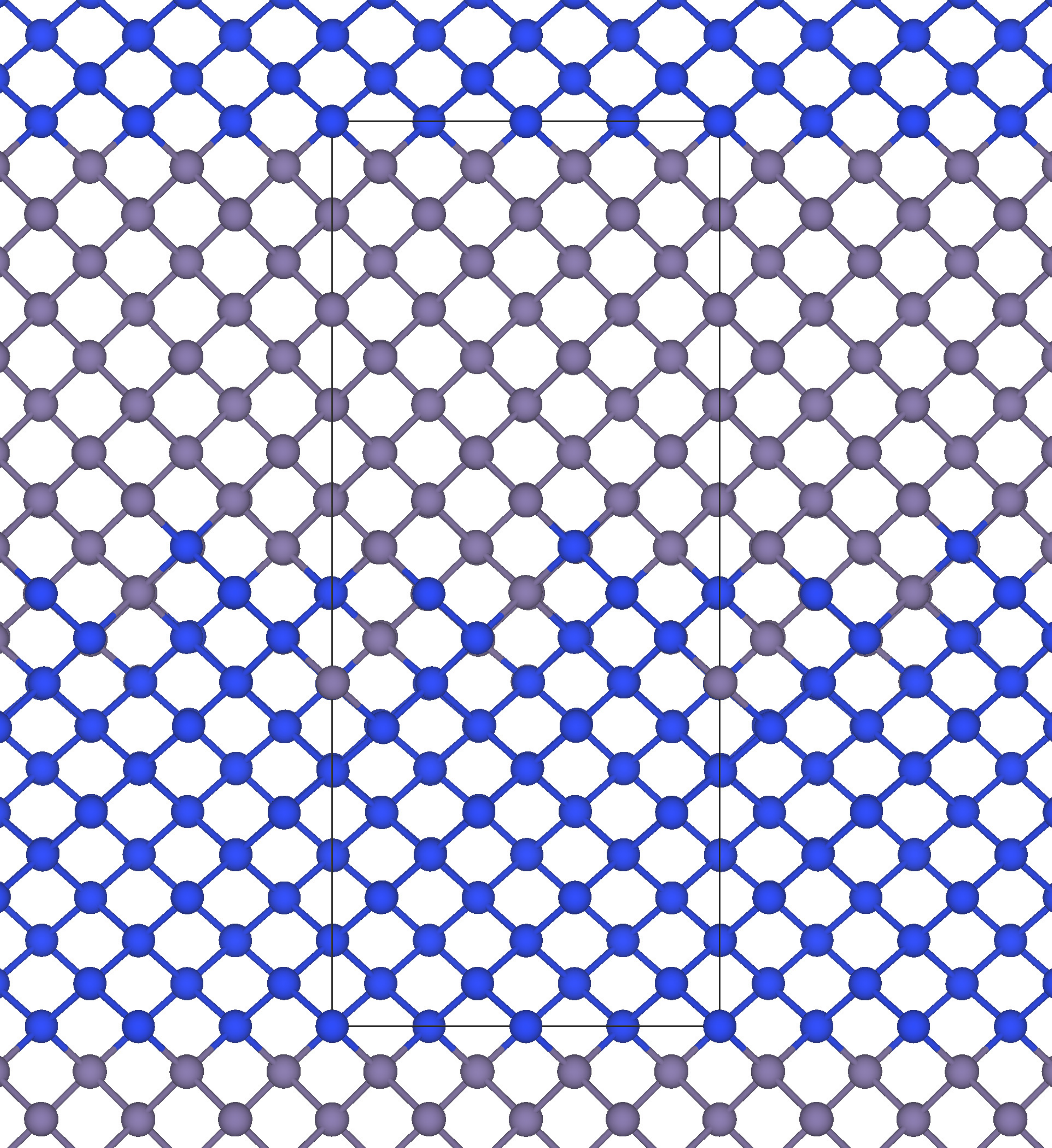}\label{fig:Si-Ge:MACE:structures:1st}}
    \hspace{1em}
    \subfloat[2nd lowest: $-151.46$~\si{\milli\electronvolt/\angstrom\squared}]{\includegraphics[width=0.3\linewidth]{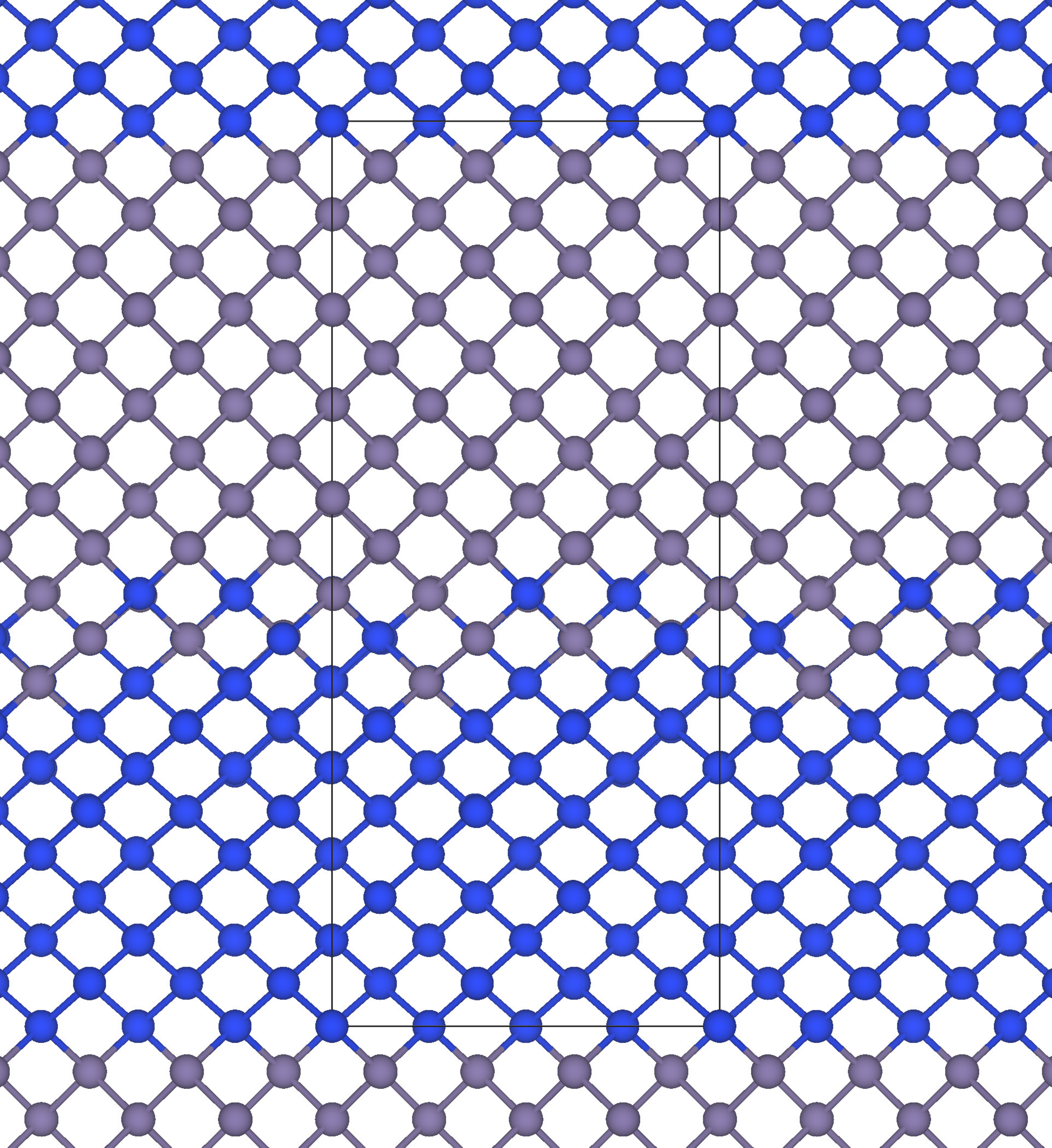}\label{fig:Si-Ge:MACE:structures:2nd}}

    \caption{
        Ball and stick models of atomically relaxed Si|Ge interfaces, where the black box represents the unit cell.
        \protect\subref{fig:Si-Ge:MACE:structures:abrupt} The abrupt Si|Ge interface, denoted using a circle in \green{Figure 8b in the main article}.
        The translucent overlay indicates the region in which atoms are removed and subsequently reinserted using RAFFLE for the structure search detailed in the main article (16 Si and 16 Ge atoms).
        The \protect\subref{fig:Si-Ge:MACE:structures:1st} lowest and \protect\subref{fig:Si-Ge:MACE:structures:2nd} 2nd lowest energy structures identified generated using RAFFLE.
        All structures undergo atomic relaxation using the FIRE optimisation method~\cite{Bitzek2006StructuralRelaxationMade} (fixed cell) and energetics calculated using the MACE-MPA-0 calculator~\cite{batatia2023FoundationModelAtomistic}.
    }
    \label{fig:Si-Ge:MACE}
\end{figure*}

\Figref{fig:Si-Ge:MACE} shows three of the relaxed structures of the Si|Ge interfaces discussed in the main article.
It can be seen that structural relaxation mostly results in significant reconstruction of the germanium region of the cell, whilst the Si region remains much more ordered in comparison.

Interface formation energy, $E_{\mathrm{form}}$ is calculated via

\begin{equation}
    E_{\mathrm{form}} = \frac{ E_{\mathrm{s}} - ( E_{\mathrm{Si,slab}} - E_{\mathrm{Ge,slab}} ) }{2A},
\end{equation}

\noindent
where $E_{\mathrm{s}}$ is the energy of the interface system, $E_{\mathrm{Si,slab}}$ and $E_{\mathrm{Ge,slab}}$ are the energies of the constituent Si and Ge slabs in isolation, respectively, and $A$ is the area of the cell parallel to the interface plane.
This is akin to the adhesion energy, normalised by area.
Due to this method of calculation, interface can be negative (i.e. more favourable than the isolated slabs) as the formation of said interface provides charge compensation for dangling/missing/broken bonds.

\subsection{CHGNet results}
\label{sec:Si-Ge:analysis:CHGNet}

\begin{figure*}[ht!]
    \centering
    \subfloat[PCA]{\includegraphics[width=0.4\linewidth]{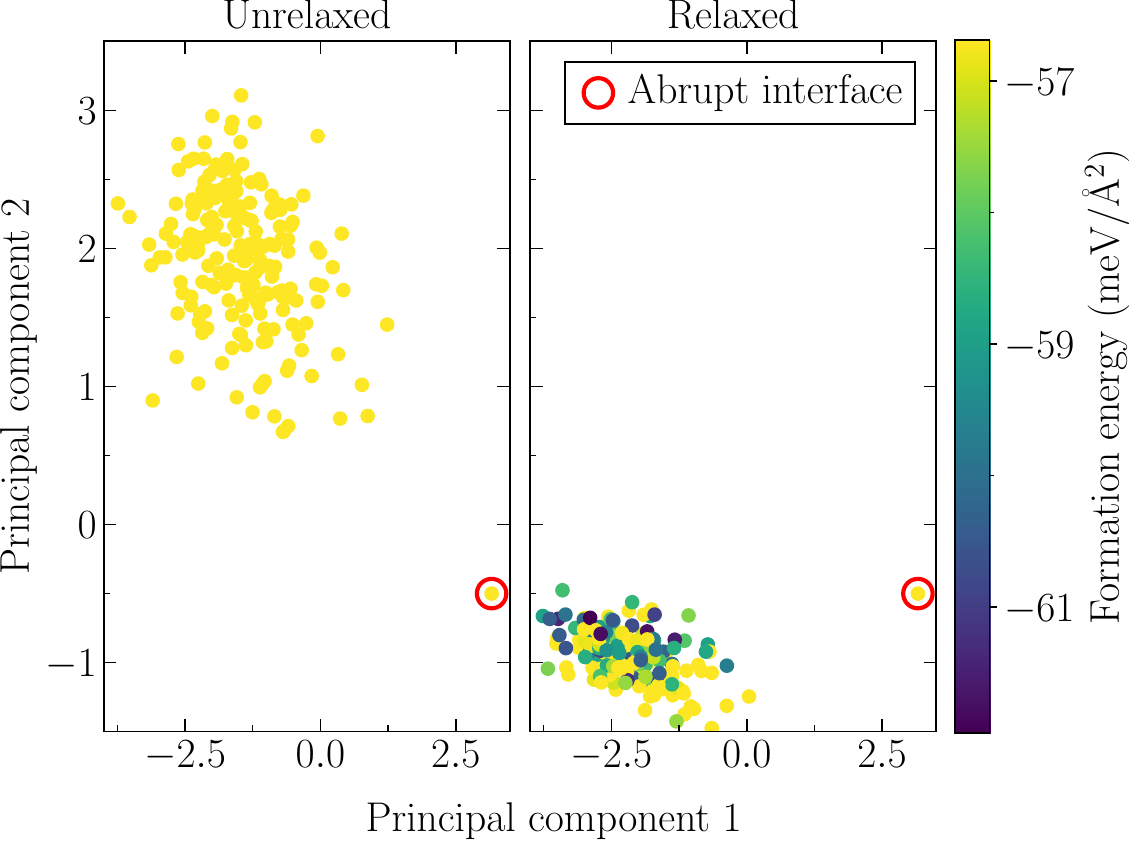}\label{fig:Si-Ge:CHGNet:pca}}
    \hspace{1em}
    \subfloat[Lowest: $-61.96$~\si{\milli\electronvolt/\angstrom\squared}]{\includegraphics[width=0.25\linewidth]{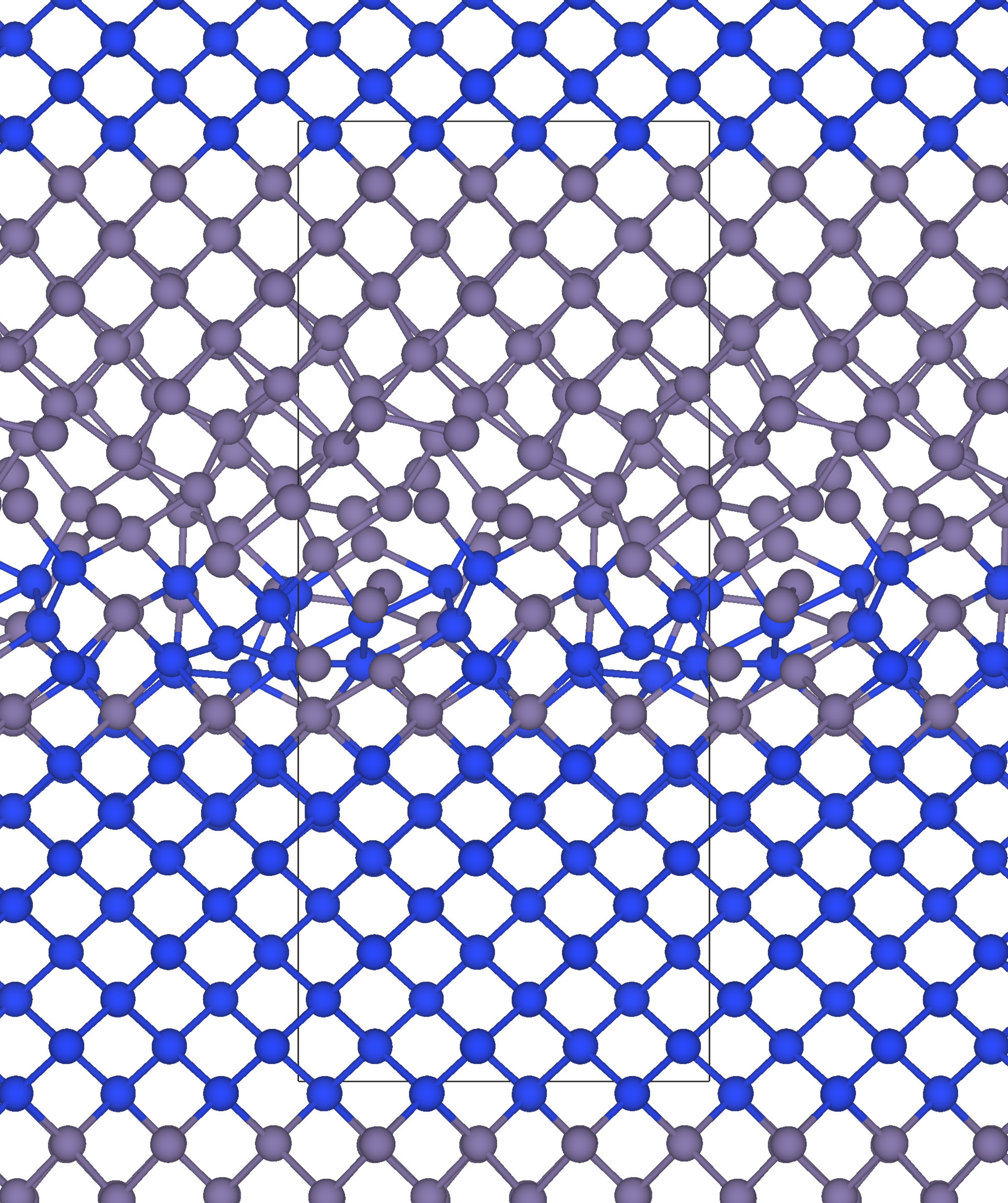}\label{fig:Si-Ge:CHGNet:structures:1st}}
    \hspace{1em}
    \subfloat[2nd lowest: $-61.79$~\si{\milli\electronvolt/\angstrom\squared}]{\includegraphics[width=0.25\linewidth]{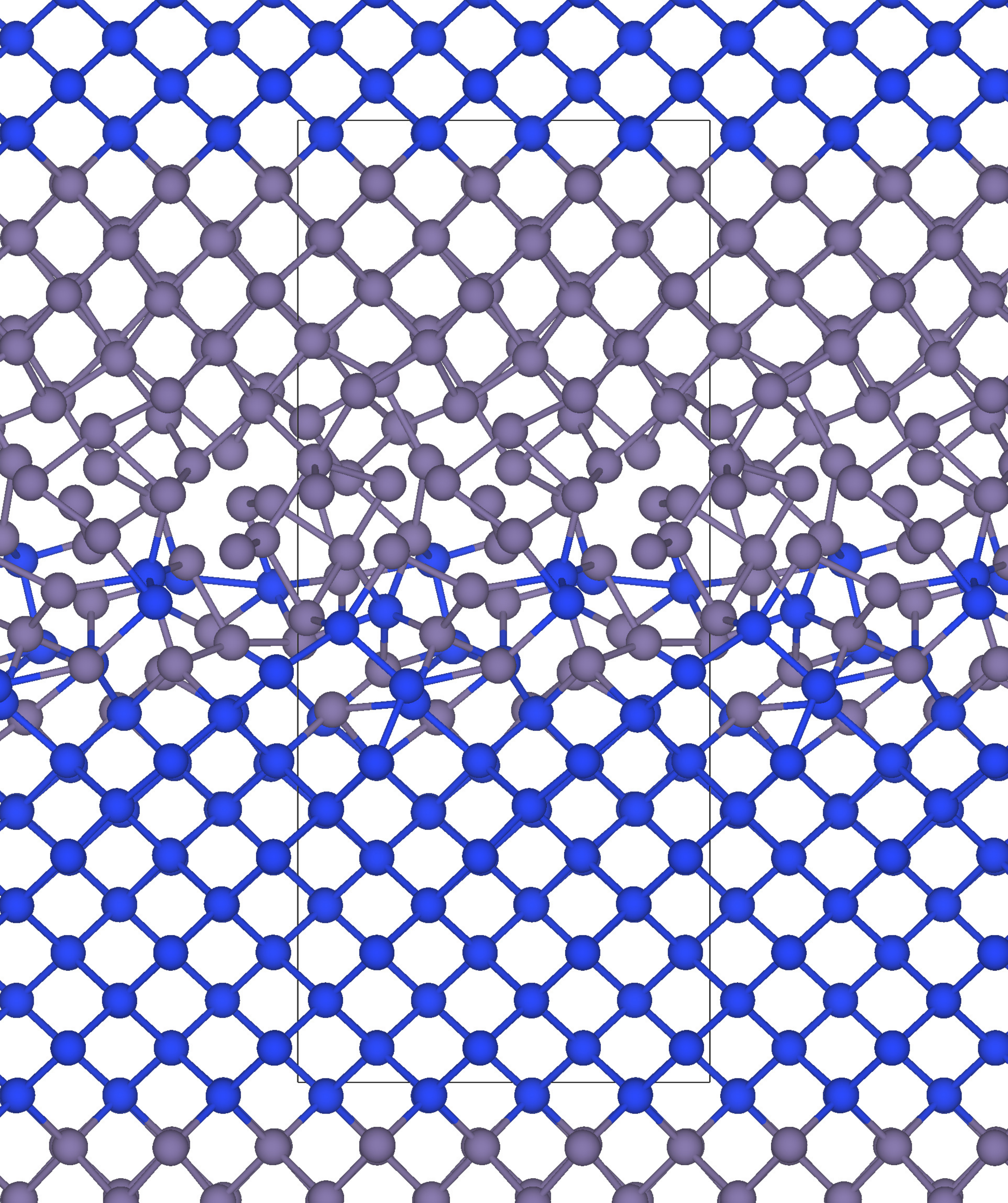}\label{fig:Si-Ge:CHGNet:structures:2nd}}

    \caption{
        Principal component analysis and ball and stick models of atomically relaxed Si|Ge interfaces using the CHGNet calculator, where the black box represents the unit cell.
        The abrupt Si|Ge interface has a formation energy of $-56.69$~\si{\milli\electronvolt/\angstrom\squared} when calculated using CHGNet.
        The abrupt interface matches that seen in \figref{fig:Si-Ge:MACE:structures:abrupt}, and the process used to generate RAFFLE structures follows the same approach outlined in the main article, but using two independent searches, each generating 100 structures.
        The \protect\subref{fig:Si-Ge:CHGNet:structures:1st} lowest and \protect\subref{fig:Si-Ge:CHGNet:structures:2nd} 2nd lowest energy structures identified generated using RAFFLE.
        All structures undergo atomic relaxation using the FIRE optimisation method~\cite{Bitzek2006StructuralRelaxationMade} (fixed cell) and energetics calculated using the CHGNet calculator~\cite{Deng2023CHGNETPretrainedUniversalNeural}.
    }
    \label{fig:Si-Ge:CHGNet}
\end{figure*}

The Si\|Ge interface structure search using the CHGNet energy calculator follows the same procedure as the MACE-MPA-0 search detailed in the main article.
The only difference is that two independent searches are run in parallel, each with a different ratio of placement methods.
Both searches undergo 20 iterations, but one places greater emphasis on known structures by assigning a higher weight to the minimum placement method (see \tabref{tab:Si-Ge:CHGNet:params} for search parameters).

The PCA analysis of all 200 generated structures is shown in \figref{fig:Si-Ge:CHGNet:pca}.
The most stable interface found in this search has an interface formation energy of $-61.96$~\si{\milli\electronvolt/\angstrom\squared}, suggesting that disordered interfaces generated using the CHGNet calculator within the RAFFLE workflow can be more energetically favourable than the abrupt, lattice-matched interface.
However, when these lowest-energy structures are recalculated using GGA-PBE in GPAW~\cite{Mortensen2024GPAWOpenPythonPackage}, the two most stable RAFFLE-generated interfaces are found to be $54.59$ and $60.73$~\si{\milli\electronvolt/\angstrom\squared} less favourable than the abrupt interface.

The two lowest-energy structures identified by RAFFLE using CHGNet are shown in \figref{fig:Si-Ge:CHGNet}.
These structures exhibit significant disorder, with most of the atomic reconstruction, likely caused by strain, occurring in the germanium region, despite all atoms being free to relax.
This contrasts sharply with the lowest-energy structures obtained using the MACE-MPA-0 calculator (\figref{fig:Si-Ge:MACE}), where the interface remains highly ordered, retaining the diamond-like phase and instead showing atomic site intermixing across the boundary.

These findings underscore the importance of thoroughly testing energy calculators within the relevant energy space for each study.

\begin{table}
    \centering
    \caption{
        RAFFLE parameters used for the Si\|Ge interface example case using the CHGNet calculator, detailed in \secref{sec:Si-Ge:analysis:CHGNet}.
        Details regarding how to set the parameters are outlined in \secref{sec:guide:parameters}.
        The $k_{\mathrm{B}}T$ parameter is set \texttt{set\_kBT()}, whilst the placement method ratio is defined during generation with the \texttt{method\_ratio} dictionary.
        The ratios are renormalised after being read in.
        Two independent searches are performed, and the results combined and presented in \figref{fig:Si-Ge:CHGNet}.
    }
    \label{tab:Si-Ge:CHGNet:params}
    \begin{tabular*}{\textwidth}{@{\extracolsep\fill}c c c c c c c}
        \toprule%
        & \multicolumn{6}{@{}c@{}}{Parameter} \\\cmidrule{2-7}%
        & \multirow{2}{*}{$k_{\mathrm{B}}T$} & \multicolumn{5}{@{}c@{}}{Placement method ratio} \\\cmidrule{3-7}%
        Example & & min & walk & grow & void & rand \\\midrule%
        \multirow{2}{*}{Si\|Ge} & \multirow{2}{*}{0.2} & 1.0 & 1.0 & 0.0 & 1.0 & 0.01 \\%
        &  & 1.0 & 0.3 & 0.0 & 0.3 & 0.01 \\%
        \botrule%
    \end{tabular*}
\end{table}

\bibliography{main}